\documentclass[12pt,nofootinbib]{iopart}
\usepackage{graphicx}
\usepackage{latexsym}
\usepackage{iopams}
\usepackage{wasysym} % for solar mass symbol
\usepackage{color}

\newcommand{\mpcoh}{\,h^{-1}\,{\rm Mpc}}

\newcommand{\lam}{\lambda}

\newcommand{\bfk}{\mbox{\boldmath$k$}}

\newcommand{\gam}{\gamma}
\newcommand{\del}{\delta}
\newcommand{\ve}{\varepsilon}
\newcommand{\al}{\alpha}

\newcommand*{\mpl}{M_{\rm Pl}}

\newcommand{\Pdd}{P_{\delta\delta}}
\newcommand{\Pdv}{P_{\delta \theta}}
\newcommand{\Pvv}{P_{\theta\theta}}
\newcommand{\DFoG}{D_{\rm FoG}}
\newcommand{\sigmav}{\sigma_{\rm v}}

\newcommand{\bfp}{\mbox{\boldmath$p$}}

\def\bea{\begin{eqnarray}}
\def\eea{\end{eqnarray}}
\def\tr{{\rm tr}\,}
\def\ba{\begin{eqnarray}}
\def\ea{\end{eqnarray}}
\def\p{\partial}
\def\2{\sqrt{2}}

\def\mc{\mathcal}

\def\be{\begin{equation}}
\def\ee{\end{equation}}
\def\bea{\begin{eqnarray}}
\def\eea{\end{eqnarray}}
\def\a{\alpha}

\def\k{\kappa}

\def\S{\Sigma}

\def\mK{{\mathcal K}}

\begin{document}

\title{Cosmological Tests of Modified Gravity}

\author{Kazuya Koyama}

\address{$^1$ Institute of Cosmology \& Gravitation, University of Portsmouth, Dennis Sciama Building, Portsmouth, PO1 3FX, United Kingdom}

\ead{Kazuya.Koyama@port.ac.uk}

\begin{abstract}
We review recent progress in the construction of modified gravity models as alternatives to dark energy as well as the development of cosmological tests of gravity. Einstein's theory of General Relativity (GR) has been tested accurately within the local universe i.e. the Solar System, but this leaves the possibility open that it is not a good description of gravity at the largest scales in the Universe. This being said, the standard model of cosmology assumes GR on all scales. In 1998, astronomers made the surprising discovery that the expansion of the Universe is accelerating, not slowing down. This late-time acceleration of the Universe has become the most challenging problem in theoretical physics. Within the framework of GR, the acceleration would originate from an unknown dark energy. Alternatively, it could be that there is no dark energy and GR itself is in error on cosmological scales.
	
In this review, we first give an overview of recent developments in modified gravity theories including $f(R)$ gravity, braneworld gravity, Horndeski theory and massive/bigravity theory. We then focus on common properties these models share, such as screening mechanisms they use to evade the stringent Solar System tests. Once armed with a theoretical knowledge of modified gravity models, we move on to discuss how we can test modifications of gravity on cosmological scales. We present tests of gravity using linear cosmological perturbations and review the latest constraints on deviations from the standard $\Lambda$CDM model. Since screening mechanisms leave distinct signatures in the non-linear structure formation, we also review novel astrophysical tests of gravity using clusters, dwarf galaxies and stars.
	
The last decade has seen a number of new constraints placed on gravity from astrophysical to cosmological scales. Thanks to on-going and future surveys, cosmological tests of gravity will enjoy another, possibly even more, exciting ten years. 
\\
\end{abstract}

%Uncomment for PACS numbers title message
%\pacs{00.00, 20.00, 42.10}
% Keywords required only for MST, PB, PMB, PM, JOA, JOB? 
%\vspace{2pc}
%\noindent{\it Keywords}: Article preparation, IOP journals
% Uncomment for Submitted to journal title message
%\submitto{\JPA}
% Comment out if separate title page not required

\maketitle

\section{Introduction}
\label{sec:introduction}

Einstein's theory of General Relativity (GR) has proven successful over many years of experimental tests \cite{Will:2014xja}. These tests range from millimetre scale tests in the laboratory to Solar System tests and consistency with gravitational wave emission by binary pulsars.
 
The standard model of cosmology assumes GR as the theory to describe gravity on all scales. In 1998, astronomers made the surprising discovery that the expansion of the Universe is accelerating, not slowing down \cite{Riess:1998cb, Perlmutter:1998np}. This discovery was the subject of the 2011 Nobel Prize in physics. This late-time acceleration of the Universe has become the most challenging problem in theoretical physics. Within the framework of GR, the acceleration would originate from an unknown “dark energy”. The simplest option is the cosmological constant, first introduced by Einstein. However, in order to explain the current acceleration of the Universe, the required value of this constant must be incredibly small. Particle physics predicts the existence of vacuum energy which provides a value for the cosmological constant, but this is typically more than 50 orders of magnitude larger than the observed values that assume GR. Alternatively, there could be no dark energy if GR itself is in error on cosmological scales.  

The standard model of cosmology is based on a huge extrapolation of our limited knowledge of gravity. GR has not been tested independently on galactic and cosmological scales. This discovery of the late time acceleration of the Universe may require us to revise the theory of gravity on cosmological scales and the standard model of cosmology based on GR. 

It is extremely timely to tackle this challenge now. Over the next five years, a number of vast astronomical surveys of the galaxy distribution will be underway, such as the Dark Energy Survey (DES, 2012-2017) \cite{DES}, the extended Baryon Oscillation Spectroscopic Survey (eBOSS, 2014-2018) \cite{eBOSS}, and the Mapping Nearby Galaxies at APO (MaNGA, 2014-2018) \cite{MaNGA}. These surveys will dramatically transform our measurements of the cosmic expansion and the large scale structure of the Universe thus providing a new opportunity to test gravity on astrophysical and cosmological scales. Future surveys such as the Euclid mission (2020- )\cite{Euclid}, the Dark Energy Spectroscopic Instrument (DESI, 2018-) \cite{DESI} and the Subaru Measurements of Images and Redshifts (SuMIRe, 2018-) \cite{SuMIRE} will provide an opportunity to perform ultimate tests of gravity on the largest scales in our Universe. 

Recently, there has been significant progress in developing modified gravity theories that act as an alternative to dark energy (see a review \cite{Clifton:2011jh}). Modified gravity models provide interesting theoretical ideas to tackle the cosmological constant problem and explain the late time acceleration of the Universe. One of the challenges for modified gravity models is to satisfy the stringent Solar System constraints whilst modifying gravity significantly on cosmological scales. Screening mechanisms have been developed to hide modifications of gravity on small scales (see a review \cite{Joyce:2014kja}). This research has been further developed into tests of GR itself via cosmological observations (see reviews \cite{Jain:2010ka, Jain:2013wgs}). It is in principle possible to construct non-parametric consistency tests of GR on cosmological scales by combining various probes of large-scale structure, because the Einstein equations enforce a particular relation between observables. Non-linear structure formation is complicated by screening mechanisms and new techniques are required to analyse it. Analytic methods and N-body simulation techniques have been developed to study non-linear clustering of dark matter under their influence. Furthermore, screening mechanisms have also opened up new possibilities to test gravity using astrophysical objects. 

We will review recent developments in models of modified gravity and explain why the realisation of a screening mechanism is important in these models. We then discuss how we test gravity on cosmological scales and emphasise the importance of understanding the non-linear structure formation. Based on the latest N-body simulations, we study the effects of screening mechanisms on the structures in our Universe. Finally we summarise the novel approach to test gravity using astrophysical objects. 

This review is not meant to cover all theoretical models nor observational tests of modified gravity models systematically. Rather, we focus on key ideas behind cosmological tests of gravity whilst emphasising an interplay between theoretical physics, numerical astrophysics, and cosmological and astrophysical observations. For more thorough reviews, we recommend the reader to refer to excellent recent reviews on the subject Refs.~\cite{Clifton:2011jh,Joyce:2014kja, Jain:2010ka,Jain:2013wgs}.  

This review is organised as follows. In section 2, we explain the motivation to consider modifications of GR and why it is difficult to construct such models. Then we introduce several examples of models of this type. In section 3, we explain why a screening mechanism is important for the recently developed modified gravity models and explain how it works to hide the modifications on small scales. In section 4, we discuss cosmological tests of gravity. We explain how one can construct a consistency test of the GR model with a cosmological constant, the $\Lambda$ Cold Dark Matter ($\Lambda$CDM) model and why it is important to develop model independent parametrisations of deviations from $\Lambda$CDM for linear cosmological perturbations. We then explain the complexities that arise from non-linear structure formation. We introduce analytic methods based on perturbation theory to describe quasi non-linear perturbations and emphasise the importance of non-linear physics in extracting information about modifications of gravity on linear scales. We then discuss N-body simulations to study fully non-linear structures. In section 5, we discuss how screening mechanisms operate in dark matter simulations by using two representative screening mechanisms. Section 6 is devoted to novel astrophysical tests of gravity using clusters, galaxies and stars. We discuss the future outlook in section 7.

\section{Modified gravity models}
\subsection{Problems with the standard model of cosmology}
The standard cosmology is highly successful at explaining a number of observations in a simple framework. It is based on two main assumptions:
\begin{itemize} 
\item{Our universe is homogeneous and isotropic.}
\item{Gravity is described by General Relativity (GR).} 
\end{itemize}
We use the first assumption to write down the metric for the background universe as the Friedman-Robertson-Walker (FRW) metric 
\begin{equation}
ds^2 = - dt^2 + a(t)^2 
\Big(\frac{dr^2}{1- K r^2} + r^2 d \Omega^2 \Big),
\end{equation}
and from the Einstein equations we obtain the Friedman equation 
\begin{equation}
H^2 = \left(\frac{\dot{a}}{a}  \right)^2 + \frac{K}{a^2} = \frac{8 \pi G}{3} \rho_m + \frac{\Lambda}{3},
\end{equation}
where $K$ is the curvature of the 3-space and $\Lambda$ is the cosmological constant. By defining the density parameters, 
\begin{equation}
\Omega_m = \frac{ 8 \pi G \rho_{m0}}{3 H_0^2}, \;
\Omega_K = \frac{K}{a_0^2 H_0^2},  \;
\Omega_{\Lambda} = \frac{\Lambda}{3 H_0^2},
\end{equation}
the Friedman equation today can be rewritten as  
\begin{equation}
1= \Omega_m + \Omega_K + \Omega_{\Lambda}. \;
\end{equation}
Here $H_0$ is the present-day Hubble parameter. Many independent datasets intersect on the $\Omega_m$ and $\Omega_{\Lambda}$ plane (see Fig.~1). However, ordinary matter occupies only $5\%$ of the energy density of the Universe. We need to assume that $27\%$ of the energy density of the Universe is made of dark matter and $68\%$ of the energy density is made of the cosmological constant. This surprising result comes from the fact that the expansion of the Universe today is accelerating. The accelerating expansion of the Universe was found by measuring the distance to supernovae (SNe). Due to the accelerated expansion of the Universe, supernovae look dimmer than what we would expect in the Universe without a positive cosmological constant. See Ref.~\cite{Silvestri:2009hh, Linder:2008pp} for reviews. 

%%%%%%%%%%%%%%%%%%%%%%%%%%%%%%%%%%%%%%%%%%%%%%%%%%%%%%%%%%%%
\begin{figure}[h]
  \centering{
  \includegraphics[width=13cm]{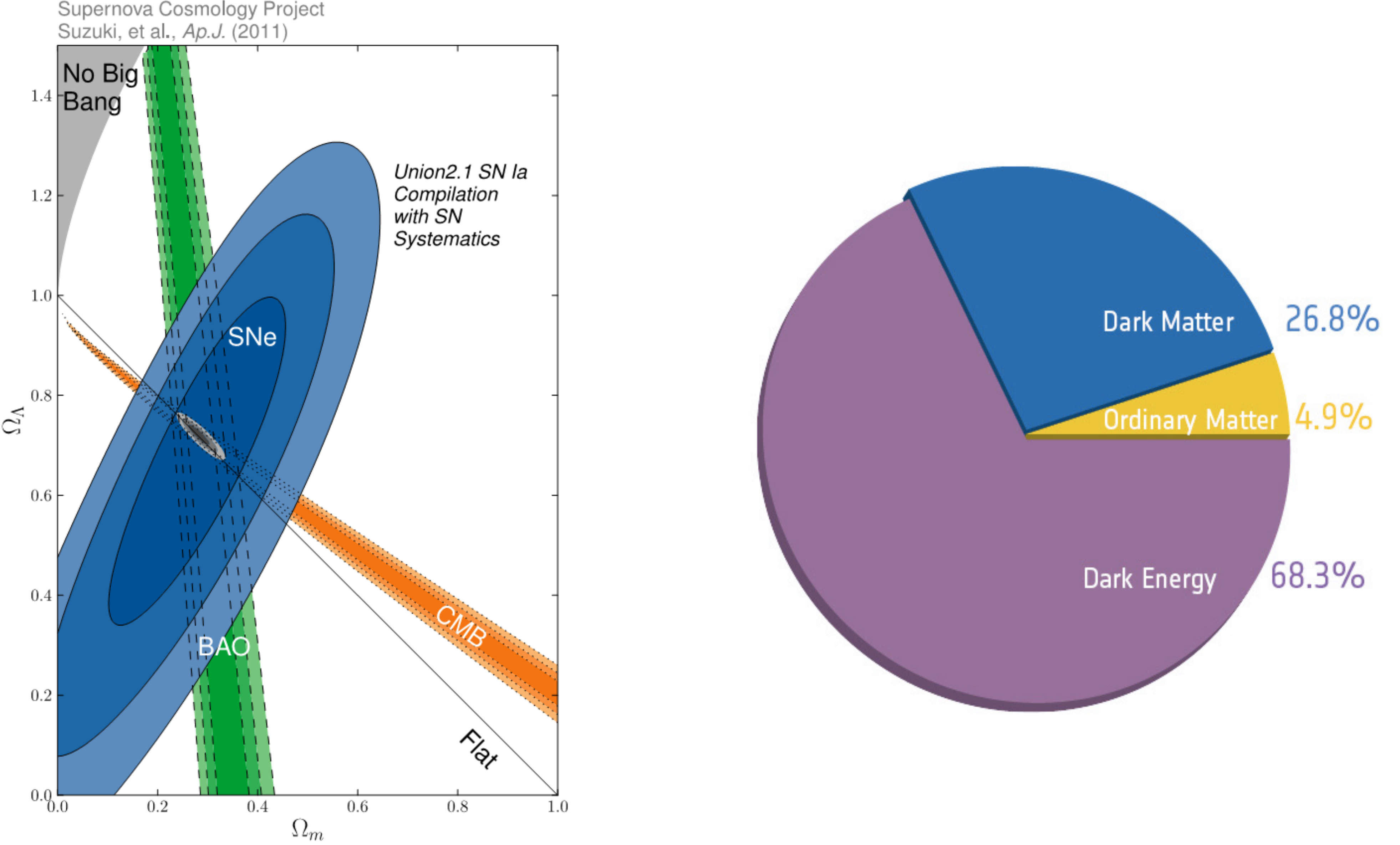}
  }
  \caption{Left: Constraints on $\Omega_{\lambda}$ and $\Omega_m$ from Supernovae (SNe), Cosmic Microwave Background (CMB) and Baryon Acoustic Oscillations (BAO) \cite{Suzuki:2011hu}. From \cite{Suzuki:2011hu, SCP} (published on 10 February 2012 \copyright AAS. Reproduced with permission). 	
  	Right: The relative amounts of the different constituents of the Universe \cite{Ade:2013zuv}. From \cite{Planck} (\copyright ESA and the Planck Collaboration).}
\end{figure}
%%%%%%%%%%%%%%%%%%%%%%%%%%%%%%%%%%%%%%%%%%%%%%%%%%%%%%%%%%%%%

The energy density associated with the cosmological constant is extremely small. Using the fact that the present day Hubble parameter $H_0$ is given by $H_0 =2.13 h \times 10^{-42}$ GeV, where $H_0 = 100 h$(km/s)/Mpc, the energy density associated with the cosmological constant is incredibly small
\begin{equation}
\rho_{\Lambda} = \frac{\Lambda}{8 \pi G}  \sim \Big( 10^{-3} {\mbox eV} \Big)^4. 
\label{Lambdaenegy}
\end{equation}
At the moment we do not have any compelling theory that allows such a small cosmological constant. The cosmological constant problem is arguably the most challenging problems in theoretical physics \cite{Weinberg:1988cp}. Quantum field theory predicts the existence of vacuum energy. Without gravity, this is the zero-point energy and it does not affect the dynamics. However, in GR, the vacuum energy gravitates {and therefore} the value of this vacuum energy has an important meaning. The vacuum energy is determined by the cut-off scale of the theory and this is typically many orders of magnitude larger than the cosmological constant that we need to explain the late time acceleration of the Universe (\ref{Lambdaenegy}). For example, for a massive particle with mass $m$, we expect that the contribution to the vacuum energy is $O(m^4)$. Even the electron $m_e = 0.5$ MeV gives a huge contribution to the vacuum energy. It is possible to add a classical contribution to the cosmological constant in GR. By tuning this piece, it is in principle possible to realise the small cosmological constant that is required for the late time acceleration of the Universe. However, the problem is that this tuning is unstable under quantum corrections. Any new additions to the matter sector or higher order loop contributions spoil the tuning (see a recent review \cite{Padilla:2015aaa}). This cosmological constant problem was already a serious problem before the discovery of the accelerated expansion of the Universe, thus it is sometimes called the {\it old} cosmological constant problem. The discovery of the accelerated expansion of the Universe has worsened the problem. Now not only do we need to explain why we do not see a large cosmological constant but also we require an explanation for why we observe such a tiny cosmological constant. 

Even if we could solve the {\it old} cosmological constant problem, we still need to explain the origin of the accelerated expansion of the Universe. There are mainly two approaches to this problem. One possibility is to abandon the assumption of homogeneity. Although the isotropy of the universe has been proven to high accuracy by the Cosmic Microwave Background (CMB) anisotropies whose temperature is isotropic to roughly one part in $100,000$, the homogeneity of the Universe has not been tested well. If the accelerated expansion of the universe is connected to the inhomogeneity of the Universe, this provides a compelling explanation for the problem known as {\it ``why now"}. If the accelerated expansion of the Universe is driven by the cosmological constant, it is indeed quite a remarkable coincidence that $\Omega_m$ and $\Omega_{\Lambda}$ are roughly the same order {\it today}. The ratio between $\Omega_m$ and $\Omega_{\Lambda}$ is nearly zero until recently and it approaches one rapidly in the future. If the acceleration is driven by the inhomogeneity caused of the growth of structure in our Universe, it can explain this coincidence between our existence and the accelerated expansion of the Universe. Unfortunately, recent studies indicate that it is not possible to explain the accelerated expansion of the Universe from the back-reaction of the structure formation in the Universe. Although, it is still possible to explain the acceleration by large inhomogeneities if they are not related to the structure growth. For example if we are living in a large void, it is in principle possible to explain the observed accelerated expansion of the Universe measured by SNe. However, the consensus is that it is difficult to explain all the observations consistently in these void models. Also this model requires us to abandon the Copernican principle which states that we are not living in a special place in the Universe. 
It is still an open question whether more sophisticated inhomogeneous models can explain the late time acceleration or not. (See a review \cite{Clarkson:2012bg} and references therein). 

The most popular approach is to introduce {\it dark energy} to explain the acceleration within the framework of GR (see a review \cite{Copeland:2006wr}). We can replace the cosmological constant by dark energy with the energy density $\rho_{\rm de}$ and the pressure $P_{\rm de}$. In order to explain the accelerated expansion of the universe, the equation of state $w_{\rm de} = P_{\rm de}/\rho_{\rm de}$ of dark energy needs to be smaller than $-1/3$. The cosmological constant is a special case where $w_{\rm de} = -1$. The most studied candidate for dark energy is a quintessence model in which the scalar field plays the role of dark energy whose energy density and pressure are given by 
\begin{equation}
\rho_{\rm de} = \frac{1}{2} \dot{\phi}^2 + V(\phi), 
\quad 
P_{\rm de} = 
\frac{1}{2} \dot{\phi}^2 - V(\phi),
\end{equation}
where $\phi$ is a scalar field. If the scalar field's kinetic energy is sub-dominant compared to its potential energy, we can realise the negative pressure that is required to explain the accelerated expansion of the Universe. It is also possible to find a tracking solution where the scalar field energy density follows that of matter $\rho_{\rm de}/\rho_{\rm m} =$const. for an exponential potential $V \propto \exp(-\lambda \phi)$. This means that the final density of the dark energy is insensitive to the initial conditions, which could solve the {\it why now} problem. However, for this simple example, the scalar field never dominates the Universe and it cannot explain the accelerated expansion of the Universe. In order to achieve the acceleration, we need to add another exponential potential and the time when dark energy dominates over the matter energy density is controlled by the parameters in the potential. This is exactly the same situation in the $\Lambda$CDM model and we can therefore no longer solve the {\it why now} problem. Another difficulty of the quintessence models is that the parameters in the potential are unstable under radiative corrections. In order for the scalar field to be a successful candidate for dark energy, the mass of the scalar field needs to be of the order $H_0$ and this mass is not protected under quantum corrections. Due to these theoretical difficulties, dark energy is often treated as a fluid with a given equation of state as a function of scale factor $w_{\rm de}(a)$ without specifying its physical origin. 

\subsection{Why modified gravity?}
The cosmological constant problem is of fundamental origin thus it is important to reconsider all the assumptions that we make in the standard model of cosmology. One of the assumptions is that gravity is described by GR on all scales. However we should bear in mind that we tested gravity only in our local universe. There is an interesting example in history of where unexpected observations lead to a revolution in gravitational physics. When the precession of perihelion of Mercury was found, a French mathematician, Urbain Le Verrier hypothesised that there existed a planet called Vulcan between the Sun and Mercury that caused the anomaly. In fact he succeeded in predicting the existence of Neptune using the same technique. Vulcan was never found and we now know that it was Newton's gravity that needed to be revised by Einstein. This story tells us that sometimes we need to be open-minded when we encounter unexpected observations. 

Fig.~2 summarises where we have tested gravity  \cite{Baker:2014zba}. Gravity is parametrised by the two quantities, the gravitation potential $GM/r$ and the curvature of space $GM/r^3$ for a spherical object with mass $M$ and the radius $r$. GR is very well tested in the Solar System and also by binary pulsars. However it is especially not well tested in the low curvature regime. We need dark matter to explain the rotation curves of galaxies, clusters of galaxies and the formation of large scale structure of the Universe. There have been many attempts to explain dark matter by modifications of gravity (see  \cite{Milgrom:2014usa} for a review). In this review, we do not cover these developments. Given the myriad of observations, it is becoming almost impossible to explain all the observed evidence for dark matter by modifications of gravity. In the next decades, progress in direct and indirect detection experiments will shed light on the origin of dark matter (see a review \cite{Drees:2012ji}). In this review, we assume that dark matter is the dominant constituent of the Universe at late times. At the lowest curvature, we require dark energy to explain the accelerated expansion of the Universe. Here only cosmology provides a means to test gravity. 

%%%%%%%%%%%%%%%%%%%%%%%%%%%%%%%%%%%%%%%%%%%%%%%%%%%%%%%%%%%%
\begin{figure}[h]
  \centering{
  \includegraphics[width=15cm]{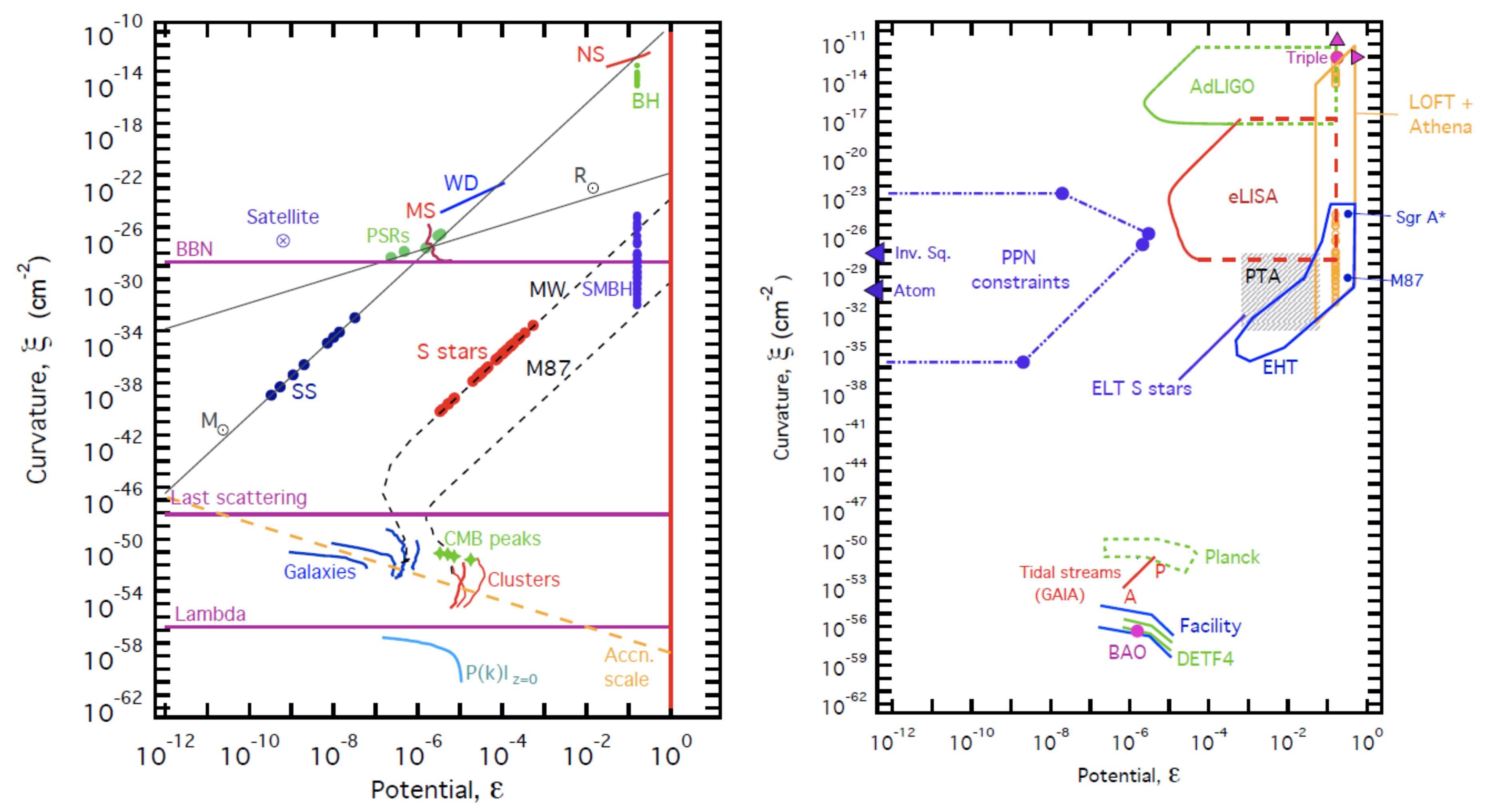}
  }
  \caption{Left: A parameter space for gravitational fields. Right: The experimental version of the parameter space. See Ref.~\cite{Baker:2014zba} for details. The horizontal lines in the left figure indicate the background curvature of the Universe at Big Bang Nucleosynthesis (BBN) and Last scattering, and the curvature associated with $\Lambda$.  Some of the label abbreviations are: SS=planets of the Solar System MS=Main Sequence stars, WD=white dwarfs, PRSs=binary pulsars, NS=Neutron stars, BH= stellar mass black holes, MW=the Milky Way, SMBH=supermassive black holes. PPN= Parameterised Post-Newtonian regime, Inv.Sq.=laboratory tests of the inverse square law of the gravitational force, Atom=atom interferometry experiments, EHT=Event Horizon Telescope, ELT=the Extremely Large Telescope, DEFT4=a hypothetical stage 4 experiment of dark energy, Facility=a futuristic large radio telescope such as the Square Kilometer Array. 
  From~\cite{Baker:2014zba} (published on 20 March 2015 \copyright AAS. Reproduced with permission). 
   }
\end{figure}
%%%%%%%%%%%%%%%%%%%%%%%%%%%%%%%%%%%%%%%%%%%%%%%%%%%%%%%%%%%%%

Modifications to gravity could provide an interesting solution to the cosmological constant problem. The first idea is to modify the way in which gravity responds to the cosmological constant \cite{Dvali:2007kt}. An example is provided by a {\it braneworld} model in 6D (see \cite{Burgess:2004ib, Koyama:2007rx} for reviews). Braneworld theories postulate that we are living in a 4D membrane embedded in a higher dimensional spacetime (see \cite{Maartens:2010ar} for a review). In 6D, there is an interesting property of gravity that causes the cosmological constant on a brane not to gravitate but merely curve the extra two dimensions. Thus even if there is a huge contribution to the cosmological constant in the 4D brane where standard model particles live, the 4D spacetime remains flat. In this way we can avoid the cosmological constant problem. Although it is difficult to find a concrete model that realises this idea in practice, this provides an interesting insight into a possibility that modifications of gravity might address the fundamental problem of the cosmological constant. 

Modified gravity models also provide interesting ideas to explain the late time acceleration of the Universe. One possibility to explain the acceleration is to add a tiny mass to the graviton (see \cite{deRham:2014zqa} for a review). The smallness of the mass can be protected by the restoration of a symmetry. A mass term breaks the diffeomorphism invariance that is present in GR in the massless limit. Due to this restoring of the symmetry in the massless limit, it is {\it technically natural} to have a small mass. More precisely if we compute the quantum loop corrections to the mass, the mass is shifted by $m^2 \to m^2 + O(1) m^2$ because the quantum corrections arise only from the mass term that breaks the symmetry. Thus it is natural to tune the mass to be small to account for the late time acceleration $m^2 \sim H_0^2$ where $H_0$ is the present day Hubble scale as this tuning is not spoiled by the quantum corrections unlike the cosmological constant. 

Another interesting idea is {\it self-acceleration}. Even without the cosmological constant, the expansion of the Universe can accelerate due to the modification of gravity. A typical example is again provided by a braneworld model. In this case, a 4D brane is embedded in a 5D spacetime. Gravity in the 5D bulk spacetime is described by the 5D Einstein gravity \cite{Dvali:2000hr}. On the 4D brane there is an induced gravity described by the 4D Einstein-Hilbert action. Moreover, there is a solution where the expansion of the Universe is determined by the ratio between the 4D and 5D Newton constant $H \propto G_4/G_5$ without the cosmological constant \cite{Deffayet:2001pu}.  

Unfortunately, we still do not have a model that realises these new ideas in a consistent way. However, clearly it is worthwhile challenging GR in order to address the fundamental problems in theoretical physics. Modified gravity models also provided motivation for developing cosmological tests of gravity (see reviews \cite{Jain:2010ka,Jain:2013wgs,Joyce:2014kja}). It is the right time to challenge GR on cosmological scales. As mentioned in the introduction, there will be a number of astronomical surveys aiming to reveal the nature of dark energy. Although it is possible to construct model independent tests of gravity on large scales, it is important to study theoretical models to know what kind of deviations from GR one should search for. Also in order to exploit the vast information available on non-linear scales, we need concrete theoretical models.  

\subsection{Why is it difficult to modify GR?}
We immediately face many problems when we modify GR. There is a powerful theorem known as Lovelock's theorem. Lovelock's theorem proves that Einstein's equations are the only second-order local equations of motion for a metric derivable from the action in 4D. This indicates that if we modify GR, we need to have one or more of these:
\begin{itemize}
\item{Extra degrees of freedom}
\item{Higher derivatives}
\item{Higher dimensional spacetime}
\item{Non-locality}
\end{itemize}
Once we introduce these additional ingredients into the theory beyond GR, we need to check the theoretical consistency of the model. First we need to check that the solutions are stable. There are several kinds of instabilities which can be illustrated by a simple scalar field example described by the action
\begin{equation}
S= \int dt d^3 x ( K_t \dot{\phi}^2 -  K_x (\partial_i \phi)(\partial^i \phi) - m^2 \phi^2).
\end{equation}
The tachyonic instability arises when the scalar field has a negative mass squared $m^2<0$. This instability is not necessarily catastrophic if the instability time scale $|m|^{-1}$ is long enough. Another instability arises when the gradient term has a wrong sign $K_x <0$. This instability is more severe than the tachyonic instability as the instability time scale is determined by the wavenumber of the mode that we are interested in, so the time scale becomes shorter and shorter on small scales. On the other hand, the ghost instability, which arises when the time kinetic term of the scalar field has a wrong sign $K_t <0$, is more severe. At the quantum level, the vacuum is unstable and decays instantaneously. To avoid the instability, it is required to introduce a non-Lorentz-invariant cut-off in the theory (see \cite{Sbisa:2014pzo} for a review).  
 
Another common problem is known as the strong coupling problem. In addition to the kinetic term, the scalar field can have non-linear interaction terms. An example of the non-linear term that appears later is 
\begin{equation}
S_{\rm non-linear}= \int d^4 x \frac{1}{\Lambda_3^3}  \Box \phi (\partial \phi)^2.
\label{strong}
\end{equation} 
This non-linear interaction becomes important at energy scale higher than $\Lambda_3$ and quantum corrections generate terms that are suppressed by $\Lambda_3$. Then in general we loose control of the theory beyond $\Lambda_3$. This strong coupling scale is often associated with the energy scale related to the accelerated expansion of the Universe, $H_0$, which is extremely small compared with the scale of gravity $\mpl$. Thus we will often find that the strong coupling scale is rather low in modified gravity models. This means that we need to treat these theories as an {\it effective theory}, which is valid only at energy scales lower than $\Lambda_3$. 

Once the theory satisfies the requirements for theoretical consistency, it also needs to pass observational tests. First of all, the theory needs to satisfy the stringent Solar System constraints. The {deflection angle $\theta$} of stars due to the Sun is observed to be \cite{Shapiro:2004zz} 
\begin{equation}
\theta = (0.99992 \pm 0.00023) \times 1.75'',
\label{solar1}
\end{equation}
where $1.75''$ is the prediction of GR. Another prediction of GR is time delation due to the effect of the Sun's gravitational field. This was measured very accurately using the Cassini satellite as \cite{Bertotti:2003rm}
\begin{equation}
\Delta t = (1.00001 \pm 0.00001) \Delta t_{GR}.
\label{solar2}
\end{equation}
Any modified theory of gravity needs to satisfy these stringent constraints on deviations from GR in the Solar System. On cosmological scales, observations become better and better and at the moment, the standard $\Lambda$CDM model survives all these improved measurements. In the background, the expansion of the Universe should look very similar to that of the $\Lambda$CDM model. The equation of state for dark energy is constrained to be \cite{Planck:2015xua} 
\begin{equation}
w_{\rm de} = -1.006 \pm 0.045.
\end{equation}
Modified gravity models need to pass all these observational tests. 

\subsection{Examples of models}
In this section, we discuss several representative modified gravity models that have been studied recently. As we mentioned in the introduction, the purpose of this review is not to cover all possible modified gravity models in a systematic way. We only review models that are relevant for later discussions. For a more complete review of various modified gravity models, see Ref~\cite{Clifton:2011jh}.  

\subsubsection{Brans-Dicke gravity}  
Let's first start from the simplest modified gravity model, Brans-Dicke gravity \cite{Brans:1961sx}. This theory includes a scalar field non-minimally coupled to gravity
\begin{equation}
S = \int d^4 x \sqrt{-g} \Big(
\psi R - \frac{\omega_{BD}}{\psi} (\partial \psi)^2 \Big)+ \int d^4 x \sqrt{-g}
{\cal L}_m.
\label{BD}
\end{equation}
We consider a non-relativistic source $T^{0}_0 = -\rho$. Using the quasi-static approximation to ignore time derivatives of perturbations compared with spatial derivatives, the perturbations of the metric 
\begin{equation}
ds^2 = -(1+ 2 \Psi) dt^2 + (1 - 2 \Phi) \delta_{ij} dx^i dx^j,
\end{equation}
and the scalar field perturbation $\psi = \psi_0 + \varphi$ obey the following equations 
\begin{eqnarray}
 \nabla^2 \Psi = 4 \pi G \rho - \frac{1}{2} \nabla^2 \varphi, 
 \label{BDeq1}\\
(3 + 2 \omega_{BD}) \nabla^2 \varphi =- 8 \pi G \rho, \\
  \Phi - \Psi =  \varphi
  \label{BDeq2}.  
\end{eqnarray}
Note that the scalar field that is non-minimally coupled to gravity gives an effective anisotropic stress through its perturbations, modifying the relation between $\Phi$  and $\Psi$. These equations can be rewritten as
\begin{eqnarray}
\nabla^2 \Psi = 4 \pi G \mu \rho, \quad \Psi = \eta^{-1} \Phi,
\label{BD1}
\end{eqnarray}
where 
\begin{equation}
\mu = \frac{4 + 2 \omega_{BD}}{3 + 2 \omega_{BD}}, \quad 
\eta = \frac{1 + \omega_{BD}}{2 + \omega_{BD}}.
\label{BD2}
\end{equation}
We recover GR in the large $\omega_{BD}$ limit. Indeed, imposing the Solar System constraints Eqs.~(\ref{solar1}) and (\ref{solar2}), we obtain $|\eta-1| = (2.1 \pm 2.3) \times 10^{-5}$. Then the constraint on $\omega_{BD}$ is given by $\omega_{BD} > 40,000$. Once we impose this constraint on the parameter of the model $\omega_{BD}$, the theory is basically indistinguishable from GR on all scales. This is one of the main problems of modified gravity models. The Solar System tests are so stringent that if they are imposed on the parameters of the model, there is virtually no room to modify gravity on cosmological scales. 

\subsubsection{$f(R)$ gravity} 
Another popular model is $f(R)$ gravity where the Einstein-Hilbert action is generalised to be a function of the Ricci curvature \cite{Buchdahl:1983zz, Starobinsky:1980te}
\begin{equation}
S= \int d^4 x \sqrt{-g} f(R) +  \int d^4 x \sqrt{-g} {\cal L}_m.
\end{equation}
See \cite{Nojiri:2008nt, Sotiriou:2008rp, DeFelice:2010aj} for a review. The equation of motion for the metric is fourth order thus this model can be classified as a higher derivative theory. However, it is possible to introduce a scalar field and make the equation of motion second order. The action is equivalent to 
\begin{equation}
S = \int d^4 x \sqrt{-g} 
\left(f(\phi) + (R- \phi) f'(\phi) \right).
\end{equation}
By taking a variation with respect to $\phi$, we obtain $(R - \phi) f''(\phi)=0$. As long as $f''(\phi) \neq 0$, and $R= \phi$, we recover the original action. By defining $\psi=f'(\phi)$ and $V=f(\phi) - \phi f'(\phi)$, the action can be written the same as the one for Brans-Dicke gravity with a potential 
\begin{equation}
S = \int d^4 x \sqrt{-g} \Big(\psi R - V(\psi) \Big).
\end{equation}
Comparing this with the action (\ref{BD}), we notice that the BD parameter is given by $\omega_{BD}=0$. As such, if we ignore the potential, this model is already excluded by the Solar System constraints. However, by choosing the potential, i.e. the form of the $f(R)$ function appropriately, it is possible to incorporate a screening mechanism known as the chameleon mechanism to evade the Solar System constraint as we will see later \cite{Khoury:2003aq, Khoury:2003rn, Hu:2007nk, Brax:2008hh}. 

In general, the scalar tensor theory is described by the following action 
\begin{equation}
S = \int d^4 x \sqrt{-g} \Big(\psi R - \frac{\omega_{BD}(\psi)}{\psi} (\partial \psi)^2- V(\psi) \Big) + \int d^4 x \sqrt{-g} {\cal L}_m (g_{\mu \nu} ).
\end{equation}
By a conformal transformation $g_{\mu \nu} =A(\phi)^2 \bar{g}_{\mu \nu} $ and a redefinition of the scalar field, we can transform the action to the Einstein frame
\begin{equation}
S = \int d^4 x \sqrt{-\bar{g}}\Big(R - \frac{1}{2} (\partial \phi)^2- \bar{V}(\phi) \Big)
 + \int d^4 x \sqrt{-g} {\cal L}_m \Big(A(\phi)^2 \bar{g}_{\mu \nu} \Big).
\label{Eframe}
\end{equation}
In this frame, the scalar field is directly coupled to matter.

Like the quintessence models, a choice of the function $f(R)$ leads to many models but the successful models for the late time cosmology share the same features \cite{Hu:2007nk,Starobinsky:2007hu, Appleby:2007vb}. In the high curvature limit, the function looks like 
\begin{equation}
f(R) = R- 2 \Lambda + |f_{R0}| \frac{\bar{R}^{n+1}}{R^{n}},
\label{HS}
\end{equation}
where $\bar{R}$ is the curvature today. This model requires an effective cosmological constant to explain the observed accelerated expansion of the universe. However, it is possible to find a function $f(R)$ so that this constant disappears in the low curvature limit. The correction to the $\Lambda$CDM disappears in the high curvature limit $R \gg \bar{R}$. As we will see later, the Solar System constraint imposes the condition $|f_{R0}| < 10^{-6}$. The background cosmology is indistinguishable from $\Lambda$CDM if $|f_{R0}|$ is small. In the Einstein frame, the potential and the coupling function take the following form 
\begin{equation}
\bar{V} = \Lambda - M^4 \left( \frac{\phi}{\mpl} \right)^{\frac{n}{1+n}}, \;\;
A(\phi)^2 = e^{\sqrt{2/3} \phi/\mpl},
\label{HS-E}
\end{equation}
and the scalar field in the original Jordan frame $\psi= f_R \equiv df/dR$ is related to $\phi$ as 
\begin{equation}
\phi =- \sqrt{\frac{3}{2}} \mpl \log(1 + f_R). 
\label{fr-trans}
\end{equation}
Cosmology of $f(R)$ models have been studied intensively. See references in \cite{Sotiriou:2008rp, DeFelice:2010aj}. 

\subsubsection{Braneworld gravity} 
The idea of braneworld was inspired by the discovery of {\it D-branes} in string theory (see \cite{Maartens:2010ar} for a review). In this model, we are living on a 4D membrane (brane) in a higher dimensional spacetime (bulk). The standard model particles are confined to the brane while gravity can propagate throughout the whole spacetime. The simplest model known as the Dvali-Gabadadze-Porrati (DGP) model is a 5D model described by the action \cite{Dvali:2000hr}
\begin{equation}
S= \frac{M_5^3}{2} \int d^5 x \sqrt{-{}^{(5)}g} {}^{(5)} R 
+ \frac{M_4^2}{2} \int d^4 x \sqrt{-g} \Big( 
{}^{(4)} R + {\cal L}_m \Big), 
\label{nDGP}
\end{equation}
where ${}^{(5)} R$ and ${}^{(4)} R$ are 5D and 4D Ricci curvature respectively. The ratio between the 5D and 4D Newton constant $r_c = M_4^2/2 M_5^3$ is called the cross-over scale and it is a parameter of the model. This model provides an interesting example of {\it self-acceleration}. The Friedman equation in this model is given by \cite{Deffayet:2001pu}
\begin{equation}
H^2 =\pm \frac{H}{r_c} + \frac{8 \pi G}{3} \rho.
\end{equation}
At early times $H r_c \gg 1$, we recover the usual 4D Friedmann equation. On the other hand, at late times, the Hubble parameter approaches a constant $H \to 1/r_c$ in the upper branch of the solution. Thus the expansion of the Universe accelerates without the cosmological constant. This is known as the self-accelerating branch. On the other hand, the lower branch solution, often called the normal branch solution requires the cosmological constant to realise the accelerated expansion of the Universe. In order to recover the standard cosmology at early times, the cross-over scale needs to be tuned as $r_c \sim H_0^{-1}$ so that the modification of gravity appears only at late times. See references in \cite{Maartens:2010ar} for cosmological studies of the model. 

If we study the perturbations around this background under the quasi-static approximations, we find that gravity is described by the BD theory with an additional non-linear interaction term \cite{Lue:2004rj, Lue:2002sw, Koyama:2007ih} 
\begin{eqnarray}
 \nabla^2 \Psi = 4 \pi G a^2 \rho - \frac{1}{2} \nabla^2 \varphi, \\
(3 + 2 \omega_{BD}(a)) \nabla^2 \varphi 
+ \frac{r_c^2}{a^2} \Big[ 
(\nabla^2 \varphi)^2 - (\nabla_i \nabla_j \varphi)^2
  \Big] 
=- 8 \pi G a^2 \rho, 
\label{DGP}
\\
  \Phi - \Psi =  \varphi,  
\end{eqnarray}
where the BD parameter $\omega_{BD}(a)$ is given by 
\begin{equation}
\omega_{BD} = \frac{3}{2} (\beta(a)-1), \quad 
\beta(a)= 1- 2 r_c H \Big( 1 + \frac{\dot{H}}{3 H^2}\Big).
\end{equation}
The scalar field originates from the bending of the brane in the 5D bulk spacetime. Note that the BD parameter is always negative in the self-accelerated branch and the scalar field is mediating a repulsive force. This is the manifestation of the problem that this solution suffers from a ghost instability \cite{Luty:2003vm, Nicolis:2008in, Koyama:2005tx, Gorbunov:2005zk, Koyama:2007zz, Charmousis:2006pn}. In addition, at late times $\omega_{BD} \sim O(1)$ thus by linearising the equation, we again find that this theory would be excluded by the Solar System constraints. However, the coefficient of the non-linear interaction term is very large, $r_c^2 \sim H_0^{-2}$. This non-linear term becomes important even if gravity is weak. This is responsible for the screening mechanism known as the Vainshtein mechanism as we will see later. Interestingly, the Solar System constraints impose the constraint $r_c > 0.1 H_0^{-1}$. This constraint is independent of the requirement to recover standard cosmology at early times $r_c \sim H_0^{-1}$. Thus any improvement on this constraint will give severe constraints on the model as we will see in section 3.5. 

An interesting extension of the DGP model is provided by 6D models. As we explained in section 2.2, 6D spacetime has the property that the cosmological constant on a 4D brane does not gravitate.  However, the singular structure of the brane is far more complicated compared with the 5D case. There have been many attempts to extend the 5D DGP to 6D \cite{Dubovsky:2002jm, Kaloper:2007ap,Kaloper:2007qh,  Corradini:2007cz, Corradini:2008tu, deRham:2008zz, deRham:2007rw, deRham:2009wb, Agarwal:2009gy, deRham:2010rw, Minamitsuji:2008fz, Sbisa':2014uza, Sbisa:2014vva}. 

\subsection{Galileons, Horndeski theory and beyond}
The non-linear interaction that appears in the DGP braneworld model has a very special property \cite{Luty:2003vm, Nicolis:2004qq}. The equation of motion in the flat spacetime can be derived from the action (\ref{strong}). It looks like this action leads to an equation of motion that contains higher order derivatives, but this is not the case as can been seen from the fact that the equation of motion (\ref{DGP}) contains only second order derivatives. Furthermore,this action has the following symmetry in field space
\begin{equation}
\partial_{\mu} \phi \to \partial_{\mu} \phi + c_{\mu},
\label{gal-sym}
\end{equation}
where $c_{\mu}$ is a constant vector. Due to its similarity with the Galilean symmetry in Newtonian gravity, the scalar field described by this action is called  the {\it galileon} \cite{Nicolis:2008in}. Interestingly, in 4D spacetime, there are only three Lagrangians, in addition to the canonical kinetic term, that lead to second order equations of motion with the galileon symmetry;
\begin{eqnarray}
{\cal L}_3^{\rm gal}=-\frac{1}{2}(\partial \phi)^2 \Box \phi, \\
{\cal L}_4^{\rm gal} =-\frac{1}{2} (\partial \phi)^2 \left[(\Box\phi)^2-(\partial_\mu\partial_\nu\phi)^2\right],  \\
{\cal L}_5^{\rm gal} = -\frac{1}{4}(\partial \phi)^2 
\bigl[(\Box\phi)^3 -3\Box\phi(\partial_\mu\partial_\nu\phi)^2
+2(\partial_\mu\partial_\nu\phi)^3 \bigr].
\label{galileon}
\end{eqnarray}
There are many other expressions for the same action that are related by integrations by parts. The most convenient form to understand why the equation of motion contains only second derivatives can be constructed by making use of the $\mathit{Levi-Civita}$ epsilon tensor to write the Lagrangian for the galileons in a compact form.  
Using the following property:
\begin{equation}
\ve_{\gam_1 \dots \gam_{4-n} \al_{1}\dots \al_{n}}\ve^{\gam_1 \dots \gam_{4-n} \beta_{1}\dots \beta_n} = -(4-n)!\,n!\,\delta^{[\beta_{1}\dots \beta_n]}_{\al_{1}\dots\al_n},
\end{equation}
where the square brackets represent normalised anti-symmetric permutations, we can write the galileon Lagrangians as:
\begin{eqnarray}
\bar{\mc{L}}_3^{\rm gal} & =& \frac{1}{2!}\ve^{\mu_1 \mu_3 \nu \lam}\ve^{\mu_2 \mu_4}_{\,\,\,\,\,\,\,\,\,\,\,\, \nu \lam} \phi_{\mu_1}\phi_{\mu_2}(\phi_{\mu_3\mu_4}) := \mc{E}_{(4)}\phi_1 \phi_2 (\phi_{34}), \nonumber\\ 
\bar{\mc{L}}_4^{\rm gal} & =&  \ve^{\mu_1 \mu_3 \mu_5 \nu} \ve^{\mu_2 \mu_4 \mu_6}_{\,\,\,\,\,\,\,\,\,\,\,\,\,\,\,\,\,\,\, \nu} \phi_{\mu_1}\phi_{\mu_2}(\phi_{\mu_3\mu_4}\phi_{\mu_5 \mu_6}) := \mc{E}_{(6)}\phi_1 \phi_2 (\phi_{34}\phi_{56}), \nonumber\\ 
\bar{\mc{L}}_5^{\rm gal} & =&  \ve^{\mu_1 \mu_3 \mu_5 \mu_7} \ve^{\mu_2 \mu_4 \mu_6 \mu_8}\phi_{\mu_1}\phi_{\mu_2}(\phi_{\mu_3\mu_4}\phi_{\mu_5 \mu_6}\phi_{\mu_7 \mu_8}) := \mc{E}_{(8)}\phi_1 \phi_2(\phi_{34}\phi_{56}\phi_{78}),\nonumber\\
\label{galileon-as}
\end{eqnarray} 
Where we have defined $\mathcal{E}^{1234 \dots}_{2n} = \frac{1}{(4-n)!}\ve^{135 \ldots \nu_1\nu_2 \ldots \nu_{4-n}}\ve^{246 \ldots}_{\,\,\,\,\,\,\,\,\,\,\,\,\,\,\,\nu_1\nu_2 \ldots \nu_{4-n}}$ which has been written in short hand as $ \mc{E}_{(2n)}$ and the numbers are short hand for labeled indices: $\{\mu_1 \mu_2 \dots\}$. Furthermore, we have that $\phi_{\mu_1\dots\mu_n} \equiv \p_{\mu_n}\dots \p_{\mu_1}\phi$.\\ 
With this notation it is very easy to see that the variation of these Lagrangians would never have higher than two derivatives. For instance, taking the variation of $\mc{L}_5$ gives us: 
\begin{eqnarray}
0 &=& \,\del \mc{S}_5 = \int \mathrm{d}^4 x \,\del \bar{\mc{L}}_5^{\rm gal} \nonumber \\ &=& \int \mathrm{d}^4 x \,\mc{E}_{(8)}\Big[ 2\del\phi_1\phi_2(\phi_{34}\phi_{56}\phi_{78}) + 3\phi_1\phi_2 (\del\phi_{34}\phi_{56}\phi_{78})\Big] \nonumber \\ &=& \int \mathrm{d}^4 x \, \mc{E}_{(8)} \Big[-2\p_1\big(\phi_{2}\phi_{34}\phi_{56}\phi_{78}\big)-3\p_3\p_4\big(\phi \phi_{12} \phi_{56}\phi_{78}\big)\Big]\del\phi \nonumber \\ &=& -5 \int \mathrm{d}^4 x \,  \mc{E}_{(8)}(\phi_{12}\phi_{34}\phi_{56}\phi_{78}).
\end{eqnarray}
Where we have integrated by parts and found that the only term to survive the summation with the totally antisymmetric tensor $\mc{E}_{(8)}$ has, indeed, only derivatives of second order. 

As shown by Ref.~\cite{Nicolis:2008in}, these galileon interactions can be constructed as the short-distance limit of a conformally invariant scalar field Lagrangian. Due to the conformal invariance, we generally expect that the maximally symmetric solutions including the self-accelerating de Sitter solution to describe the vacuum. These conformal invariant theories can be written in terms of a particular combination of the curvature invariants built out of an effective metric $g_{\mu \nu} = e^{2 \phi} \eta_{\mu \nu}$ \cite{Nicolis:2008in, Creminelli:2013fxa}. These galileon interaction terms can be also constructed from the action of a probe brane floating in a higher dimensional spacetime in the limit in which only a scalar field degree of freedom remains \cite{deRham:2010eu}. If we consider a probe brane in a 5D Anti-de Sitter spacetime, the resultant galileon action was shown to be equivalent to the conformal invariant theories constructed from $g_{\mu \nu} = e^{2 \phi} \eta_{\mu \nu}$ \cite{Creminelli:2013fxa}. Also it is possible to construct galileon interactions for multiple scalar fields. See \cite{Trodden:2011xh, Trodden:2015qua, deRham:2012az} for reviews and references therein. The galileon interactions can be also extended to vectors \cite{Tasinato:2014eka, Heisenberg:2014rta, Hull:2014bga}. 

In curved space, if we covariantise the action naively by replacing the partial derivatives by covariant derivatives, the equation of motion now contains the third derivative of the metric for the quartic (${\cal L}_4$) and quintic (${\cal L}_5$) galileons \cite{Deffayet:2009mn}. In general, higher derivative theories lead to the so-called Ostragradsky ghost. $f(R)$ gravity avoids this instability as the equation of motion can be rewritten in the second order form by introducing a scalar field. In fact, it has been claimed that in the case of a naive covariantisation of Eq.~(\ref{galileon-as}), the equations of motion can be written in the second order form despite the appearance of the higher derivative terms due to a hidden constraint \cite{Gleyzes:2014dya} (see also \cite{Zumalacarregui:2013pma, Bettoni:2015wta}). Here we first discuss a way to remove the third derivatives by introducing a counter term. The covariant action that leads to the second order equations of motion for metric and the scalar field is given by \cite{Deffayet:2009wt}
\begin{eqnarray}
{\cal L}_3 = -\frac{1}{2}(\nabla \phi)^2 \Box \phi, 
\label{covgalileon}
\\
{\cal L}_4 = \frac{1}{8} (\nabla \phi)^4 R - \frac{1}{2}(\nabla \phi)^2
\left[(\Box\phi)^2-(\nabla_\mu\nabla_\nu\phi)^2\right], \\
{\cal L}_5 = -\frac{3}{8} (\nabla \phi)^4 G^{\mu\nu} \nabla_\mu\nabla_\nu\phi
\nonumber\\
\quad \quad - \frac{1}{4}  (\nabla \phi)^2
\bigl[(\Box\phi)^3 -3\Box\phi(\nabla_\mu\nabla_\nu\phi)^2
+2(\nabla_\mu\nabla_\nu\phi)^3 \bigr].
\end{eqnarray}  
The non-minimal coupling terms for ${\cal L}_4$ and ${\cal L}_5$ are the counter terms that are necessary to remove higher derivatives in the equations of motion. 

These actions can be generalised further, leading to the Horndeski action \cite{Deffayet:2009mn}:
\begin{eqnarray}
{\cal L}_2 = K(\phi, X), \\
{\cal L}_3 =-G_3(\phi, X)\Box\phi, \\
{\cal L}_4 = G_4(\phi, X)R
+G_{4X}(\phi, X)\left[(\Box\phi)^2-(\nabla_\mu\nabla_\nu\phi)^2\right],
\\
{\cal L}_5 =
G_5(\phi, X)G^{\mu\nu}\nabla_\mu\nabla_\nu\phi \nonumber \\
\quad \quad -\frac{1}{6}G_{5X}(\phi, X)\bigl[(\Box\phi)^3
-3\Box\phi(\nabla_\mu\nabla_\nu\phi)^2
+2(\nabla_\mu\nabla_\nu\phi)^3\bigr], 
\label{Horndeski}
\end{eqnarray}
where $X = - \partial^\mu \partial_{\mu} \phi/2 $ and $K, G_3, G_4$ and $G_5$ are free function of the scalar field and $X$. This action was originally found by Horndeski in 1974 \cite{Horndeski:1974wa}. The Horndeski action was revisited by Ref.~\cite{Charmousis:2011bf} and it was shown to be equivalent to the action (\ref{Horndeski}) by Ref.~\cite{Kobayashi:2011nu}. 

Expanding the metric and scalar field around the Minkowski background, we obtain the galileon interactions. In addition, there appear additional couplings between tensors and scalars due the non-minimal coupling \cite{Narikawa:2013pjr, Koyama:2013paa}. 
We consider a Minkowski background with a constant scalar field $\phi=\phi_0$ and study the deviations around it, namely
\begin{equation}
\phi = \phi_0 + \pi, \qquad g_{\mu \nu} = \eta_{\mu \nu} + h_{\mu \nu}\,.\label{pertcon}
\end{equation}
Notice that for the background to be a solution, one needs $K=dK/d\phi =0$ at $\phi=\phi_0$. We expand the Horndeski action (\ref{Horndeski}) in terms of the fluctuations (\ref{pertcon}), using the following assumptions: the fields $\pi$ and $h_{\mu \nu}$ are small, hence we neglect higher order interactions containing the tensor perturbations $h_{\mu \nu}$, as well as terms containing
higher order powers of the scalar fluctuation $\pi$ and its first derivatives only. On the other hand, we keep all terms with second order derivatives of $\pi$ that preserve the galileon symmetry (\ref{gal-sym}). The kinetic term for metric perturbations is contained in $G_4(\phi, x) R$. Thus we introduce the Planck scale as $G_4= \mpl^2/2$ and canonically normalise $h_{\mu \nu}$ as $\bar{h}_{\mu\nu} = \mpl h_{\mu\nu}$. We further introduce a new mass dimension $\Lambda_3$ so that terms involving second order derivatives of $\pi$ have the appropriate dimensions. By explicitly expanding the Horndeski action (\ref{Horndeski}) under these prescriptions, we find \cite{Koyama:2013paa}
\begin{eqnarray}\label{decaction}
{\cal L}^{\rm eff} &=& - \frac{1}{4} \bar{h}^{\mu \nu} {\cal E}^{\alpha \beta}_{\;\;\;\;\; \mu \nu} \bar{h}_{\alpha \beta} + \frac{\eta}{2} \pi \Box \pi  
+ \frac{\mu}{\Lambda_3^3}{\cal L}^{\rm gal}_3
+ \frac{\nu}{\Lambda_3^6} {\cal L}^{\rm gal}_4 + \frac{\varpi}{\Lambda^9} {\cal L}^{\rm gal}_5 \nonumber\\
&-&  \xi \bar{h}^{\mu \nu} X^{(1)}_{\mu \nu} - \frac{1}{\Lambda^3} \alpha \bar{h}^{\mu \nu}
X^{(2)}_{\mu \nu} + \frac{1}{2 \Lambda^6} \beta \bar{h}^{\mu \nu} X^{(3)}_{\mu \nu}
+ \frac{1}{2 \mpl} \bar{h}^{\mu \nu} T_{\mu \nu},
\end{eqnarray}
where ${\cal L}^{\rm gal}$ are the galileon terms (\ref{galileon})
and the tensor-scalar couplings $X_{\mu \nu}^{(i)}$ are given by
\begin{eqnarray}
X^{(1)}_{\mu \nu} &=& \eta_{\mu \nu} [\Pi] - \Pi_{\mu \nu}, \nonumber\\
X^{(2)}_{\mu \nu} &=& \Pi^2_{\mu \nu} - [\Pi] \Pi_{\mu \nu} +\frac{1}{2} \eta_{\mu \nu}
([\Pi]^2 - [\Pi^2]), \nonumber\\
X^{(3)}_{\mu \nu} &=& 6 \Pi^3_{\mu \nu} - 6 \Pi^2_{\mu \nu} [\Pi] + 3 \Pi_{\mu \nu}
([\Pi]^2 - [\Pi^2]) \nonumber\\
&&  - \eta_{\mu \nu}([\Pi]^3 - 3 [\Pi][\Pi^2]+2 [\Pi^3]),
\end{eqnarray}
where the shorthand notations $\Pi_{\mu \nu}= \partial_{\mu} \partial_{\nu} \pi$, $\Pi^n_{\mu \nu}= \Pi_{\mu \alpha} \Pi^{\alpha \beta}...\Pi_{\lambda \nu}$ and
$[\Pi^n] = \Pi^{n \; \mu}_{\;\;\;\;\;\;\mu}$ are introduced and ${\cal E}^{\alpha \beta}_{\;\;\;\;\; \mu \nu} \bar{h}_{\alpha \beta}$ is the linearised Einstein tensor. We define seven dimensionless parameters $\xi, \eta, \mu, \nu, \varpi,\alpha,\beta$ as follows \cite{Narikawa:2013pjr}
\begin{eqnarray}
G_{4\phi} =\mpl \xi, \quad
K_X-2G_{3\phi} =\eta, 
-G_{3X}+3G_{4\phi X} = \frac{\mu}{\Lambda_3^3}, \nonumber\\
G_{4X}-G_{5\phi} =\frac{\mpl}{\Lambda_3^3}\alpha,
\quad 
G_{4XX}-\frac{2}{3}G_{5\phi X}=\frac{\nu}{\Lambda_3^6}, \nonumber\\
G_{5X}=-\frac{3\mpl}{\Lambda_3^6}\beta, 
G_{5XX} = - \frac{3 \varpi}{\Lambda_3^9},
\end{eqnarray}
where all functions are evaluated at the background, $\phi=\phi_0$ and $X=0$, and $G_{3 \phi}=d G_3/d \phi, G_{3X} = dG_3/dX$ etc.  

The above action respects the galileon symmetry (\ref{gal-sym}). The coupling between the metric and scalar perturbations $\bar{h}^{\mu \nu} X^{(1)}_{\mu \nu}$ and $\bar{h}^{\mu \nu} X^{(2)}_{\mu \nu}$ can be eliminated by the local field redefinition \cite{deRham:2010tw}
\begin{equation}
\bar{h}_{\mu \nu} = \hat{h}_{\mu \nu} - 2 \xi  \pi \eta_{\mu\nu}
+ \frac{2 \alpha}{\Lambda_3^3} \partial_{\mu} \pi \partial_{\nu} \pi,
\label{trans}
\end{equation}
so that the action (\ref{decaction}) now becomes
\begin{eqnarray}
{\cal L}^{\rm eff} = - \frac{1}{4} \hat{h}^{\mu \nu} {\cal E}^{\alpha \beta}_{\;\;\;\;\; \mu \nu} \hat{h}_{\alpha \beta} +
\frac{\eta + 6 \xi^2}{2} \pi \Box \pi \nonumber\\
+ \frac{\mu + 6 \alpha \xi}{\Lambda_3^3}{\cal L}^{\rm gal}_3
+
\frac{\nu+2 \alpha^2+4 \beta \xi}{\Lambda_3^6} {\cal L}^{\rm gal}_4 
+ \frac{\varpi+ 10 \alpha \beta}{\Lambda_3^9} {\cal L}^{\rm gal}_5 \nonumber\\
+  \frac{1}{2 \Lambda_3^6} \beta \hat{h}^{\mu \nu} X^{(3)}_{\mu \nu}
+ \frac{1}{2 \mpl} \hat{h}^{\mu \nu} T_{\mu \nu} 
- \frac{2 \xi}{\mpl} \pi T + \frac{2 \alpha}{ \mpl \Lambda_3^3}
\partial_{\mu} \pi \partial_{\nu} \pi T^{\mu \nu}. 
\label{decoupling}
\end{eqnarray}
However, the coupling $\hat{h}^{\mu \nu} X^{(3)}_{\mu \nu}$ cannot be removed by a local field redefinition \cite{deRham:2010tw}. The transformation (\ref{trans}) introduces a coupling between $\pi$ and $T_{\mu\nu}$ via the de-mixing of scalar from gravity. This provides the effective action that describes the scalar-tensor interactions that satisfies the galileon symmetry. 

Although we started from the Horndeski action, the non-linear completion of this effective action is not necessarily the Horndeski action. As we have already seen, the cubic galileon term appears in the decoupling limit of the 5D braneworld model. Massive gravity is another example with a decoupling limit described by this effective action, as we will see later. Another simple example is given by an extension of the Horndeski action called {\it beyond Horndeski} \cite{Gleyzes:2014dya}. 

In order to introduce beyond Horndeski gravity, first we formulate the Horndeski theory in the unitary gauge \cite{Gleyzes:2014dya, Gleyzes:2014qga}. In the unitary gauge, the scalar field is only the function of time $\phi=\phi(t)$. We perform the 3+1 ADM decomposition of space-time described by the line element
\begin{equation}
ds^{2}=g_{\mu \nu }dx^{\mu }dx^{\nu}
=-N^{2}dt^{2}+h_{ij}(dx^{i}+N^{i}dt)(dx^{j}+N^{j}dt)\,,  
\label{ADMmetric}
\end{equation}
We introduce the extrinsic curvature defined by  
\begin{equation}
K_{\mu \nu}=h^{\lambda}_{\mu} n_{\nu;\lambda},
\end{equation} 
where $n_{\mu}=(-N,0,0,0)$ is a unit vector orthogonal to 
the constant $t$ hyper-surfaces $\Sigma_t$ and a semicolon represents a covariant derivative. We also introduce the three-dimensional Ricci tensor 
${\cal R}_{\mu \nu}={}^{(3)}R_{\mu \nu}$ on $\Sigma_t$
Then we can construct a number of geometric scalar quantities:
\begin{equation}
K \equiv {K^{\mu}}_{\mu},\quad
{\cal S} \equiv K_{\mu \nu} K^{\mu \nu}, \quad
{\cal R} \equiv
{{\cal R}^{\mu}}_{\mu},\quad
\mathcal{R}^{\mu \nu}, \quad
{\cal U} \equiv {\cal R}_{\mu \nu} K^{\mu \nu}\,. 
\label{ADMscalar}
\end{equation}
The Lagrangian that leads to second order equations of motion can be expressed in terms of the geometric scalars introduced above as 
 \cite{Gleyzes:2014dya, Gleyzes:2014qga}
\begin{eqnarray}
L =
A_2(N,t)+A_3(N,t)K+A_4(N,t) (K^2-{\cal S}) 
+B_4(N,t){\cal R} \nonumber\\ 
+ A_5(N,t) K_3 
+B_5(N,t) \left( {\cal U}-K {\cal R}/2 \right)\,,
\label{LGLPV}
\end{eqnarray}
where $K_3 \equiv K^3-3KK_{\mu \nu}K^{\mu \nu}
+2K_{\mu \nu}K^{\mu \lambda}{K^{\nu}}_{\lambda}$. 

By expressing the Horndeski action using the geometric scalars in the unitary gauge, we can find the relation between the coefficients $G_i$ in Eq.~(\ref{Horndeski}) and $A_i,B_i$ in Eq.~(\ref{LGLPV})
\begin{eqnarray}
& & A_2=G_2-XF_{3,\phi}\,,\qquad
A_3=2(-X)^{3/2}F_{3,X}-2\sqrt{-X}G_{4,\phi}\,,\label{A3}
\nonumber \\
& & A_4=-G_4+2XG_{4,X}+XG_{5,\phi}/2\,,\qquad
B_4=G_4+X(G_{5,\phi}-F_{5,\phi})/2\,,\label{B4} 
\nonumber \\
& & A_5=-(-X)^{3/2}G_{5,X}/3\,,\qquad
B_5=-\sqrt{-X}F_{5}\,,\label{B5}
\label{AB}
\end{eqnarray}
where $F_3$ and $F_5$ are auxiliary functions obeying
the relations $G_3=F_3+2XF_{3,X}$ and $G_{5,X}=F_5/(2X)+F_{5,X}$. 
Since $X=-\dot{\phi}^2(t)/N^2$ in unitary gauge, 
the functional dependence of $\phi$ and $X$ can translate to 
that of $t$ and $N$.

The Horndeski theories satisfy the following two conditions 
\begin{equation}
A_4=2XB_{4,X}-B_4\,,\qquad
A_5=-\frac13 XB_{5,X}\,.
\label{ABcon}
\end{equation}
It is possible to go beyond the Horndeski theory without imposing
the two conditions (\ref{ABcon}) \cite{Gleyzes:2014dya}.
This generally gives rise to derivatives higher than second order, but 
it does not necessarily mean that an extra propagating degree 
of freedom is present \cite{Deffayet:2015qwa, Langlois:2015cwa}. Interesting, if we naively covariantise the galileon action of the form given by Eq.~(\ref{galileon-as}), this is included in the beyond Horndeski theory. 

The Horndeski theory does not specify how matter couples to gravity. Another possibility for {\it beyond Horndeski} theories is to consider a metric
\begin{equation}
\bar{g}_{\mu \nu} = \Omega^2(\phi, X) g_{\mu \nu} + \Gamma(\phi, X)
\partial_{\mu} \phi \partial_{\nu} \phi,
\end{equation}
and couple matter to gravity using $\bar{g}_{\mu \nu}$ \cite{Bekenstein:1992pj,Zumalacarregui:2013pma,Bettoni:2015wta}. The disformal coupling without the kinetic term dependence $\Gamma(\phi)$ has attracted a lot of interest recently (see for example \cite{Sakstein:2014isa} and references therein) whereas the kinetically dependent disformal coupling has been used to show the connection between beyond Horndeski theories given by Eq.~(\ref{LGLPV}) and the Horndeski theory \cite{Gleyzes:2014qga}.   

Given the number of free functions in these theories it is difficult to discuss cosmology in a general way. One interesting feature of this model is that it is possible to explain the accelerated expansion of the Universe without introducing a potential. A simple example is given by the Horndeski theory with 
\cite{Deffayet:2010qz, Kimura:2010di}
\begin{equation}
K(X) =- X, \quad G_3(X)=  \left(\frac{r_c^2}{\mpl^2}  X \right)^n. 
\end{equation} 
Note that the sign for the kinetic term is opposite to the normal scalar field. However, this does not imply that the perturbations are ghostly when expanding around the non-trivial cosmological background. This theory admits the tracker solution for the scalar field given by
\begin{equation}
\dot{\phi} = \frac{1}{3 G_{X} H},
\end{equation}
and this leads to the Friedmann equation of the form 
\begin{equation}
\left( \frac{H}{H_0} \right)^2 
= (1- \Omega_m) \left( \frac{H}{H_0} \right)^{-\frac{2}{2n-1}} 
+ \Omega_m a^{-3}.
\end{equation}
This Friedman equation is similar to the one that we found in the DGP braneworld which realised self-acceleration. The covariant galileon model described by Eq.~(\ref{covgalileon}) corresponds to $n=1$. The perturbations around this solution is stable despite the wrong sign for the kinetic term thanks to the cubic galileon term. See Refs.~\cite{Chow:2009fm, Silva:2009km, Kobayashi:2009wr, Kobayashi:2010wa, DeFelice:2010pv, Gannouji:2010au, DeFelice:2010jn, DeFelice:2010gb, Mota:2010bs, DeFelice:2010as, Nesseris:2010pc, Deffayet:2010qz, Ali:2010gr, DeFelice:2010nf,  Pujolas:2011he, DeFelice:2011bh, Hirano:2011wj,deRham:2011by, DeFelice:2011aa, Barreira:2012kk, Appleby:2012rx, Barreira:2013jma, Kase:2014yya, DeFelice:2015isa} for an incomplete list of studies of cosmology in this class of models. Galileons have also been used to construct general models of inflation. See references in \cite{Joyce:2014kja, deRham:2012az, Regan:2014hea}. 

\subsection{Massive gravity/bigravity}
Another theory that is closely related to galileons is massive gravity (see \cite{deRham:2014zqa} for a review). This section is based largely on a review  Ref.~\cite{Tasinato:2013rza}. Can the graviton have a mass? Attempts to answer this question date back to the work by Fierz and Pauli (FP) in 1939 \cite{Fierz:1939ix}. They considered a mass term for linear gravitational perturbations, which is uniquely determined by requiring the absence of ghost degrees of freedom. The mass term breaks the gauge invariance of GR, leading to a graviton with five degrees of freedom instead of the two found in GR. There have been intensive studies into what happens beyond the linearised theory of FP. In 1972, Boulware and Deser found a scalar ghost mode at the non-linear level, the so called sixth degree of freedom in the FP theory \cite{Boulware:1973my}. This issue has been re-examined using an effective field theory approach \cite{ArkaniHamed:2002sp}. They introduced scalar fields known as St\"uckelberg fields to restore gauge invariance so that the theory is manifestly convariant. In this formulation, the St\"uckelberg fields describe the additional degrees of freedom in massive graviton. 
They acquire non-linear interactions containing more than two time derivatives, signalling the existence of a ghost. In order to construct a consistent theory, non-linear terms should be added to the FP model, which are tuned so that they remove the ghost order by order in perturbation theory.

Interestingly, this approach sheds light on another famous problem with FP massive gravity; due to contributions from the scalar degree of freedom, solutions in the FP model do not continuously connect to solutions in GR, even in the limit of zero graviton mass. This is known as the van Dam, Veltman, and Zakharov (vDVZ) discontinuity \cite{vanDam:1970vg, Zakharov:1970cc}. Observations such as light bending in the Solar System would exclude the FP theory, no matter how small
the graviton mass is. In 1972, Vainshtein \cite{Vainshtein:1972sx} proposed a mechanism to avoid this conclusion; in the small mass limit, the scalar degree of freedom becomes strongly coupled and the linearised FP theory is no longer reliable. In this regime, higher order interactions, which are introduced to remove the ghost degree of freedom, should shield the scalar interaction and recover GR on sufficiently small scales realising the Vainshtein mechanism.  

In order to avoid the presence of a ghost, interactions have to be chosen
in such a way that the equations of motion for the scalar degrees of freedom contain no more than two time derivatives, i.e the galileon terms. Therefore, one expects that any consistent non-linear completion of FP contains these galileon terms in the limit in which the scalar mode decouples from the tensor modes, the so-called {\it decoupling limit}. This turns out to be a powerful criteria for building higher order interactions with the desired properties. Indeed, following this route,  de Rham and Gabadadze constructed a family of ghost-free extensions to the FP theory, which reduce to the galileon terms in the decoupling limit \cite{deRham:2010ik}. Adopting an effective field theory approach, the basic building block is a tensor $H_{\mu\nu}$, corresponding to the covariantization of metric perturbations, namely:
\be\label{H}
g_{\mu \nu} = \eta_{\mu \nu} +h_{\mu\nu}\,\equiv\,H_{\mu \nu}+\Sigma_{\mu\nu},
\quad
\Sigma_{\mu\nu}\equiv\partial_\mu \phi^\alpha \partial_\nu \phi^\beta \eta_{\alpha \beta}.
\ee
The St\"uckelberg fields $\phi^\alpha$ are introduced to restore  reparameterisation invariance, hence
transforming as scalar from the point of view
of the physical metric \cite{ArkaniHamed:2002sp}. The internal metric $\eta_{\alpha \beta}$ corresponds to a  non-dynamical reference
metric,  usually assumed to be Minkowski space-time.
Therefore, around flat space,  we can rewrite $H_{\mu \nu}$ as
\bea\label{defhmn}
H_{\mu \nu}&=&h_{\mu \nu}+\eta_{\beta \nu}\partial_\mu \pi^\beta+\eta_{\alpha \mu}\partial_\nu
\pi^\alpha-
\eta_{\alpha \beta} \partial_\mu \pi^\alpha \partial_\nu \pi^\beta.
\eea
From now on, indices are raised/lowered with the dynamical metric $g_{\mu\nu}$, unless otherwise stated.
For example, $H^\mu_{\,\,\nu}\,=\,g^{\mu \rho} H_{\rho \nu}$.
Moreover, the Lagrangian constructed from $H_{\mu \nu}$ is invariant under coordinate transformations $x^{\mu} \to x^{\mu} + \xi^{\mu}$, provided $\pi^{\mu}$ transforms as
\be\label{pitransf}
\pi^{\mu} \to \pi^{\mu} + \xi^{\mu}.
\ee

The dynamics of the St\"uckelberg fields $\phi^\alpha$ are at the origin of the two features: the BD ghost excitation and the vDVZ discontinuity. With respect to the first issue, as noticed by Fierz and Pauli, one can remove the ghost excitation, to linear order in perturbations, by choosing the quadratic structure $H_{\mu\nu}H^{\mu\nu}-H^2$.
When expressed in the St\"uckelberg field language, one of the four St\"uckelberg fields becomes non-dynamical. However, when going beyond linear order, this constraint disappears, signalling the emergence of an additional ghost mode \cite{ArkaniHamed:2002sp}. Remarkably, Ref.~\cite{deRham:2010ik} has shown how to construct a potential, tuned at each order in powers of $H_{\mu\nu}$,
to hold the constraint and remove one of the St\"uckelberg fields. Even though the potential is expressed in terms of  an infinite series of terms for $H_{\mu\nu}$, it can be re-summed into the following finite form \cite{deRham:2010kj}
\be
{\cal U}= -m^2\left[{\cal U}_2+\alpha_3\, {\cal U}_3+\alpha_4\, {\cal U}_4\right],
\label{potentialU}
\ee
where $\alpha_n$ are free dimensionless parameters, ${\cal U}_n=n!\det_{n}(\mK)$ and the tensor ${\cal K}_{\mu}^{\ \nu}$ is defined as
\bea
{\mathcal K}_{\mu}^{\ \nu} &\equiv &\delta_{\mu}^{\ \nu}-\left(\sqrt{g^{-1} \Sigma}\right)_{\mu}^{\ \nu}.
\eea
The square root is formally understood as $\sqrt{{\cal K}}_{\mu}^{\ \alpha}\sqrt{\cal K}_{\alpha}^{\ \nu}={\cal K}_{\mu}^{\ \nu}$.
The determinant can be written in terms of traces as
\bea
\det_{2}(\mK)&=&(\tr\mK)^2-\tr (\mK^2),\nonumber \\
\det_{3}(\mK)&=&(\tr \mK)^3 - 3 (\tr \mK)(\tr \mK^2) + 2 \tr \mK^3,\nonumber \\
\det_{4}(\mK)&=& (\tr \mK)^4 - 6 (\tr \mK)^2 (\tr \mK^2)
+ 8 (\tr \mK)(\tr \mK^3) + 3 (\tr \mK^2)^2 - 6 \tr \mK^4\,. \nonumber
\eea
All terms $\det_{n} (\mK) $ with $n > 4$ vanish in four dimensions. Therefore, the massive gravity theory can be written as
\be\label{genlag}
 {\cal L} = \frac{M_{Pl}^2}{2}\,\sqrt{-g}\left( R  - m^2{\cal U}\right),
\ee
where ${\cal U}$ is given by (\ref{potentialU}). 

The theory defined by (\ref{genlag}) has Minkowski spacetime as a trivial solution, hence one can rewrite the metric
$g_{\mu\nu}$ and the scalars $\phi^\mu$ as deviations from flat space, namely
\be\label{phi-pi}
g_{\mu\nu}=\eta_{\mu\nu}+h_{\mu\nu},\qquad \qquad \phi^\a=x^\a-\pi^\a,
\ee
where $x^\a$ are the usual cartesian coordinates spanning $\eta_{\alpha\beta}$.

We focus on a convenient limit of Lagrangian (\ref{genlag}) which captures most of the dynamics of the helicity-0 mode, but keeps the linear behaviour of the helicity-2 (tensor)  mode \cite{ArkaniHamed:2002sp}. 
The limit, called the decoupling limit, is defined as
\be\label{declimit}
m\to 0\,,\hskip1cm M_{Pl}\to \infty\,,\hskip1cm \Lambda_3\,\equiv\,m^2 M_{Pl}=\,{\rm{fixed}}.
\ee
In order to obtain canonically normalized kinetic terms for the helicity 2 and helicity 1 modes, together with the relevant couplings for the helicity 0 modes, when this limit is taken one needs to canonically normalise the fields in the following way
\be\label{cannorm}
h_{\mu\nu}\,\to\,M_{Pl}\, h_{\mu\nu}
\,, \hskip 0.5cm A_{\mu}\,\to\,m M_{Pl}\, A_\mu
\,, \hskip 0.5cm \pi\,\to\,m^2 M_{Pl}\, \pi,
\ee
where we have split the St\"uckelberg fields $\pi^\mu$ into a scalar component $\pi$ and a divergenceless vector $A^\mu$, namely
\be\label{vecscal}
\pi^\mu=\eta^{\mu\nu}(\partial_\nu \pi +A_\nu).
\ee
The resulting Lagrangian in the decoupling limit is \cite{deRham:2010ik}
\bea\label{lagdec}
{\cal L} &=&-\frac12 \,h^{\mu \nu} {\cal E}_{\mu \nu}^{\alpha \beta} h_{\alpha\beta} +h^{\mu\nu} \left( {X}^{(1)}_{\mu\nu} +{X}^{(2)}_{\mu\nu} +{X}^{(3)}_{\mu\nu} \right),
\eea
where we have ignored the vector. This decoupling limit theory is a special case of the effective theory derived from the Horndeski theory Eq.~(\ref{decoupling}). Thus in this limit the theory shares the same properties as the Horndeski theory. 

Cosmology in this massive gravity theory is rather peculiar. If we assume that the fiducial metric $\Sigma^{\mu}_{\nu}$ obeys the FRW symmetry only the open FRW solution is allowed \cite{D'Amico:2011jj, Gumrukcuoglu:2011ew}. We can allow an inhomogeneous fiducial metric by choosing the St\"uckelberg fields of the form \cite{Koyama:2011xz, Koyama:2011yg, Gratia:2012wt, Volkov:2012cf, Volkov:2012zb}
\be\label{stuckatz}
\phi^0=f(t,r), \quad 
\phi^i=g(t,r)\frac{x^i}{r}.
\ee
Then there exists a class of solution characterised by 
\be
g(t,r)= c_0^{-1} a(t,r) r, \; c_0 =
 \frac{1 + 6 \a_3 + 12 \a_4 \pm\sqrt{1 + 3 \a_3 + 9 \a_3^2 - 12
\a_4}}{3 (1+3 \a_3 +4 \a_4)}.
\ee
The Einstein equations are given by
\be
G^{\mu}_{\nu} = - 3 H^2 \delta^{\mu}_{\nu}, \quad H^2=\frac{1 + 3 \alpha_3 \pm 2 \alpha_5}{ 3(1 + 3 \alpha_3 \pm \alpha_5)^2} m^2,
\label{Einstein}
\ee
where \be \alpha_5^2\equiv1+3\alpha_3+9\alpha_3^2-12\alpha_4\,.\label{defda5}
\ee
The graviton mass plays the role of an effective cosmological constant. This model could realise the idea of technical naturalness. Since the massless limit has more symmetry (i.e. GR), the small mass is natural in the sense that the quantum corrections only renormalise the mass as $m^2 \to m^2 + {\cal O}(1) m^2$ \cite{deRham:2013qqa}. Unfortunately, however, perturbations around these solutions are found to be unstable \cite{Gumrukcuoglu:2011zh, DeFelice:2012mx, Khosravi:2013axa, Motloch:2014nwa}. 

A natural extension of massive gravity is realised by promoting the fiducial metric, $\Sigma^{\mu}_{\nu}$, to a dynamical metric described by another Einstein-Hilbert action. This is known as bigravity. 
The action for the massive gravity model (\ref{genlag}) can be extended to the bigravity model without introducing a ghost \cite{Hassan:2011zd}. 
In this case, it is possible to find FRW solutions for both of the metrics \cite{Volkov:2011an, Comelli:2011zm, Akrami:2012vf}. Unfortunately, again these solutions are found to be unstable under linear perturbations \cite{Comelli:2012db, Comelli:2014bqa, Lagos:2014lca, Cusin:2014psa, Amendola:2015tua, Solomon:2014dua, Konnig:2014xva} unless we tune the Planck scale for the second metric to decouple the massive graviton from matter \cite{Akrami:2015qga}. It is also possible to modify the coupling to matter by introducing an effective metric that is made of two metrics. Generally this brings back the BD ghost \cite{Yamashita:2014fga}. There is a coupling which removes the BD ghost in the decoupling limit \cite{deRham:2014naa, deRham:2014fha}. However, linear perturbations around the cosmological solutions are again unstable \cite{Enander:2014xga, Solomon:2014iwa, Comelli:2015pua, Gumrukcuoglu:2015nua}. These massive/bi-gravity theories can be formulated elegantly using the tetrad formalism \cite{Hinterbichler:2012cn}. It has been suggested that there is a matter coupling that does not introduce the BD ghost in the full theory although it is not easy to find a corresponding metric formulation \cite{Hinterbichler:2015yaa} (see also \cite{deRham:2015cha, DeFelice:2015yha}). 
See Ref.~\cite{deRham:2014zqa} for other extensions of the model and a more complete list of references. It remains to be seen whether there exist stable solutions where the acceleration is driven by the mass and are also distinguishable from $\Lambda$CDM observationally.  

\subsection{Outlook}
Although there has been significant progress in developing modified gravity models and many interesting new ideas have been explored to tackle the cosmological constant and the late time acceleration problem, we still do not have a consistent theory that is a true alternative to $\Lambda$CDM. Our quest for modified gravity models as an alternative to $\Lambda$CDM still continues. 

However, many different models that have been reviewed so far have common properties. There are usually three regimes of gravity: 

\begin{itemize}
\item
There is a length scale above which gravity is modified. This can be simply determined by the cosmological constant, or the Compton wavelength of a massive graviton, or the cross-over scale in a braneworld model. 

\item
Even below the modification scale, gravity is still modified due to the extra scalar degree of freedom. In this regime, gravity is described by scalar tensor gravity with a $O(1)$ Brans-Dicke parameter.

\item
On small scales there appears a scale below which GR is restored due to the non-linear interaction of the scalar field. This is often called the screening mechanism. This is essential to ensure that the theory passes the stringent Solar System constraints.    
\end{itemize} 

This general picture is useful to develop a strategy to construct cosmological and astrophysical tests of gravity even though the examples that we have seen suffer from various theoretical problems \cite{Hu:2009ua}. One of the major developments of modified gravity is the screening mechanism, which enables us to modify gravity significantly on cosmological scales yet to satisfy the Solar System constraints. We will review the screening mechanisms in the next section. 

\section{Screening mechanism}

\subsection{Why do we need a screening mechanism?}
As we learnt from several examples of modified gravity models, one of the difficulties in explaining the accelerated expansion of the Universe by modifying gravity is that GR is tested to high accuracy in the Solar System. In the simplest example of Brans-Dicke gravity, once we impose the Solar System constraints, there is essentially no room to modify gravity on cosmological scales. One possibility to avoid this problem is to break the equivalence principle. The stringent constraints in the Solar System are obtained using objects made of baryons. If the additional degree of freedom only couples to dark matter not to baryons, we can evade these stringent constraints while it is possible to modify gravity significantly on cosmological scales. This is known as interacting dark energy models in the Einstein frame where the scalar field is coupled only to dark matter \cite{Amendola:1999er}. 

If we keep the equivalence principle, we need to have a mechanism to suppress the modification of gravity on small scales. We loosely call this mechanism the screening mechanism. The screening mechanism is realised by the fact that the additional degree of freedom, which is often represented by a scalar field, obeys a non-linear equation driven by the density. The density varies over many orders of magnitude in our Universe. The critical density of the universe is given by $\rho_{crit} = 10^{-29}$ g cm$^{-3}$. The typical density inside galaxies is $\rho_{gal} = 10^{-24}$ g cm$^{-3}$. The density in the Sun is given by $\rho_{sun}= 10$ g cm$^{-3}$.
Thus if we expand the density around the cosmological background, the density contrast in the environments where we perform conventional tests of gravity is much larger than one. The screening mechanism utilises the non-linearity of the scalar field driven by the non-linear density contrast to change the behaviour of the scalar field from cosmology to the Solar System. 

\subsection{Screening mechanisms}
In this section we follow Ref.~\cite{Joyce:2014kja} and classify the screening mechanisms. A general Lagrangian for a scalar field can be written schematically as
\begin{equation}
{\cal L} = -\frac{1}{2} Z^{\mu \nu} (\phi, \partial \phi, \partial^2 \phi) \partial_{\mu} \phi \partial_{\nu} \phi -V(\phi) 
+ \beta(\phi) T^{\mu}_{\mu},
\end{equation}
where $Z^{\mu \nu}$ represents derivative self-interactions of the scalar field, $V(\phi)$ is a potential, $\beta(\phi)$ is a coupling function and $T^{\mu}_{\mu}$ is the trace of the matter energy-momentum tensor. In order to avoid the Ostrogradsky ghost associated with higher time derivatives, $Z^{\mu \nu}$ contains up to the second derivative of the field. In the presence of non-relativistic matter $T^{\mu}_{\mu} = -\rho$, the scalar field's dynamics depends on the local density of the system. Let's consider the background field $\bar{\phi}$ which depends on the local density. Around this background, the dynamics of fluctuations is determined by three parameters: the mass $m(\bar{\phi})$, the coupling $\beta(\bar{\phi})$ and the kinetic function $Z^{\mu \nu}(\bar{\phi})$. Screening can be realised mainly in three different ways utilising these three parameters:

\begin{itemize}
\item  Large mass \\
If the mass of fluctuations $m^2(\bar{\phi})$ is large in dense environments, the scalar field does not propagate above the Compton wavelength $m(\bar{\phi})^{-1}$ and the additional force mediated by the scalar field is suppressed. On the other hand, in low density environments such as cosmological background, the mass can be light and the scalar field mediates the fifth force modifying gravity significantly. This idea is realised in the chameleon type screening mechanism \cite{Khoury:2003aq, Khoury:2003rn}. 

\item Small coupling \\
If the coupling to matter $\beta(\bar{\phi})$ is small in the region of high density, the strength of the fifth force generated by the scalar field is weak and modifications of gravity is suppressed. On the other hand, in low density environments, the fifth force strength can be of the same order of gravity. This idea is realised in the dilaton \cite{Brax:2010gi} and symmetron mechanism \cite{Hinterbichler:2010es}. 

\item Large kinetic term \\
If we make the kinetic function $Z^{\mu \nu}(\bar{\phi})$ large in dense environments, the coupling to matter is effectively suppressed. There are two possibilities to make the kinetic term large. One is to assume that the first derivative of the field becomes large. This idea is realised in the k-mouflage type mechanism \cite{Babichev:2009ee}. On the other hand, in the Vainshtein mechanism \cite{Vainshtein:1972sx}, the second derivative of the field becomes large in the region of high density. 

\end{itemize}
In the next section, we look at examples of models that accommodate these screening mechanisms. 

\subsection{Examples}
\subsubsection{Chamleon/Symmetron/Dilaton mechanism}

This class of model is represented by an action (\ref{Eframe}) in the Einstein frame 
\begin{equation}
S= \int d^4 x \sqrt{-g} \left[
\frac{1}{16 \pi G} R - \frac{1}{2} (\nabla \phi)^2 - V(\phi)
 \right]
 + S_{m} ( A^2(\phi) g_{\mu \nu}).
\label{Einsteinframe}
\end{equation}
The matter fields couple to a metric $A^2(\phi) g_{\mu \nu}$. Due to this coupling, a test particle feels the fifth force $\nabla \ln A(\phi)$ generated by the scalar field. In these models, the dynamics of the scalar field is determined by the local density dependent effective potential
\begin{equation}
V_{\rm eff} = V(\phi) -[A(\phi)-1] T^{\mu}_{\mu}.
\end{equation}
The dynamics of the scalar field is characterised by the mass of the scalar field around the minimum of the potential $\phi=\bar{\phi}$ and the coupling function
\begin{equation}
m^2 = \left. V_{\rm eff}''(\bar{\phi}) \quad \beta=\mpl \frac{d \ln A}{d \phi} \right |_{\phi=\bar{\phi}}.
\end{equation}

Below we show typical choices of the potential and the coupling function to realise the screening mechanisms: 
\begin{eqnarray}
A(\phi) &=& 1 + \xi \frac{\phi}{M_{\rm pl}}, 
\quad V(\phi) = \frac{M^{4 +n}}{\phi^n} \quad \mbox {chamaleon}, \\
A(\phi) &=& 1 + \frac{1}{2 M} 
(\phi - \bar{\phi})^2, 
\quad V(\phi) = V_0 e^{- \phi/M_{\rm pl}} 
\quad \mbox{dilaton}, \\
A(\phi) &=& 1 + \frac{1}{2 M^2} 
\phi^2, 
\quad V(\phi) = -\frac{\mu^2}{2} \phi^2 + \frac{\lambda}{4} \phi^4 \quad \mbox{symmetron},
\end{eqnarray}
where the mass scale $M$ is a parameter of these models. 
In the chameleon mechanism, the mass $m^2$ becomes large in the region of high densities while in the dilaton and symmetron models, the coupling $\beta$ becomes small in high dense environments. Note that the second derivative of the effective potential does not depend on the enegy density explicitly in chameleon models as $A(\phi)$ is lienar in $\phi$. The density dependence of the mass comes from the fact that the minimum of the potential $\bar{\phi}$ depends on the density. 

In all these models, if we consider an object with the gravitational potential $|\Psi_{N}|$, the screening of the fifth force happens when the gravitational potential $|\Psi_{\rm N}|$ exceeds some critical value $|\Psi_{N}| > \psi_c$. We will show this in the case of the chameleon mechanism below. 

\subsubsection{K-moflouge/Vainshtein mechanism}
The other class of the model utilises the non-linear derivative interactions to screen the fifth force. Typical choices of the functions in the Horndeski action (\ref{Horndeski}) 
\begin{equation}
S=\int d^4 x \sqrt{-g} \left[
\frac{1}{16 \pi G} R  + K(\phi, X)  - G_3(\phi, X) \Box \phi
 \right]
 + S_{m} (g_{\mu \nu}).
\end{equation}
to realise the screening mechanisms are given below:  
\begin{eqnarray}
K(\phi, X) &=& X
+ \frac{\alpha}{4 \Lambda^4} X^2, 
\quad \mbox{k-mouflage},  \\
K(\phi, X) &=&  X,
\quad G_3 (\phi, X) = \frac{1}{\Lambda^3} X
\quad \mbox{Vainshtein}.
\end{eqnarray}

The screening condition for a spherically symmetric object with $M$ and radius $R$ is determined by the solution for the scalar field $\phi$. Again let us consider an object with the gravitational potential $|\Psi_{N}|= G M/R$. In the k-mouflage mechanism, the screening of the fifth force happens when the first derivative of the potential, i.e. the gravitational acceleration exceeds some critical value $|\partial \Psi_{N}| > \Lambda_c$. On the other hand, in the Vainshtein mechanism the screening operates when the second derivative of the potential, i.e. the spatial curvature, exceeds some critical value $|\partial^2 \Psi_{N}| > \Lambda_c^2$. We will show this explicitly for the Vainshtein mechanism below by solving the scalar field equation. 

\subsection{Chameleon screening}
Using a simple example, we show how the chameleon mechanism suppresses the scalar field \cite{Khoury:2003rn}. Let's consider a scalar field with mass $m$ that is coupled to matter. We assume that the function $A(\phi)$ can be approximated as $A(\phi)=1+\beta \phi/\mpl$. In the case of BD gravity 
\begin{equation}
\beta^2 =\frac{1}{2 (3 + 2 \omega_{BD})},
\end{equation}
which gives $\beta^2=1/6$ for $f(R)$ gravity.  
In the Einstein frame, the geodesic equation of a test particle is modified due to the coupling between the scalar field and matter. From the geodesic equation, the acceleration of the particle is obtained as  
\begin{equation}
{\bf a} = - \nabla \Psi - \frac{d \log A}{d \phi} \nabla \phi 
= - \nabla \Psi - \beta \nabla \phi,  
\label{acceleration}
\end{equation}
and the equation of motion for the scalar field is given by
\begin{equation}
\nabla^2 \phi - m^2 \phi =  8 \pi G \beta \rho,
\end{equation}
where we assume $\phi$ does not depend on time. A spherically symmetric solution outside the matter source is given by
\begin{equation}
\phi(r) = -\frac{2 \beta G M}{r} e^{- m r},
\label{unscreened}
\end{equation}
where $M$ is the mass of the object. For scales smaller than the Compton wavelength, we can ignore the Yukawa suppression in Eq.~(\ref{unscreened}). The acceleration of the particle can be obtained by substituting Eq.~(\ref{unscreened}) into Eq.~(\ref{acceleration}) as $a= -G_{\rm eff} M/r^2$ where 
\begin{equation}
G_{{\rm eff}} = G \Big(1 + 2 \beta^2\Big) = G \Big(\frac{4 + 2 \omega_{BD}}{3 + 2 \omega_{BD}}\Big). 
\end{equation}
Thus the test particle feels the modified Newton constant as in BD gravity without potential (see Eqs.~(\ref{BD1}) and (\ref{BD2})). 
On the other hand, due to the Yukawa suppression at length scales larger than the Compton wavelength $m^{-1}$, the scalar field does not propagate and the test particle feels normal gravity. The idea of chameleon screening is to make the mass dependent on the density so that the scalar field does not propagate in the dense environment. 

To realise this idea, we need to introduce a potential to the scalar field. The equation of motion is generalised to 
\begin{equation}
\nabla^2 \varphi = V_{{\rm eff}}'( \varphi)=  V'(\phi) + 8 \pi G \beta \rho.
\end{equation}
By choosing the potential $V(\phi)$ appropriately, we can realise the situation where the mass of the scalar field is large for larger densities.

%%%%%%%%%%%%%%%%%%%%%%%%%%%%%%%%%%%%%%%%%%%%%%%%%%%%%%%%%%%%
\begin{figure}[ht]
  \centering{
  \includegraphics[width=13cm]{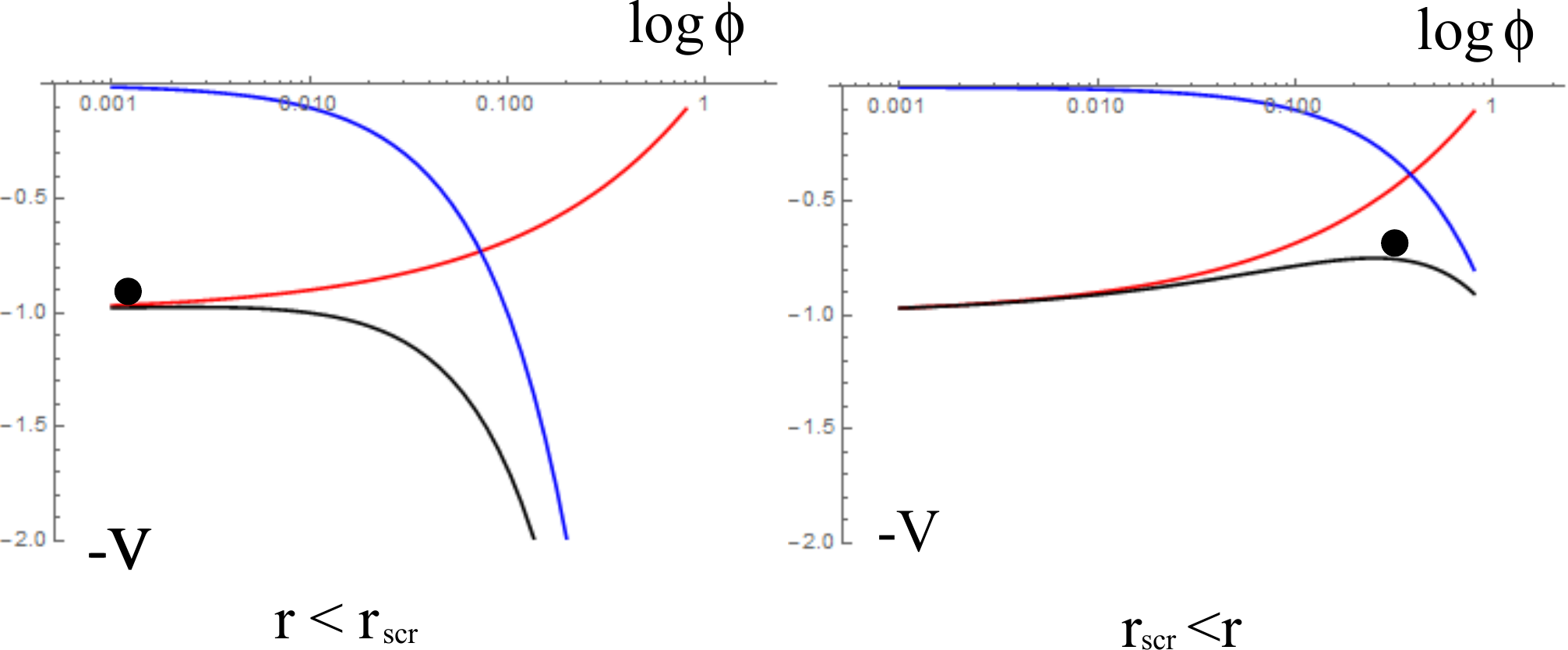}
  }
  \caption{The effective potential $- V_{\rm eff}$ inside and outside the matter source. We can understand the spherically symmetric solution as a dynamically time dependent field governed by $- V_{\rm eff}$. The red curve is the bare potential $-V$ and the blue curve represents the density dependent contribution. The back curve is the total effective potential $-V_{\rm eff}$.  
   }
\end{figure}
%%%%%%%%%%%%%%%%%%%%%%%%%%%%%%%%%%%%%%%%%%%%%%%%%%%%%%%%%%%%%

Let's consider a spherically symmetric matter source with a constant density within the radius $r=R$. Inside the source, the scalar field is trapped at the minimum of the effective potential $V'_{eff}=0$, $\phi=\phi_s$. Outside the source, we assume that we have the background density $\rho_{\infty} \ll \rho_s$ and the scalar field is again trapped by the minimum $\phi=\phi_{\infty}$ with mass $m_{\infty}$. The dynamics of the scalar field is governed by the equation
\begin{equation}
\phi(r)'' + \frac{2}{r} \phi'(r) = V'(\phi) + 8 \pi G \beta \rho. 
\end{equation}
By identifying $r$ as time $t$, this is equivalent to the time evolution of the scalar field with a potential $-V_{{\rm eff}}$ (Fig.~3). The scalar field starts from the ``maximum'' of the potential $\phi=\phi_{s}$ and then rolls to another ``maximum'' $\phi=\phi_{\infty}$. The solution for the scalar field outside the matter source is obtained as $\phi(r) = - (C/r) e^{- m_{\infty} r} + \phi_{\infty}$. From the radius $r=R$ until the scalar field settles down to $\phi=\phi_s$ at radius $r=r_{\rm scr}$, the scalar field dynamics is driven by the change of the density so we can ignore the contribution from the potential. Then the equation of motion for the scalar in this regime is given by 
\begin{equation}
\phi(r)'' + \frac{2}{r} \phi'(r) = 8 \pi G \beta \rho,
\end{equation}
which admits the solution $\phi = (\alpha \rho_c/3 \mpl) r^2/ 2
+ A/r +B$. The integration constants $A, B$ and $C$, and the radius $r_{{\rm scr}}$ are determined by matching the field and its first derivative at $r=R$ and $r=r_{{\rm scr}}$. We obtain the solution outside the source as 
\begin{equation}
\phi(r) = -
\left( \frac{3 \Delta R}{R}  \right) \frac{2 G M \beta }{r} e^{- m_{\infty} r} + \phi_{\infty},
\label{chameleon-force}
\end{equation}
where 
\begin{equation}
\frac{\Delta R}{R}  = \frac{\phi_{\infty} - \phi_{s}}{6 \beta \mpl |\Psi_{N}| }
\sim \frac{R - r_{\rm scr}}{R},
\end{equation}
and $\Psi_{N}$ is the gravitational potential of the object $|\Psi_{N}| = G M/R$. If $3 \Delta R/R > 1 $ this solution is not valid and the solution becomes Eq.~(\ref{chameleon-force}) with  $3 \Delta R/R =1$. The condition $\Delta R/R \ll 1$ is called the thin shell condition. If this condition is satisfied only the mass within the thin-shell of size $\Delta R$ contributes to the force because in the interior of the source the scalar field is heavy and the scalar force is Yukawa suppressed. Note that we naively say that the scalar field is screened in dense environments. This is in fact not a very accurate statement. What matters is the gravitational potential of the object not the density. See Fig.~4 for a schematic picture of the thin shell solution. 

%%%%%%%%%%%%%%%%%%%%%%%%%%%%%%%%%%%%%%%%%%%%%%%%%%%%%%%%%%%%
\begin{figure}[ht]
  \centering{
  \includegraphics[width=8cm]{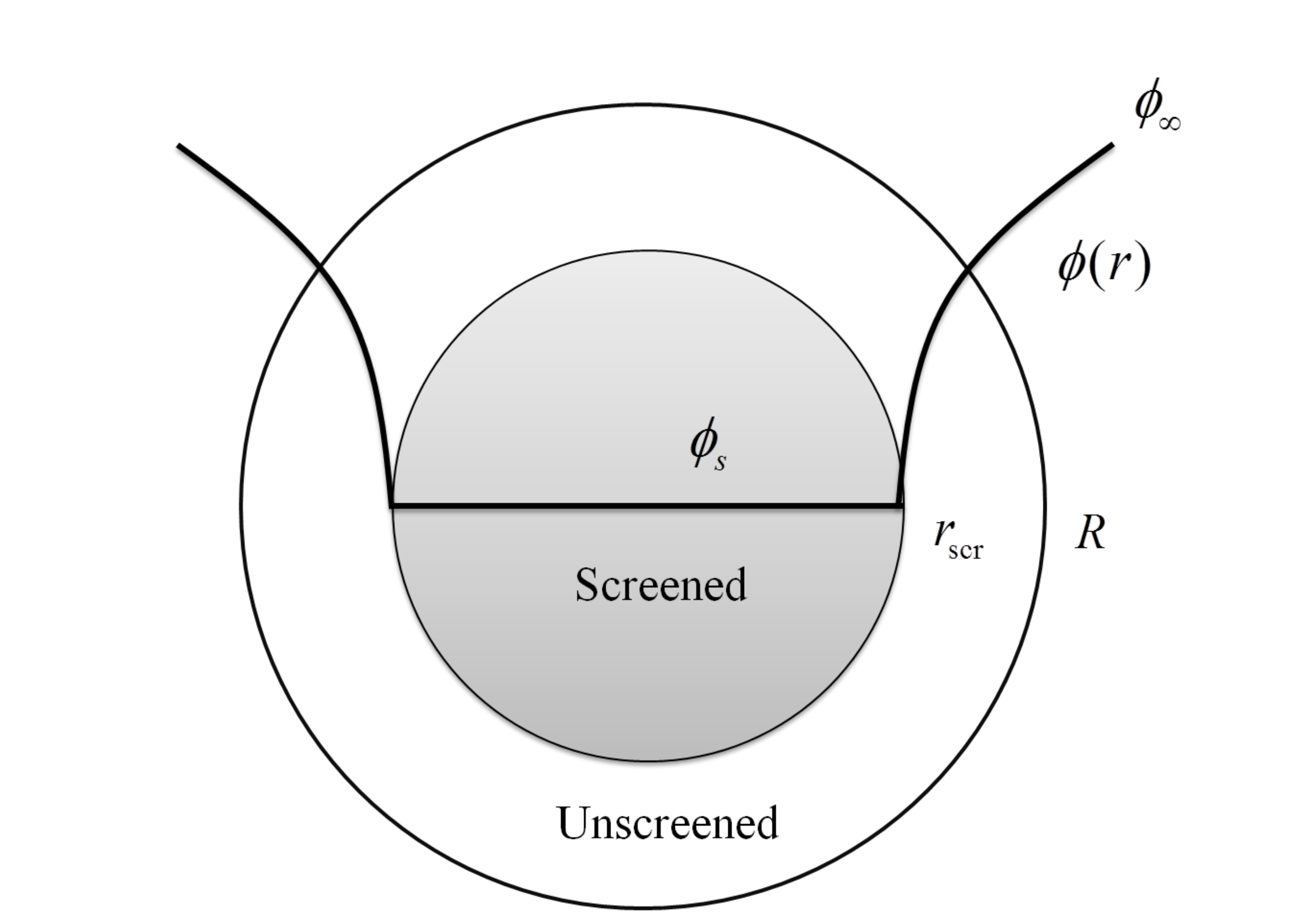}
  }
  \caption{A schematic picture of the thin shell profile of the scalar field.   
   }
\end{figure}
%%%%%%%%%%%%%%%%%%%%%%%%%%%%%%%%%%%%%%%%%%%%%%%%%%%%%%%%%%%%%

Although we derived the solutions in the Einstein frame where the dynamics of the scalar field is easier to understand, it is straightforward to translate the solutions into those found in the original (Jordan) frame where the scalar field is not coupled to matter directly while it couples to gravity non-minimally (see Eq.~(\ref{BD})).  The solutions for the metric and scalar field perturbation are given by 
\begin{eqnarray}
\nabla^2 \Psi = 4 \pi G \mu \rho, \quad \Psi = \eta^{-1} \Phi,
\end{eqnarray}
\begin{equation}
\mu = 1 + 2 \beta_{\rm eff}^2, \quad \eta = \frac{1 - 2 \beta_{\rm eff}^2}{1 + 2 \beta_{\rm eff}^2}, \quad \beta_{\rm eff}^2 = \beta^2 \frac{3 \Delta R}{R},
\end{equation}
when the thin shell condition is satisfied. The scalar field in the Jordan frame $\psi$ is related to $\phi$ as $\psi = - 2 \beta \phi/\mpl$ for $\psi \ll 1$. In particular, for $f(R)$ gravity we have, $\psi = f_R = -\sqrt{2/3} ( \phi/\mpl)$ (see Eq.~(\ref{fr-trans})) and the thin shell condition can be written as 
\begin{equation}
|f_{R \infty} - f_{R s}| < \frac{2}{3} |\Psi_{N}|.
\label{fr-thishel}
\end{equation}

Let's consider the Solar System constraint \cite{Hu:2007nk}. In the Sun, the density is $\rho = 10$ g cm$^{-3}$ while the density of the Milky way galaxy is typically $10^{-24}$ g cm$^{-3}$. Thus the scalar field in the Solar System is well suppressed compared with the one in the Milky way galaxy $\phi_{\rm solar} \ll \phi_{\rm gal}$. If the thin shell condition is satisfied, $\beta_{\rm eff}^2$ is small so we can approximate $\eta$ as $\eta -1 = - 4 \beta_{\rm eff} ^2$. 
The constraints $|\eta -1| =(2.1 \pm 2.3) \times 10^{-5}$ gives a constraint on the galaxy field 
\begin{equation}
\frac{\beta \phi_{\rm gal}}{\mpl} <  10^{-11},
\label{solar}
\end{equation}
where we used the fact that the Milky Way's potential is approximately given by $\Psi_{N \rm gal} \sim 10^{-6}$. This is a model independent constraint but in order to translate this constraint to the mode parameters, we need to interpolate this constraint to the scalar field in the cosmological background $\phi_{\rm cos}$. For this, we need to specify the potential. We consider a potential of the form
\begin{equation}
V= \Lambda - M^4 \left( \frac{\phi}{\mpl} \right)^{\frac{n}{1+n}},
\label{cham-pot}
\end{equation}
as in the $f(R)$ gravity given by Eq.~(\ref{HS}) (see Eq.~(\ref{HS-E})). 
We assume that the scalar field is at the minimum of the potential 
\begin{equation}
\left( \frac{\phi}{\mpl} \right) 
= \left(\frac{(n+1) \beta \rho}{n M^4} \right)^{-(1+n)}. 
\end{equation}
Given that the ratio of the cosmological background and the Milky way galaxy is $\rho_{\rm gal}/\rho_{\rm cos} \sim 10^5$, the ratio of the scalar field is given by 
\begin{equation}
\frac{\phi_{\rm cos}}{\phi_{\rm gal}} = \left(
\frac{\rho_{\rm cos}}{\rho_{\rm gal}}  \right)^{-(1+n)} 
=10^{5(1+n)}. 
\end{equation}
Using this relation between $\phi_{\rm gal}$ and $\phi_{\rm cos}$, we obtain the constraint on the cosmological field $\phi_{\rm cos}$ 

To derive the constraint, we assumed that the galaxy field is at the minimum of the potential. In fact, if the Milky way is not screened the dynamics of stars are disrupted. Imposing the thin shell condition for the Milky Way galaxy, we obtain the constraint $\beta \phi_{\rm cos}/\mpl < 10^{-6}$ where we again have used $\Psi_{N \rm gal} \sim 10^{-6}$ and $\phi_{\rm gal} \ll \phi_{\rm cos}$. This constraint is much more stringent than Eq.~(\ref{solar}) for $n >0$. Note that to derive the constraint that the Milky Way galaxy is screened, we assumed that the Milky way is an isolated system in the cosmological background. There is a possibility that the Milky Way galaxy is screened by the Local Group with a potential $|\Psi_{N}| \sim 10^{-4}$. In this case the constraint is relaxed to be $\beta \phi_{\infty}/\mpl \sim 10^{-4}$. 

For the potential given by (\ref{cham-pot}), laboratory tests give much weaker constraints compared with the cosmological constraints \cite{Jain:2013wgs}. See \cite{Jain:2013wgs, Joyce:2014kja} for laboratory tests of the chameleon models. 

The constraint that the Milky Way with the potential $|\Psi_{N}|=10^{-6}$ is screened has important implications. Firstly, this unfortunately excludes the possibility of {\it self-acceleration} \cite{Wang:2012kj}. The change of $\phi$ between the cosmological density and the galactic density $\rho_{\rm gal} = 10^5 \rho_{\rm cos}$ is suppressed by the thin shell condition $\beta(\phi_{\rm cos} - \phi_{\rm gal})/\mpl < 10^{-6}$. If we translate this to the cosmological evolution of the scalar field, this implies that the scalar field barely moves between $0 < z<1$ where $z$ is the redshift (see \cite{Takahashi:2015ifa} for caveats). Thus the scalar field cannot be responsible for the acceleration. This also excludes the possibility to modify gravity on cosmological scales. The mass of the scalar field around the minimum in the cosmological background is given by \cite{Wang:2012kj, Brax:2011aw}
\begin{equation}
m_{\rm cos}^2 \sim \frac{\rho_{\rm cos} }{\mpl} \left(\frac{\mpl}{\beta \phi_{\rm cos}} \right) < 10^{-6} H_0^2.
\end{equation}  
Thus there is no modification of gravity on scales larger than 1 h$^{-1}$Mpc. Another implication is that GR is modified only with objects with a shallower potential $|\Psi_{N}| < 10^{-6}$. We will discuss how we test models with the Chameleon mechanism using these objects later. 

Finally we discuss briefly about the quantum corrections. The one-loop Coleman-Weinberg correction to the potential grows as $m_{\rm eff}^4$. Imposing the condition that the quantum corrections are under control at laboratory density $\rho_{\rm lab} \sim 10$ g cm$^{-3}$, the constraint on the mass of the chameleon is given by \cite{Upadhye:2012vh}
\begin{equation}
m_{\rm lab} < \left( \frac{48 \pi^2 \beta^2 \rho_{\rm lab}^2}{\mpl} \right)^{1/6}
 = 7.3 \times 10^{-3}  \left( \frac{\beta \rho_{\rm lab}}{10 \mbox{\; g cm}^{-3} }\right)^{1/3} \mbox{eV}. 
\end{equation}
This upper bound is already in tension with the laboratory constraints on the fifth force. This also means that the cut-off scale of this theory is rather low $\Lambda_{\rm UV} \sim 10^{-3}$eV $\sim \rho_{\Lambda}^{1/4}$ . The coupling to matter also generates corrections to the potential but this contribution is suppressed by the cut-off scale of the theory $\Lambda_{\rm UV}$ as $\Delta m/m = \Lambda_{\rm UV}/\mpl^2$ so this is totally negligible.

\subsection{Vainshtein mechanism} 
We consider the simplest cubic galileon model to understand how the Vainshtein mechanism operates. We use the equation of motion for the scalar field in the DGP model, Eq.~(\ref{DGP}): 
\begin{equation}
3 \beta \nabla^2 \phi + r_c^2
\Big[(\nabla^2 \phi)^2 - (\nabla_i \nabla_j \phi)(\nabla^i \nabla^j \phi) \Big]
 = 8 \pi G \delta \rho.
 \label{V-scalar}
\end{equation}
For the spherically symmetric case, it is possible to integrate the equation once. Outside the matter source, the solution obtained is \cite{Koyama:2007ih} 
\begin{equation}
\frac{d \phi}{d r }= \frac{2 r_g}{ 3 \beta r^2} 
\left(\frac{r}{r_V} \right)^3 \left( 
\sqrt{1+ \left(\frac{r_V}{r} \right)^3} -1 \right), \quad
r_V = \left(\frac{8 r_c^2 r_g}{9 \beta^2}   \right)^{1/3},
\label{DGPsol}
\end{equation}
where $r_g= 2 G M$. Below the Vainshtein radius $r_V$, the non-linear interaction suppresses the fifth force and we recover GR solutions with small corrections \cite{Lue:2002sw, Lue:2004rj}. The solutions for the metric perturbations can be obtained by subsituting the scalar field solution (\ref{DGPsol}) in the limit $r \ll r_V$ into Eqs.~(\ref{BDeq1}) and (\ref{BDeq2}) as 
\begin{equation}
\Phi = \frac{G M}{r} \pm  \sqrt{\frac{r_g r }{2 r_c^2}},
\quad 
\Psi = -\frac{G M}{r} \pm \sqrt{\frac{r_g r }{2 r_c^2}}.
\label{Vain}
\end{equation}
where the upper (lower) sign corresponds to a positive (negative) $\beta$, i.e the normal (self-accelerating) branch. On the other hand, outside the Vainshtein radius, the linear solution is recovered 
\cite{Lue:2002sw, Lue:2004rj, Koyama:2005kd} 
\begin{equation}
\Phi = \frac{G M}{r} \left(1- \frac{1}{3 \beta}  \right),
\quad 
\Psi = -\frac{G M}{r}  \left(1+ \frac{1}{3 \beta}  \right).
\end{equation}
See Fig.~5 for a schematic picture of the Vainshtein solution.
%%%%%%%%%%%%%%%%%%%%%%%%%%%%%%%%%%%%%%%%%%%%%%%%%%%%%%%%%%%%
\begin{figure}[ht]
  \centering{
  \includegraphics[width=8cm]{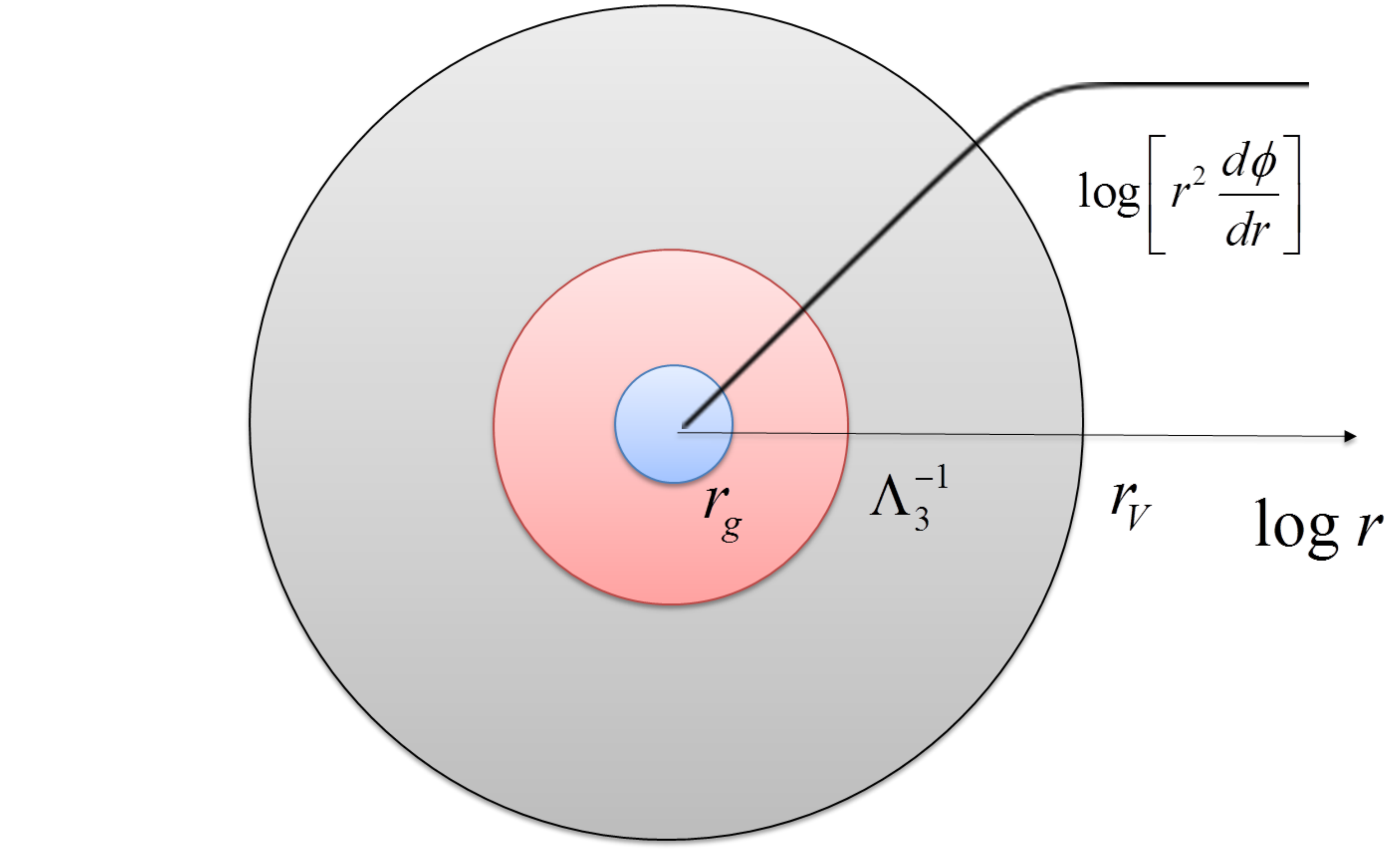}
  }
  \caption{A schematic picture of the Vainshtein solution (\ref{V-scalar}). The Vainshtein radius $r_V$ at which the non-linear interaction of the scalar field becomes important is much larger than the gravitational radius $r_g$ at which the spin-2 graviton becomes non-linear. The cut-off scale of the theory $\Lambda_3^{-1}$ is typically very large.   
   }
\end{figure}
%%%%%%%%%%%%%%%%%%%%%%%%%%%%%%%%%%%%%%%%%%%%%%%%%%%%%%%%%%%%%
Note that the condition to realise the Vainshtein mechanism $r\ll r_V$ can be written as $|\partial^2 \Psi_N| > r_c^{-2}$, confirming that the Vainstein mechanism operates when the spatial curvature exceeds a critical value.   

The Vainshtein radius $r_V$ is much larger than the gravitational length for $r_c \sim H_0$. For the Sun, $r_g=2.95$km while the Vainshtein radius is $r_V=130$ pc. Thus inside the Solar System, the solutions for the metric are well approximated by Eqs.~(\ref{Vain}). The perihelion precession per orbit $\Delta \phi$ is calculated as \cite{Lue:2002sw}
\begin{equation}
\Delta \phi = 2 \pi + \frac{3 \pi r_g}{r} \pm \frac{3 \pi}{2} 
\left(\frac{r^3}{2 r_c^2 r_g} \right)^{1/2},
\end{equation}
where the second term is the usual Einstein precession and the last term is due to the fifth force. The angle of perihelion advance due to the fifth force during one orbital period is then given by
\begin{equation}
\frac{\Delta \phi}{P} = \frac{3}{8 r_c} = 5 \times 10^{-4} 
\left(\frac{r_c}{5 \mbox{Gpc}}  \right)^{-1} 
{\rm arcseccond \; per \;century}.
\label{precession}
\end{equation}
The interesting feature of this prediction is that the precession rate is independent of mass so it is universal. Combining the observed precession of planets in the Solar System, the constraint was obtained as 
$\Delta \phi/P < 0.02$ arcsecond per century, which gives the constraint $r_c> 130$ Mpc \cite{Battat:2008bu}. Another constraint comes from Lunar Laser Ranging experiments. These experiments put constraints on the correction to the Newton potential $\delta \Psi/\Psi <2.4 \times 10^{-11}$. Using Eq.~(\ref{Vain}), $r_g=0.886$ cm for the Earth and $r=3.84 \times 10^{10}$ cm for the Earth-Moon distance we obtain $r_c > 162$ Mpc \cite{Dvali:2002vf}. This constraint will be improved by a factor of 10 by the Apache Point Observatory Lunar Laser-ranging Operation (APOLLO). See Ref.~\cite{Murphy:2013qya} for a review. These results are obtained by treating the planets or the Moon as a test body. This assumption is not necessarily valid. For example for the Earth-Moon system, it was shown that there is a correction to the universal precession rate (\ref{precession}) due to the non-superimposability of the field that depends on the mass ratio of the two bodies, (although the effect is small (4$\%$) for the Earth-Moon system) \cite{Hiramatsu:2012xj}. See Ref.~\cite{Brax:2011sv} for the constraints from laboratory tests. 

The most general action describing the Vainshtein mechanism can be obtained from the Horndeski action as we described in section 2.5. From the effective action (\ref{decoupling}), it is easy to find static spherically symmetric solutions \cite{Koyama:2013paa, Narikawa:2013pjr}. We consider the following configuration
\begin{equation}
\hat{h}_{tt}=-2 \Phi, \quad \hat{h}_{ij} = -2 \Psi \delta_{ij},
\quad \phi = \phi_0+\pi(r).
\end{equation}
The field equations yield
\begin{eqnarray}
P(x, A)  \equiv  \xi A(r)+\left(\frac{\eta}{2}+3\xi^2\right)x
+\Big(\mu+6\alpha\xi-3\beta A(r)\Big)x^2 \nonumber\\
+\Big(\nu+2\alpha^2+4\beta\xi\Big)x^3-3\beta^2x^5=0,\\
y(r) = \beta x^3+A(r),
\label{basic-equation}
\end{eqnarray}
where we define
\begin{eqnarray}
x(r) &=& \frac{1}{\Lambda_3^3}\frac{\pi'}{r}, \quad
y(r) = \frac{\mpl}{\Lambda_3^3}\frac{\Phi'}{r} \label{ydef}
 = \frac{\mpl}{\Lambda_3^3}\frac{\Psi'}{r}, \\
A(r)&=&\frac{1}{\mpl\Lambda_3^3}\frac{M(<r)}{8\pi r^3}.
\end{eqnarray}
The function $M(<r)$ represents the mass of a spherically symmetric, pressure-less matter source, up to a radius $r$. Outside the
surface $r_s$ of  the  matter source, $A(r)$ can be written as
\begin{equation}
A(r) = \Big( \frac{r_V}{r} \Big)^3, \quad r_V= \Big( \frac{M}{8 \pi \mpl \Lambda_3^3} \Big)^{1/3},
\end{equation}
where $r_V$ is the Vainshtein radius depending on the total mass $M$ of the source.
 
Stability of these solutions can be studied by considering small perturbations around the background
$ \pi = \pi_0(r) + \varphi(t,r,\Omega), \Phi=\Phi(r) + \delta \Phi(t,r,\Omega),
\Psi=\Psi(r) + \delta \Psi(t,r,\Omega)$ where $\Omega$ represents angular coordinates. By expanding the effective action up to the second order in these small perturbations we find \cite{Koyama:2013paa}
\begin{equation}
S_{\varphi}= \frac{1}{2} \int d^4 x
\Big[
K_t(r) (\partial_t \varphi)^2 - K_r(r) (\partial_r \varphi)
-K_{\Omega}(r) (\partial_{\Omega} \varphi)^2
\Big],
\label{vain-per}
\end{equation}
where
\begin{eqnarray}\label{Ks}
K_r(r) &=& 2 \partial_x P(x,r) |_{x=x_0},
\end{eqnarray}
\begin{eqnarray}
K_t(r) = \frac{1}{3 r^2} \frac{d}{dr}
\Big[r^3 \Big\{
\eta+6 \xi^2 + 6 (\mu + 6 \alpha \xi) x 
+ 12 \alpha A  \nonumber\\
\quad \quad \quad   + 18 (\nu + 2 \alpha^2 + 4 \xi) x^2 
+ 12 (10 \alpha \beta + \varpi) x^3 + 36 \beta x y
\Big\}
\Big ],\\
%\end{eqnarray}
%\begin{eqnarray}
K_{\Omega}(r) = \frac{1}{2 r} \frac{d}{dr}
\Big[r^2
\Big\{
\eta+6 \xi^2 + 4 (\mu + 6 \alpha \xi) x \nonumber\\
\quad \quad \quad  + 6 (\nu + 2 \alpha^2 + 4 \xi) x^2 - 12 \beta x y
\Big\}
\Big].
\end{eqnarray}
In addition to the quadratic action for scalar perturbations, there is also a coupling between metric perturbations and the scalar perturbations. The stability of fluctuations depend strongly on the presence of two particular couplings in the Lagrangian (\ref{decoupling}), one given by $\partial_{\mu} \pi \partial_{\nu} \pi T^{\mu \nu}$ and the other by $\hat{h}^{\mu \nu} X^{(3)}_{\mu \nu}$ , which have $\alpha$ and $\beta$ as coefficients, respectively. We will consider their consequences in what follows, by analysing each relevant case separately \cite{Koyama:2013paa}.

\begin{itemize} 
\item
$\alpha=\beta=0$. \\
In this case
 the action (\ref{decoupling}) reduces to the galileon action introduced by Ref.~\cite{Nicolis:2008in}. Interestingly, it was found that the stability conditions force the the speed of sound for fluctuations propagating radially to be greater than one. On the other hand, if the fourth order galileon term ${\cal L}^{\rm gal}_{4}$ is included, then $K_{\Omega} \ll K_{t}, K_{r}$ and the propagation of angular fluctuations is extremely subluminal, invalidating a quasi-static approximation.

\item $\alpha  \neq 0$ and $\beta$=0. \\
A new feature of this class of models is the disformal coupling between matter and the scalar, namely $\partial_{\mu} \pi \partial_{\nu} \pi T^{\mu \nu}$. As pointed out in Ref.~\cite{Berezhiani:2013dw}, this coupling has deep implications for the stability. Inside a matter source, the coupling introduces a kinetic term proportional to $\alpha \rho (\partial_t \varphi)^2$. If $\alpha <0$, the fluctuations behave as ghosts, forcing us to choose $\alpha >0$. In the case of massive gravity, there is only one dimensionless parameter, $\alpha\equiv 1+3\alpha_3$ (in the theory where $\beta\equiv\alpha_3+4\alpha_4=0$), leading to the so-called restricted galileon \cite{Berezhiani:2013dw}. In this case, there is no Vainshtein solution that can be connected to the asymptotically flat solution. Around the non-flat asymptotic solution,  the radial propagation is subluminal, even though the extreme subluminality of the angular propagation persists \cite{Berezhiani:2013dw}.

\item $\alpha \neq 0$, $\beta \neq 0$ \\
In this case, the coupling between the metric and the scalar $\hat{h}^{\mu \nu} X^{(3)}_{\mu \nu}$ gives a different picture from the previous cases. Inside the Vainshtein radius the solution that reduces to GR near a matter source only exists if $\beta>0$, and the solution is given by \cite{Chkareuli:2011te,Sbisa:2012zk}
\begin{equation}\label{solx0}
x_0=\pm \sqrt{\frac{\xi}{3 \beta}}.
\end{equation}
Since $x$ is constant and $A(r)\gg 1$ inside the Vainshtein radius, all functions $K_i$ in (\ref{Ks}) are dominated by  the $y\sim A(r)\gg 1$ contribution in the limit $r \ll r_V$, as shown from its definition (\ref{ydef}). In the limit $A(r) \gg 1$, we can also  ignore the coupling between the scalar and metric perturbations. By taking into account only the contributions depending on $A= (r_V/r)^3$ outside the surface of the source $r>r_s$, we find in this limit
\begin{equation}
K_t=0, \quad K_r = -12 \beta A(r) x_0, \quad K_{\Omega}= 6 \beta A(r) x_0.
\end{equation}
Thus inside the Vainshtein radius $r \ll r_V$, but outside the source surface, the speed of the fluctuations are always superluminal, for both radial and angular directions. Moreover, given the fact that $K_\Omega$ and $K_r$ have opposite sign, all solutions are unstable.

\end{itemize}

These results indicate that it is not always possible to have a successful Vainshtein mechanism. For example, there is no stable Vainshtein solution that connects to the asymptotically flat spacetime in massive gravity theory discussed in section 2.6. See Refs~\cite{DeFelice:2011th, Kimura:2011dc, Babichev:2011iz, Babichev:2011iz, Kase:2013uja, Kobayashi:2014ida, Babichev:2013pfa} for the studies of the Vainshtein mechanism in the Horndeski, beyond Horndeski and bigraviy theories. 

Away from the decoupling limit, numerical methods are required to examine the solutions. Refs.~\cite{Babichev:2009jt, Babichev:2010jd} found solutions featuring the Vainshtein mechanism in the complete theory in the framework of a non-linear extension of Fierz-Pauli massive gravity. See \cite{Babichev:2013usa} and references therein. There have been several studies of the Vainshtein mechanism away from the spherically symmetric case (see for example \cite{deRham:2012fw, Chu:2012kz}) and weak gravity \cite{Kaloper:2011qc, Chagoya:2014fza}. 

There are also various theoretical problems associated with the non-linear interaction terms. The first problem is the {\it superluminality}. As we discussed earlier, if we consider perturbations around spherically symmetric solutions, the radial speed of propagation is larger than the speed of light in galileon models. This is a generic feature of galileons as the derivative interactions change the structure of light-cone for perturbations. The consequence of this superluminality is still debated \cite{Burrage:2011cr,Deser:2013qza, Deser:2014fta} (see \cite{deRham:2014zqa} for a summary of discussions). Another problem is the {\it strong coupling} problem. The quantum corrections become important at energy scales larger than $\Lambda_3$. In the DGP case, $\Lambda_3^{-1} =(\mpl r_c^{-2})^{-1/3}\sim 1000$km to satisfy the Solar System constraint. This strong coupling scale depends on the background as seen from Eq.~(\ref{vain-per}). Around the Vainshtein solution, the strong coupling scale is renormalised \cite{Nicolis:2004qq}
\begin{equation}
\Lambda_3^{\rm eff} = \left(\frac{r_V}{r} \right)^{3/4} \Lambda_3,
\end{equation} 
which reduces the strong coupling scale to 1cm around the Earth. This low cut-off scale is a generic problem in theories with galileon interactions such as massive gravity \cite{Burrage:2012ja}. Although quantum corrections are not under control above $\Lambda_3$, the galileon terms themselves do not get renormalised upon loop corrections so that their classical values can be trusted quantum mechanically \cite{Hinterbichler:2010xn}. Galileons also enjoy self-duality, i.e. various galileon terms are remapped to other galileon terms by a field dependent coordinate transformation and a field redefinition \cite{deRham:2013hsa}. It is even possible to find galileon terms that can be mapped to a free theory. Galileon duality was used to argue that the strong coupling and the superluminality problem are not fundamental \cite{Keltner:2015xda} while it was also argued that the Ultra Violet (UV) completion influences the low energy theory significantly \cite{Kaloper:2014vqa}. Clearly more work seems to be warranted to understand the UV properties of galileon models.  

\section{Cosmological tests of gravity}

\subsection{Consistency test}
Armed with the theoretical knowledge of modified gravity models, we now discuss how we test gravity on cosmological scales. The first approach is to test specific modified gravity models using observations. The models discussed in section 2 have been tested intensively by observations (see \cite{DeFelice:2010aj} for $f(R)$ gravity, \cite{Maartens:2010ar} for the DGP models and \cite{Barreira:2014jha, Neveu:2013mfa, Neveu:2014vua} for covariant galileons). We will not attempt to review observational tests of individual models. 

At the background level, the Friedman equation in modified gravity models can always be recast into the Friedman equation in GR with dark energy 
\begin{equation}
H^2 = \frac{8 \pi G}{3} (\rho + \rho_{\rm de}).
\end{equation}
By tuning the equation of state for dark energy $w_{\rm de} = P_{\rm de}/\rho_{\rm de}$ it is always possible to mimic the background expansion of the Universe in modified gravity models by dark energy. The equation of state is commonly parametrised as \cite{Chevallier:2000qy}
\begin{equation}
w_{\rm de}(a) = w_0 + w_a (1-a).  
\label{CPL}
\end{equation}
This degeneracy is broken if we include observables determined by the structure formation assuming that dark energy is smooth. The most commonly used parametrisation of the structure growth is to parametrise the growth rate, the logarithmic derivative of the dark matter density perturbation with respect to the scale factor, \cite{Linder:2005in} 
\begin{equation}
f = \frac{d \log \delta_m}{d \log a} = \Omega_m(a)^{\gamma}
\end{equation}
where $\delta_m$ is the density fluctuation of dark matter. In $\Lambda$CDM, $\gamma$ is approximately given by $\gamma \sim 0.55$. In dark energy models $\gamma$ is completely determined by the equation of state $w_{\rm de}$. However, in modified gravity models, $\gamma$ is not related to the equation of state. Thus if our universe were described by a modified gravity model but we tried to fit various observations including those measuring the background expansion of the Universe and the growth of the structure, we would find inconsistencies \cite{Ishak:2005zs, Linder:2005in}. Fig.~6 demonstrates this fact \cite{Shapiro:2010si}. 

%%%%%%%%%%%%%%%%%%%%%%%%%%%%%%%%%%%%%%%%%%%%%%%%%%%%%%%%%%%
\begin{figure}[h]
  \centering{
  \includegraphics[width=15cm]{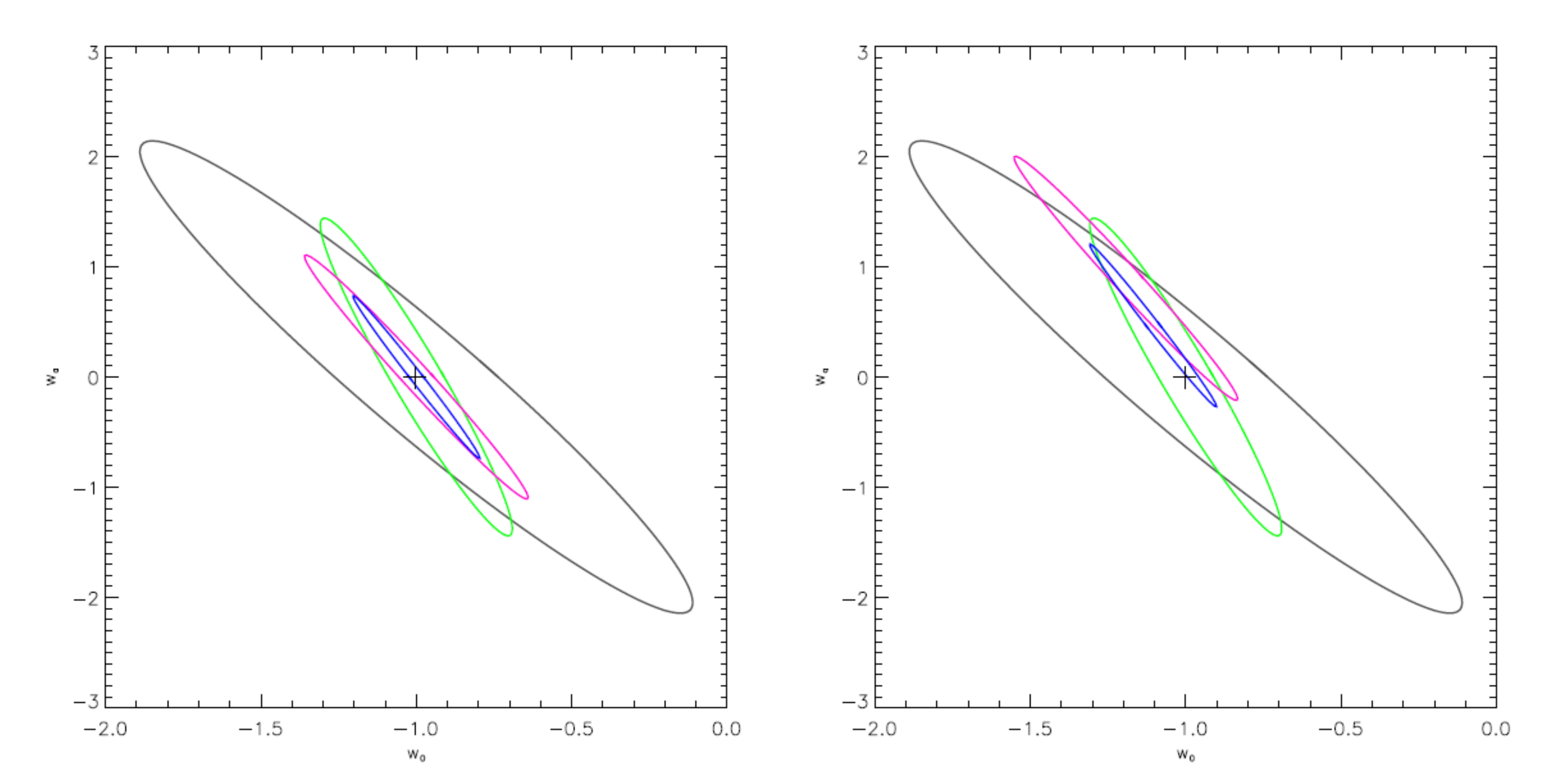}
  }
  \caption{Left: Forecasted 1 $\sigma$ constraints on the dark energy equation of state (\ref{CPL}) for the Dark Energy Survey \cite{DES}. From the largest to the smallest ellipse, the probes considered here are Baryon Acoustic Oscillations (black), Supernovae (green), cluster counts (magenta) and weak lensing (blue). The true model is assumed to be $\Lambda$CDM. Right: Same as the left figure but the true model is assumed to be a modified gravity model with $\gamma=0.68$. Due to the incorrect attempt to fit a GR + dark energy model to the data, the centres of the weak lensing and cluster counts ellipses have moved. From \cite{Shapiro:2010si}.}
\end{figure}
%%%%%%%%%%%%%%%%%%%%%%%%%%%%%%%%%%%%%%%%%%%%%%%%%%%%%%%%%%%%%

We can formulate the consistency test of $\Lambda$CDM formally using linear cosmological perturbation theory \cite{Jain:2007yk, Song:2008vm}. In $\Lambda$CDM the basic equations that govern the expansion of the universe and the growth of structure are given by
\begin{eqnarray}
H^2 &=& \frac{8 \pi G}{3} \rho_T, \\
\frac{k^2}{a^2} \Psi &=& 4 \pi G \rho_T \delta_T,
\end{eqnarray}
where $\rho_T$ is the total energy density and $\delta_T$ is its fluctuation in the comoving gauge. Here we are working in the Fourier space for perturbations. 
These equations are supplemented by the conservation of energy momentum tensor. Let's assume that the dominant component of the energy density in the late-time universe is dark matter, $\rho_T=\rho_m$, $\delta_T= \delta_m$. The continuity and Euler equations give
\begin{eqnarray}
\dot{\delta}_m + \frac{1}{a} \theta_m - 3 \dot{\Phi} =0, \\
\dot{\theta}_m + H \theta_m - \frac{k^2}{a} \Psi =0, 
\label{Euler}
\end{eqnarray}
where $\theta_m$ is the velocity divergence $\theta_m = \nabla \cdot {\bf v}_m$. 
Under horizon scales, we can neglect $\dot{\Phi}$ in the continuity equation and we find a second order equation describing the growth of $\delta_m$
\begin{equation}
\ddot{\delta}_m + 2 H \dot{\delta}_m = -\frac{k^2}{a^2} \Psi.
\end{equation} 
This equation describes how the density perturbation grows due to the effect of the Newtonian potential $\Psi$. Thus the modification of gravity predicts a different growth rate. 

There are a number of cosmological observations that measure different geometrical quantities (see \cite{ Joyce:2014kja, Linder:2008pp, Silvestri:2009hh} for reviews). The Cosmic Microwave Background (CMB), Supernovae (SNe) and the Baryon acoustic oscillations (BAO) measure the distances in the background universe, hence the expansion history of the Universe $H(z)$. Weak gravitational Lensing (WL) measures the distortion of galaxies due to the deflection of light, which is determined by the lensing potential $\Phi + \Psi$. The time variation of the lensing potential changes the temperature of 
the CMB photons through the Integrated Sach-Wolfe (ISW) effects. Peculiar velocities of galaxies follow the velocity of dark matter $\theta_m$, which changes the distribution of galaxies in the redshift space. This Redshift Space Distortion (RSD) can be used to measure the velocity of dark matter. The galaxy distribution is a biased tracer of the underlying dark matter distribution. Although it is very difficult, in principle, we can reconstruct the dark matter density perturbation $\delta_m$ from the distribution of galaxies.     

By combining the Friedman equation and the Poission equation, we obtain the following equation \cite{Song:2008vm, Song:2008xd}
\begin{equation}
\alpha(k, t) = \frac{2 k^2}{3 a^2 H^2} \frac{(\Phi + \Psi) - \Psi}{\delta_m}=1. 
\label{consistency}
\end{equation}
This is the identity in $\Lambda$CDM. Let's examine each term in this expression. First $(k/a)$ is the physical wave number of the perturbations that we are interested in. The background expansion $H$ can be measured from CMB, SNe and BAO. We split $\Phi$ into $\Psi + \Phi$ and $- \Psi$. $\Psi + \Phi$ is the lensing potential which can be measured from WL and ISW. On the other hand, the Newton potential $\Psi$ governs the dynamics of galaxies. The peculiar velocities of the galaxies are determined by the Newtonian potential $k^2 \Psi =d (a \theta_m)/d t$ (Eq.~(\ref{Euler})) thus using RSD, we can reconstruct $\Psi$. Finally, from the galaxy distribution $\delta_g$ we can in principle reconstruct $\delta_m$ if we know bias $b=\delta_g/\delta_m$. This means that we have just enough observables to check the identity (\ref{consistency}). This a very strong consistency condition of $\Lambda$CDM: $\alpha(k,t)$ is unity at any time and space in $\Lambda$CDM. Any modifications to $\Lambda$CDM lead to deviations of $\alpha(k, t)$ from unity. 

The problem with this consistency relation is that it requires the measurement of the dark matter density perturbation $\delta_m$. This is difficult due to the uncertainties in galaxy bias. A less ambitious test is to combine weak lensing and peculiar velocity measurements only. The $E_g$ parameter is introduced as follows \cite{Zhang:2007nk}
\begin{equation}
E_G = \frac{\nabla^2 (\Phi +\Psi)}{-3 H_0^2 a^{-1} \theta_m}
\end{equation}
In $\Lambda$CDM, $E_G=\Omega_m/f$. This estimator can be constructed directly from observations for example using the lensing-galaxy cross power spectrum and the velocity-galaxy power spectrum \cite{Reyes:2010tr}.

\subsection{Parametrisations} 
The degeneracy between dark energy models and modified gravity models exist in principle for perturbations. One can find GR models with dark energy that have anisotropic stress and variable sound speed, which can in principle mimic
modified gravity models~\cite{Kunz:2006ca}. Mathematically, this is always the case as we can define an effective energy momentum tensor to absorb any effects of modification of gravity  
\begin{equation}
G_{\mu \nu} = 8 \pi G (T_{\mu \nu}  + E_{\mu \nu} ).
\end{equation}
At the background it is enough to specify the equation of state $w_{\rm E}=P_{\rm E}/\rho_{\rm E}$ for the fluid whose energy-momentum tensor is given by $E_{\mu \nu}$. For linear perturbations, we need to specify the sound speed $c_{\rm E}^2$ and the anisotropic stress $\pi_{\rm E}$. Once we find deviations from $\Lambda$CDM, we need theoretical models which predict these unknown quantities to distinguish between dark energy models in GR and modified gravity models. For example, modified gravity models generally predict a large anisotropic stress $\pi_E \sim \rho_{\rm E} \delta_{\rm E}$, which is difficult to find in physical fluid models. Thus it is important to go beyond the consistency test of $\Lambda$CDM and find precisely how the deviations from $\Lambda$CDM appear. 

In general, deviations from $\Lambda$CDM can be parametrised by two functions of space (wavenumber $k$) and time (redshift $z$) that characterise the relation between the geometry and matter. For linear perturbations, one can parametrise the relation between the Newton potential to the density perturbations and the relation between the two metric perturbations \cite{Caldwell:2007cw, Amendola:2007rr}
\begin{eqnarray}
k^2 \Psi = 4 \pi G \mu(k, z) a^2 \rho_m \delta_m, \\
\frac{\Phi}{\Psi} =\eta(k, z).
\label{param}
\end{eqnarray}
Another useful parametrisation which can be constructed from $\mu$ and $\eta$ is to parametrise the relation between the lensing potential and the density perturbation
\begin{eqnarray}
\k^2 (\Psi + \Phi) = 8 \pi G \Sigma(k, z) a^2 \rho_m \delta_m.
\end{eqnarray}
where $\Sigma =\mu (1 + \eta)/2$. 

These functions are unity $\mu(k,z)=\eta(k,z)=\Sigma(k,z)=1$ in $\Lambda$CDM. If we modify gravity they deviate from one. For example, in the BD gravity with a mass term, these functions are given by 
\begin{equation}
\Sigma(k,a) = 1, \quad 
\mu(k,a) = \frac{2 (2 + \omega_{BD}) + m^2 a^2/k^2}{3 + 2 \omega_{BD} + m^2 a^2/k^2},
\end{equation}
under horizon scales. Even in dark energy models, if dark energy clusters but if we do not know their existence, this looks like a modification of gravity for dark matter. Assuming that dark energy has no anisotropic stress, we predict 
\begin{equation}
\mu(k,a) = \Sigma(k,a) = 1+ \frac{\rho_{DE} \delta_{DE}}{\rho_m},
\end{equation}
where $\delta_{DE}$ is the density perturbation of dark energy. In this way, theories beyond the standard $\Lambda$CDM model predict distinct paths in the ($\mu$, $\Sigma$) plane \cite{Song:2010rm}.  

The question is how well we can constrain these two functions of time and space from observations. There are many attempts to put constraints on $\mu$ and $\Sigma$ (or $\eta$) \cite{Daniel:2010ky, Zhao:2010dz, Bean:2010zq, Song:2010fg, Daniel:2010yt, Dossett:2011tn, Simpson:2012ra, Hu:2013aqa}. Given the accuracy of current observations, we are forced to assume particular time and space dependences of these functions. One parametrisation is to assume that these functions are scale invariant and their time dependence is determined by the density parameter associated with the cosmological constant \cite{Simpson:2012ra}
\begin{equation}
\mu(a, k) = \mu_0 \Omega_L(a), \quad 
\Sigma(a, k) = \Sigma_0 \Omega_L(a),
\end{equation}
The Planck 2015 paper \cite{Ade:2015rim} gave constraints on $\mu_0$ and $\Sigma_0$ as shown in Fig.~7 by combining Planck CMB observations with various external data (BAO, WL, RSD).   
%%%%%%%%%%%%%%%%%%%%%%%%%%%%%%%%%%%%%%%%%%%%%%%%%%%%%%%%%%%
\begin{figure}[ht]
  \centering{
  \includegraphics[width=15cm]{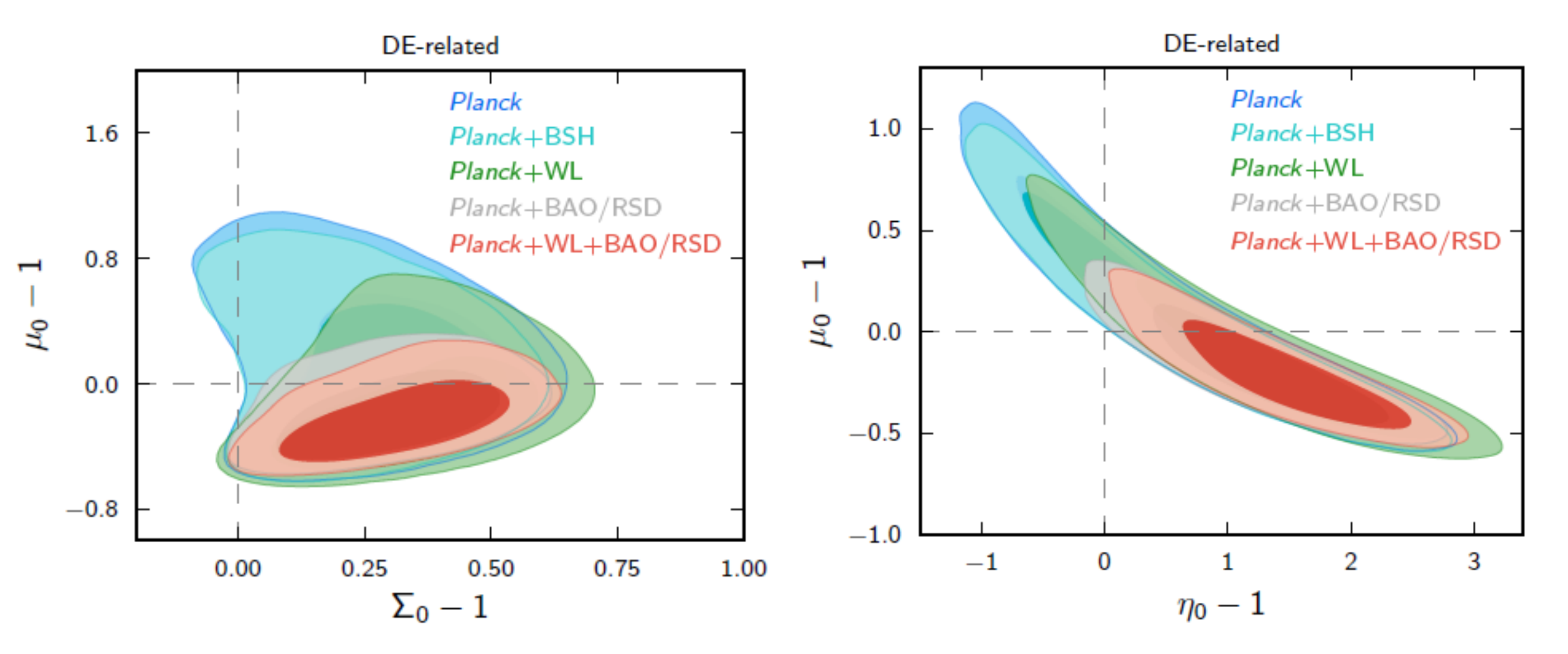}
  }
  \caption{$68\%$ and $95\%$ contour plots for the two parameters  $\Sigma_0$ and $\mu_0$ obtained from Planck CMB measurements combined with WL, BAO and RSD. From \cite{Ade:2015rim}.}
\end{figure}
%%%%%%%%%%%%%%%%%%%%%%%%%%%%%%%%%%%%%%%%%%%%%%%%%%%%%%%%%%%%%
The $\Lambda$CDM model is in tension with the data at 3$\sigma$ level if the Planck data is combined with WL and RSD/BAO. This tension is reduced to 1.7$\sigma$ level if we include the CMB lensing. The WL and RSD measurements will be dramatically improved in the next five years. It is very interesting to see whether this tension remains with improved measurements. 

In order to improve the constraints, it is important to combine RSD and WL. RSD is determined by the peculiar velocity of galaxies thus it is sensitive only to $\mu(k,z)$. On the other hand, the weak lensing is determined by the lensing potential so it is sensitive to $\Sigma(k,z)$ and also $\mu(k,z)$ as the growth of the dark matter density is determined by $\mu(k,z)$. Thus combining these two probes it is possible to break the degeneracy. Fig.~8 demonstrates this point where the time dependence of $\mu(a)$ and $\Sigma(a)$ is assumed to be 
$\mu(a) = \mu_s (1 +a^s), \Sigma(a)= \Sigma_s (1 + a^s)$ (they are assumed to be scale invariant) \cite{Song:2010fg}. This figure also demonstrates that the accuracy of the constraint strongly depends on the assumed time dependence. 

%%%%%%%%%%%%%%%%%%%%%%%%%%%%%%%%%%%%%%%%%%%%%%%%%%%%%%%%%%%
\begin{figure}[h]
  \centering{
  \includegraphics[width=15cm]{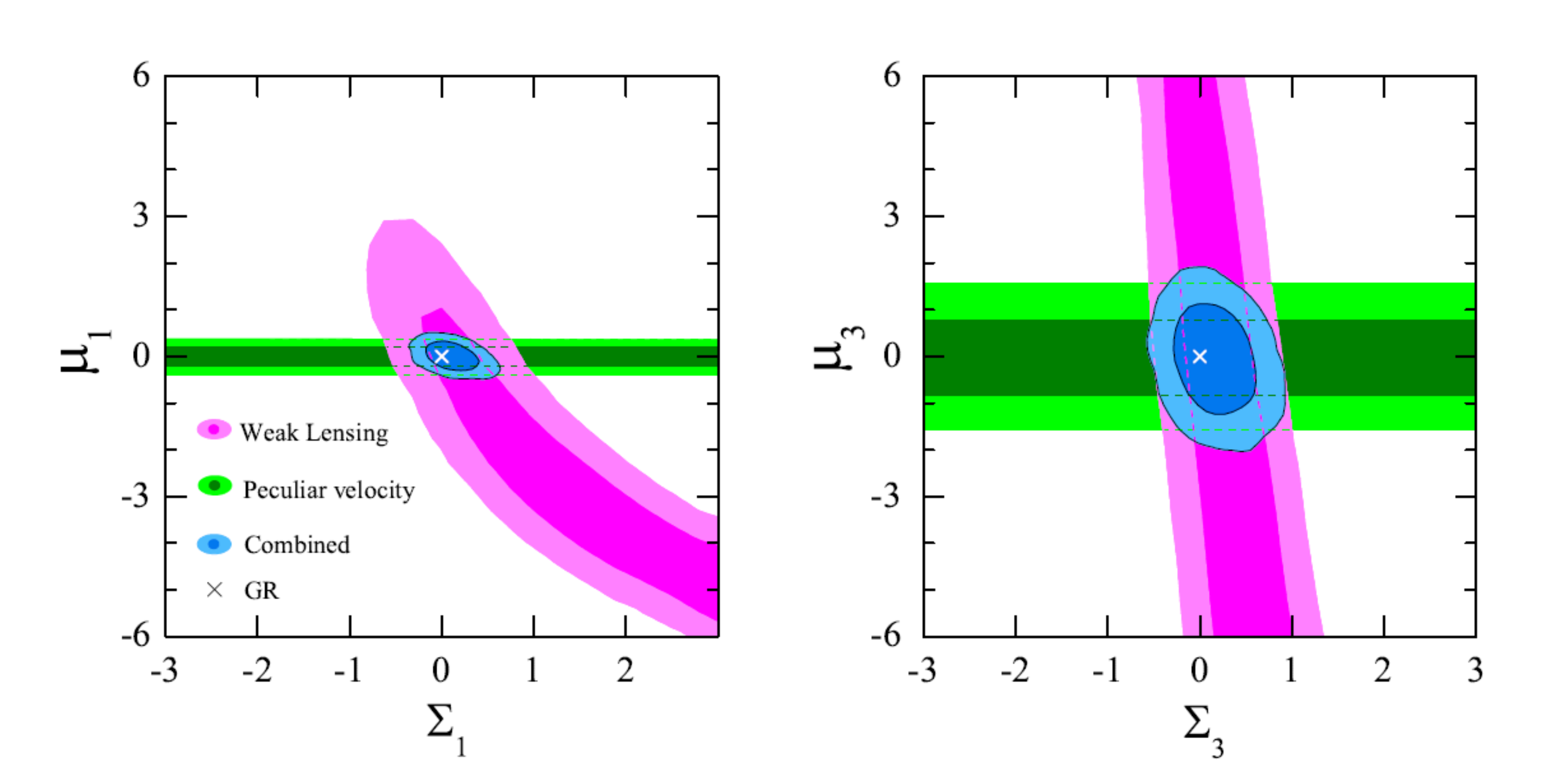}
  }
  \caption{Constraints on $\mu(a)$ and $\Sigma(a)$. In the left (right) panel, the time dependence is assumed to be $s=1$ ($s=3$). The constraints are obtained using WL (CFHTLS) and RSD (SDSS DR7 LRG). From \cite{Song:2010fg}.}
\end{figure}
%%%%%%%%%%%%%%%%%%%%%%%%%%%%%%%%%%%%%%%%%%%%%%%%%%%%%%%%%%%%%

If we do not wish to assume any theory, i.e. the functional forms of these functions, the best strategy is to simply make bins in terms of redshifts and wave-numbers and treat $\mu$ and $\eta$ as free parameters in each bin. The problem is that errors on these functions are highly correlated. Principal Component Analysis (PCA) provides a way to de-correlate the errors by creating linear combinations of these parameters \cite{Zhao:2008bn, Zhao:2009fn,Hojjati:2011xd}. Let's consider $m$ z-bins and $n$ k-bins. There are $2 \times m \times n$ parameters for $\mu$ and $\gamma$.  The covariant matrix is given by 
\begin{equation}
C_{ij} = \langle (p - \bar{p}_i) (p - \bar{p}_j) \rangle ,
\end{equation}
where $\bar{p}_i$ is the best fit values. The covariant matrix will be non-diagonal as the individual pixels of $\mu$ and $\eta$ are highly correlated. We can diagonalise the covariant matrix 
\begin{equation}
C = W^{T} \Lambda W, \quad  \Lambda_{ij} = \lambda_i \delta_{ij}.
\end{equation}
Using the matrix W, we can construct the linear combinations of $p_i$:
\begin{equation}
\alpha_i = \sum W_{ij} (p_j - \bar{p}_j).
\end{equation}
The error on $\alpha_i$ is given by $\lambda_i$ and they are uncorrelated. The parameter $p_i$ is formally expanded by the eigenvectors $W_j(k,z)$ 
\begin{equation}
p_i(k,z) = \bar{p}_i + \sum_i \alpha_j W_j (k,z).
\end{equation}

Fig.~9 shows the number of PCA eigenmodes and their uncertainties for an Euclid-like survey, which combines imaging (2D, WL) and spectroscopic (3D, RSD) surveys \cite{Asaba:2013xql}. We can clearly see that by adding the information from RSD to WL, the constraints on the eigenmodes are significantly improved. 

%%%%%%%%%%%%%%%%%%%%%%%%%%%%%%%%%%%%%%%%%%%%%%%%%%%%%%%%%%%
\begin{figure}[ht]
  \centering{
  \includegraphics[width=13cm]{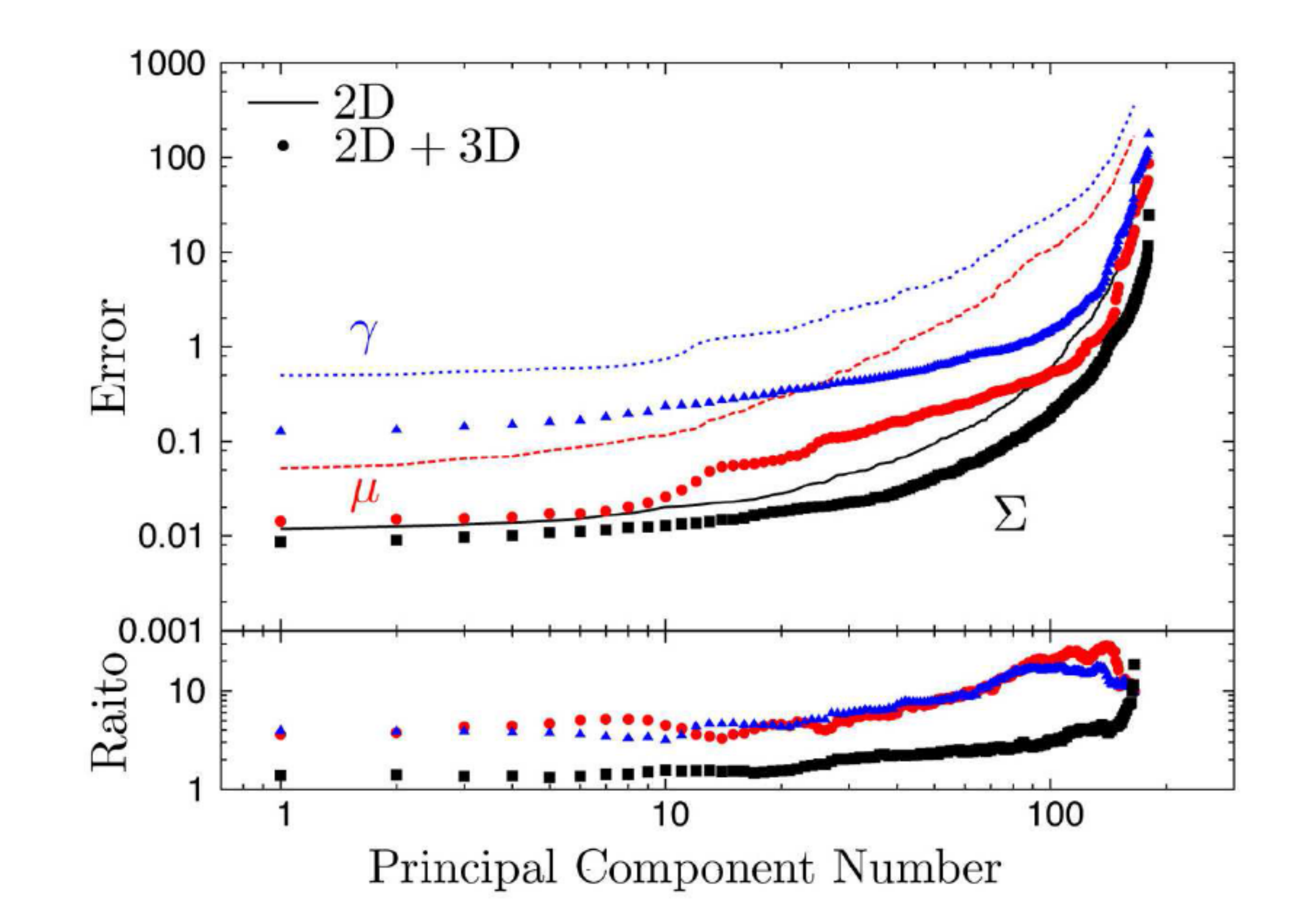}
  }
  \caption{The number of PCA eigenmodes and their uncertainties for an Euclid-like survey. Here $\gamma=\eta$. The lower panel shows the improvements of the errors when 3D (RSD) information is added to 2D (WL) measurements. From \cite{Asaba:2013xql} (published on 19 August 2013 \copyright SISSA Medialab Srl.  Reproduced by permission of IOP Publishing.  All rights reserved).}
\end{figure}
%%%%%%%%%%%%%%%%%%%%%%%%%%%%%%%%%%%%%%%%%%%%%%%%%%%%%%%%%%%%%

\subsection{Effective field theory approach}
Although the parameterisations Eq.~(\ref{param}) are the most general form for the linear perturbations, there is still a gap between the constraints on $\mu$ and $\Sigma$ and the predictions of theoretical models. Firstly, these are the parameterisations of the solutions for metric perturbations and the density perturbation. In principle we need to solve a complete set of equations for linear cosmological perturbations in a given theory to find solutions for $\mu$ and $\Sigma$. Thus these functions depend not only on the parameters of theories but also initial conditions. Secondly, in general, the modified linearised Einstein equations do not have the form given by Eq.~(\ref{param}) except for the case in which the quasi-static approximation holds. 

Several approaches have been developed to make the connection to theories more transparent. One approach is to parametrise the linearised modified Einstein equations directly \cite{Baker:2011jy, Zuntz:2011aq, Baker:2011wt, Baker:2012zs, Battye:2013aaa,Battye:2013ida,Battye:2014xna, Soergel:2014sna}. Another approach is to build the effective action \cite{Park:2010cw, Bloomfield:2011np, Bloomfield:2012ff,Gubitosi:2012hu, Gleyzes:2013ooa, Piazza:2013coa, Bloomfield:2013efa,Gergely:2014rna, Piazza:2013pua} following the effective field theory approach developed for inflationary models \cite{Weinberg:2008hq, Cheung:2007st}. Here we follow the latter approach. We consider a single scalar field $\phi(t)$ on the FRW background. The presence of the scalar field breaks the time diffeomorphism. We can use $\phi(t)$ to define a preferred time slicing $\phi=$const. We construct the action using the invariant quantities under the spatial diffeomorphisms $x^i \to x^i + \xi^i$. It is natural to use the intrinsic curvature ${}^{(3)} R$ and the extrinsic curvature $K_{\mu \nu}$ of the $\phi=$const. hypersurface. The extrinsic curvature is defined as $K_{\mu \nu} = h^{\sigma}_{\mu} \nabla_{\sigma} n_{\nu}$
where $n^{\mu}$ is the unit vector perpendicular to the hypersurface
\begin{equation}
n_{\mu} = - \frac{ \partial_{\mu} \phi}{\sqrt{-(\partial \phi)^2}}, 
\end{equation}  
and $h_{\mu \nu} = g_{\mu \nu} - n_{\nu} n_{\mu} $ is the projection tensor.
In addition, $g^{00}$ is invariant under spatial diffeomorphisms. If we demand that the equations of motion are at most second order, the most general action up to quadratic order can be written as \cite{Bloomfield:2012ff, Gubitosi:2012hu}
\begin{eqnarray}
S= \int d^4 \sqrt{-g} \Big[
\frac{M_*^2}{2} f(t) R - \Lambda(t) - c(t) g^{00} 
+ \frac{M_2(t)^4}{2} (\delta g^{00})^2 
\nonumber\\
- \frac{m_3(t)^3}{2} \delta K \delta g^{00}
- m_4^2(t) (\delta K^2 - \delta K^{\mu}_{\nu}  \delta K^{\nu} K_{\mu} )
+ \frac{\tilde{m}_4(t)^2}{2} {}^{(3)} R \delta g^{00}  \Big],
\end{eqnarray}
where $f(t), \Lambda(t), c(t), M_2(t), m_3(t), m_4(t)$ and $\tilde{m}_4 (t)$ are free functions of time. From this action, it is possible to derive the modified linearised Einstein equation. These equations have been implemented to the linear Einstein-Boltzmann code \cite{Hu:2013twa,Hu:2014oga, Raveri:2014cka}. 

Based on the effective field theory, it was found that there exists a large model space that mimics $\Lambda$CDM on all linear quasistatic subhorizon scales as well as in the background evolution \cite{Lombriser:2014ira}. The effective field theory approach allows us to go beyond the quasistatic approximations and test these models using the measurement of the relativistic contributions to galaxy clustering \cite{Lombriser:2013aj}.

It is clear that this approach has a strong connection to the ADM formulation of the Horndeski theory discussed in section 2.5. In fact in the Horndeski theory, there is a special relation between $m_4$ and $\tilde{m}_4$, $m_4=\tilde{m}_4$. This relation is broken by beyond Horndeski theories. In fact this was the way beyond Horndeski theories were discovered \cite{Gleyzes:2014dya}. 

Under the quasi-static approximations, it is possible to derive the form of $\mu$ and $\eta$ in the case of  $m_4=\tilde{m}_4$ \cite{DeFelice:2011hq,Motta:2013cwa, Silvestri:2013ne}
\begin{equation}
\eta(k,a) =  \left( \frac{p_1(a) + p_2(a) k^2 }{1 +  p_3(a) k^2}  \right)
\quad 
\mu(k, a) = \left( \frac{1 +  p_3(a)k^2}{p_4(a) +  p_5(a)k^2}  \right). 
\end{equation}
where $p_i(a)$ are free functions of time determined by the six functions in the effective action. See Refs.~\cite{Baker:2014zva, Baker:2014zva} for related approaches.

Although the theoretical prior such as the requirement that equations of motion are at most second order significantly simplified the form of the two functions $\mu(k,z)$ and $\eta(k,z)$, there are still five free functions of time. The PCA analysis has been performed for these functions. Due to the degeneracies, it was shown that these functions are not constrained individually even by a future weak lensing survey. On the other hand, about 10 eigenmodes of the combination of all $p_{\alpha}$ will be measured with uncertainty smaller than 1$\%$ of the prior by the Large Synoptic Survey Telescope (LSST) project \cite{LSST}. One approach is to reconstruct $\mu(k,a)$ and $\gamma(k,a)$ using eigenmodes of $p_i(a)$ following the approach to reconstruct the equation of state $w(a)$ using the smoothness prior on $p_i(a)$ functions \cite{Crittenden:2005wj, Crittenden:2011aa}. 

\subsection{Quasi non-linear scales} 
On linear scales, it is possible to develop model independent approaches as we described in 4.3. However, once the non-linearity becomes important, it becomes difficult to develop model independent parameterisations of the non-linear equations. On quasi non-linear scales, it is still possible to develop a general framework based on perturbation theory \cite{Koyama:2009me}. In models with a screening mechanism, the scalar field equation becomes non-linear.  Let's consider, for example, BD gravity with non-linear interactions terms 
%%%%%%%%%%%%%%%%%%%%%%%%%%%%%%%%%%%%%%%%%%%%%%%%%%%%%%%%%%%%%%%%%%%%%%
\begin{equation}
(3 +2 \omega_{\rm BD}) \frac{1}{a^2} k^2 \varphi
= - 8 \pi G \rho_m \delta - {\cal I}(\varphi),
\label{eq:BD_eq}
\end{equation}
%%%%%%%%%%%%%%%%%%%%%%%%%%%%%%%%%%%%%%%%%%%%%%%%%%%%%%%%%%%%%%%%%%%%%%
in a Fourier space. Here the interaction term ${\cal I}$ can be expanded as
%%%%%%%%%%%%%%%%%%%%%%%%%%%%%%%%%%%%%%%%%%%%%%%%%%%%%%%%%%%%%%%%%%%%%%
\begin{eqnarray}
{\cal I} (\varphi)
= M_1(k) \varphi + \frac{1}{2} \int \frac{d^3 \bfk_1 d^3 \bfk_2}
{(2 \pi)^3} \delta_D(\bfk -\bfk_{12}) M_2(\bfk_1, \bfk_2)
\varphi(\bfk_1) \varphi(\bfk_2) \nonumber\\
+ \frac{1}{6}
\int \frac{d^3 \bfk_1 d^3 \bfk_2 d^3 \bfk_3}{(2 \pi)^6}
\delta_D(\bfk - \bfk_{123}) M_3(\bfk_1, \bfk_2, \bfk_3)
\varphi(\bfk_1)\varphi(\bfk_2) \varphi(\bfk_3),
\end{eqnarray}
%%%%%%%%%%%%%%%%%%%%%%%%%%%%%%%%%%%%%%%%%%%%%%%%%%%%%%%%%%%%%%%%%%%%%%
where $\bfk_{ij}=\bfk_i+\bfk_j$ and $\bfk_{ijk}=\bfk_i+\bfk_j+\bfk_k$.

The potential $\Psi$ is coupled to $\delta$ through the BD scalar $\varphi$ in a fully non-linear way due to the interaction term ${\cal I}$. To derive the closed equations for $\delta_m$ and $\theta_m$, we must employ the
perturbative approach to Eq.~(\ref{eq:BD_eq}). By solving Eq.~(\ref{eq:BD_eq})
perturbatively assuming $\varphi<1 $, $\Psi$ can be expressed in
terms of $\delta$ as \cite{Koyama:2009me}
%%%%%%%%%%%%%%%%%%%%%%%%%%%%%%%%%%%%%%%%%%%%%%%%%%%%%%%%%%%%%%%%%%%%%%
\begin{equation}
-\left(\frac{k}{a}\right)^2\Psi=\frac{1}{2}\,\kappa^2 \,\rho_{m}\,
\left[1+ \frac{1}{3}
\frac{(k/a)^2}{\Pi(k)}\right]
\,\delta_m(\bfk) +\frac{1}{2}\left(\frac{k}{a}\right)^2\,S(\bfk),
\label{eq:modified_Poisson}
\end{equation}
where
\begin{equation}
\Pi(k) = \frac{1}{3} \left((3+2 \omega_{\rm BD}) \frac{k^2}{a^2} + M_1 \right),
\end{equation}
%%%%%%%%%%%%%%%%%%%%%%%%%%%%%%%%%%%%%%%%%%%%%%%%%%%%%%%%%%%%%%%%%%%%%%
and $\kappa^2 = 8 \pi G$.
The function $S(\bfk)$ is the non-linear source term which is
obtained perturbatively as
%%%%%%%%%%%%%%%%%%%%%%%%%%%%%%%%%%%%%%%%%%%%%%%%%%%%%%%%%%%%%%%%%%%%%%
\begin{eqnarray}
S(\bfk)&=& -\frac{(\kappa^2\,\rho_{m})^2}{54 \Pi(\bfk)}\,
\int\frac{d^3\bfk_1d^3\bfk_2}{(2\pi)^3}\,
\delta_{\rm D}(\bfk-\bfk_{12}) M_2(\bfk_1, \bfk_2)
\frac{\delta_m(\bfk_1)\,\delta_m(\bfk_2)}{\Pi(\bfk_1)\Pi(\bfk_2)}
\nonumber\\
&-& \frac{(\kappa^2\,\rho_{m})^3}{486\,\Pi(\bfk)}\,
\int\frac{d^3\bfk_1d^3\bfk_2d^3\bfk_3}{(2\pi)^6}
\delta_{\rm D}(\bfk-\bfk_{123})
\Big\{M_3(\bfk_1, \bfk_2, \bfk_3) \nonumber\\
&& -\frac{M_2(\bfk_1,\bfk_2+\bfk_3) M_2(\bfk_2, \bfk_3)}{\Pi(\bfk_{23})}\Big\}
 \frac{\delta_m(\bfk_1)\,\delta_m(\bfk_2)\delta_m(\bfk_3)}
{\Pi(\bfk_1)\Pi(\bfk_2)\Pi(\bfk_3)}. \label{eq:Perturb3}
\end{eqnarray}
%%%%%%%%%%%%%%%%%%%%%%%%%%%%%%%%%%%%%%%%%%%%%%%%%%%%%%%%%%%%%%%%%%%%%%
The expression (\ref{eq:Perturb3}) is valid up to the third-order in $\delta_m$.
Combining this with energy-momentum conservation, it is possible to calculate the density perturbation and the velocity divergence perturbatively and compute the non-linear corrections to the linear power spectrum (Fig.~10) \cite{Koyama:2009me, Taruya:2013quf, Taruya:2014faa}. 
%%%%%%%%%%%%%%%%%%%%%%%%%%%%%%%%%%%%%%%%%%%%%%%%%%%%%%%%%%%%
\begin{figure}[ht]
  \centering{
  \includegraphics[width=11cm]{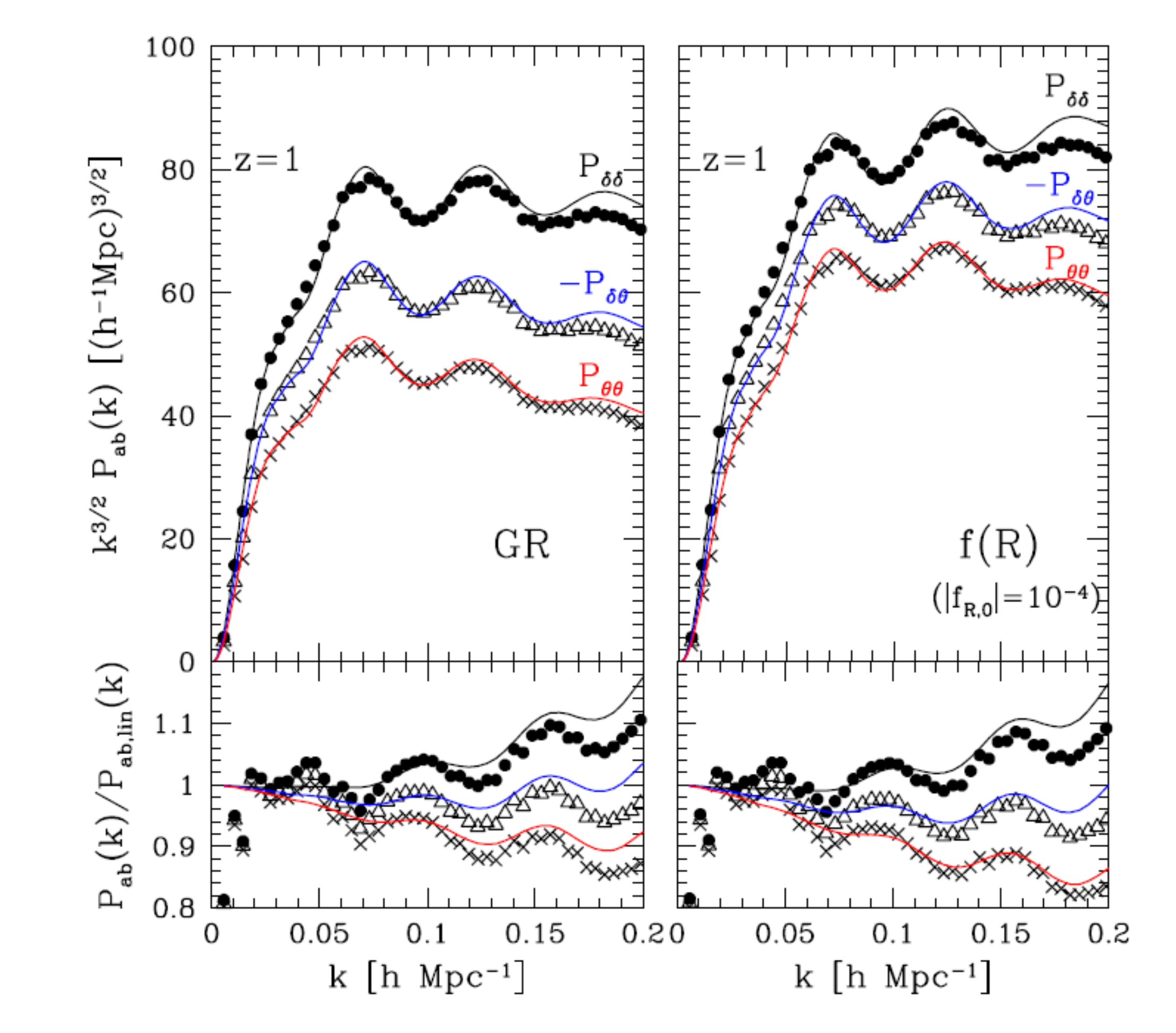}
  }
  \caption{The dark matter density-density, density-velocity, velocity-velocity power spectrum in $\Lambda$CDM (left) and $f(R)$ gravity model with $|f_{R0}|=10^{-4}$ (right). The solid lines are the results from the standard perturbation theory at the 1-loop level and symbols are results from simulations. From \cite{Taruya:2013quf}.} 
\end{figure}
%%%%%%%%%%%%%%%%%%%%%%%%%%%%%%%%%%%%%%%%%%%%%%%%%%%%%%%%%%%%%

However, we cannot observe these power spectra directly. What we observe is the power spectrum of the galaxy over-density in the redshift space \cite{Hamilton:1997zq, Scoccimarro:2004tg}. There has been significant progress is developing non-linear models of the redshift space power spectrum (see for example \cite{Reid:2011ar} and references therein). In order to demonstrate the importance of modelling the non-linear effect in the redshift space power spectrum accurately, we consider the perturbation theory based template for the redshift power spectrum \cite{Taruya:2010mx}
%%%%%%%%%%%%%%%%%%%%%%%%%%%%%%%%%%%%%%%%%%%%%%%%%%%%%%%%%%%%%%%%%%%%%%%
\begin{equation}
P^{\rm(S)}(k,\mu)=\DFoG[k\mu\,\sigmav] \times\,\Bigl\{
P_{\rm Kaiser}(k,\mu)+A(k,\mu)+B(k,\mu)
\Bigr\},
\label{eq:TNS_model}
\end{equation}
where $\mu=k_z/k$ and the observer's line-of-sight direction is taken to be the z-axis. The quantities $P_{\rm Kaiser}$, $A$, and $B$ are explicitly written as
%%%%%%%%%%%%%%%%%%%%%%%%%%%%%%%%%%%%%%%%%%%%%%%%%%%%%%%%%%%%%%%%%%%%%%%
\begin{eqnarray}
P_{\rm Kaiser}(k,\mu)=\Pdd(k)-2\,\mu^2\,\Pdv(k)+\mu^4\,\Pvv(k),
\label{eq:Kaiser}
\\
A(k,\mu) = -k\mu\,\int \frac{d^3\bfp}{(2\pi)^3} \,\,\frac{p_z}{p^2}
\left\{B_\sigma(\bfp,\bfk-\bfp,-\bfk)-B_\sigma(\bfp,\bfk,-\bfk-\bfp)\right\},
\label{eq:A_term}
\\
B(k,\mu)= (k\mu)^2\int \frac{d^3\bfp}{(2\pi)^3} F(\bfp)F(\bfk-\bfp)\,\,;
\label{eq:B_term}
\\
F(\bfp) = \frac{p_z}{p^2}
\left\{ \Pdv(p)-\frac{p_z^2}{p^2}\,\Pvv(p)\,\right\},
\nonumber
\end{eqnarray}
%%%%%%%%%%%%%%%%%%%%%%%%%%%%%%%%%%%%%%%%%%%%%%%%%%%%%%%%%%%%%%%%%%%%%%%
where $\Pdd$, $\Pvv$, and $\Pdv$ respectively denote the auto-power spectra
of the density and velocity divergence, and their cross power spectrum.
The function $B_\sigma$ is the cross bispectra defined by
%%%%%%%%%%%%%%%%%%%%%%%%%%%%%%%%%%%%%%%%%%%%%%%%%%%%%%%%%%%%%%%%%%%%%%%
\begin{eqnarray}
\left\langle \theta(\bfk_1)
\left\{\delta(\bfk_2)-\,\frac{k_{2z}^2}{k_2^2}\theta(\bfk_2)\right\}
\left\{\delta(\bfk_3)-\,\frac{k_{3z}^2}{k_3^2}\theta(\bfk_3)\right\}
\right\rangle
\nonumber\\
=(2\pi)^3\delta_D(\bfk_1+\bfk_2+\bfk_3)\,B_\sigma(\bfk_1,\bfk_2,\bfk_3).
\label{eq:def_B_sigma}
\end{eqnarray}
%%%%%%%%%%%%%%%%%%%%%%%%%%%%%%%%%%%%%%%%%%%%%%%%%%%%%%%%%%%%%%%%%%%%%%%
%%%%%%%%%%%%%%%%%%%%%%%%%%%%%%%%%%%%%%%%%%%%%%%%%%%%%%%%%%%
$\DFoG$ describes the damping of power due to non-linear velocity dispersions known as the Finger of God effect
\begin{equation}
D_{\rm FoG}(k\mu\,\sigmav) = \exp\left[-(k\mu\,\sigmav)^2\right].
\end{equation}
%%%%%%%%%%%%%%%%%%%%%%%%%%%%%%%%%%%%%%%%%%%%%%%%%%%%%%%%%%%
We treat $\sigmav$ as a free parameter and marginalise over it when obtaining the constraint on model parameters. 

%%%%%%%%%%%%%%%%%%%%%%%%%%%%%%%%%%%%%%%%%%%%%%%%%%%%%%%%%%%
\begin{figure}[ht]
  \centering{
  \includegraphics[width=11cm]{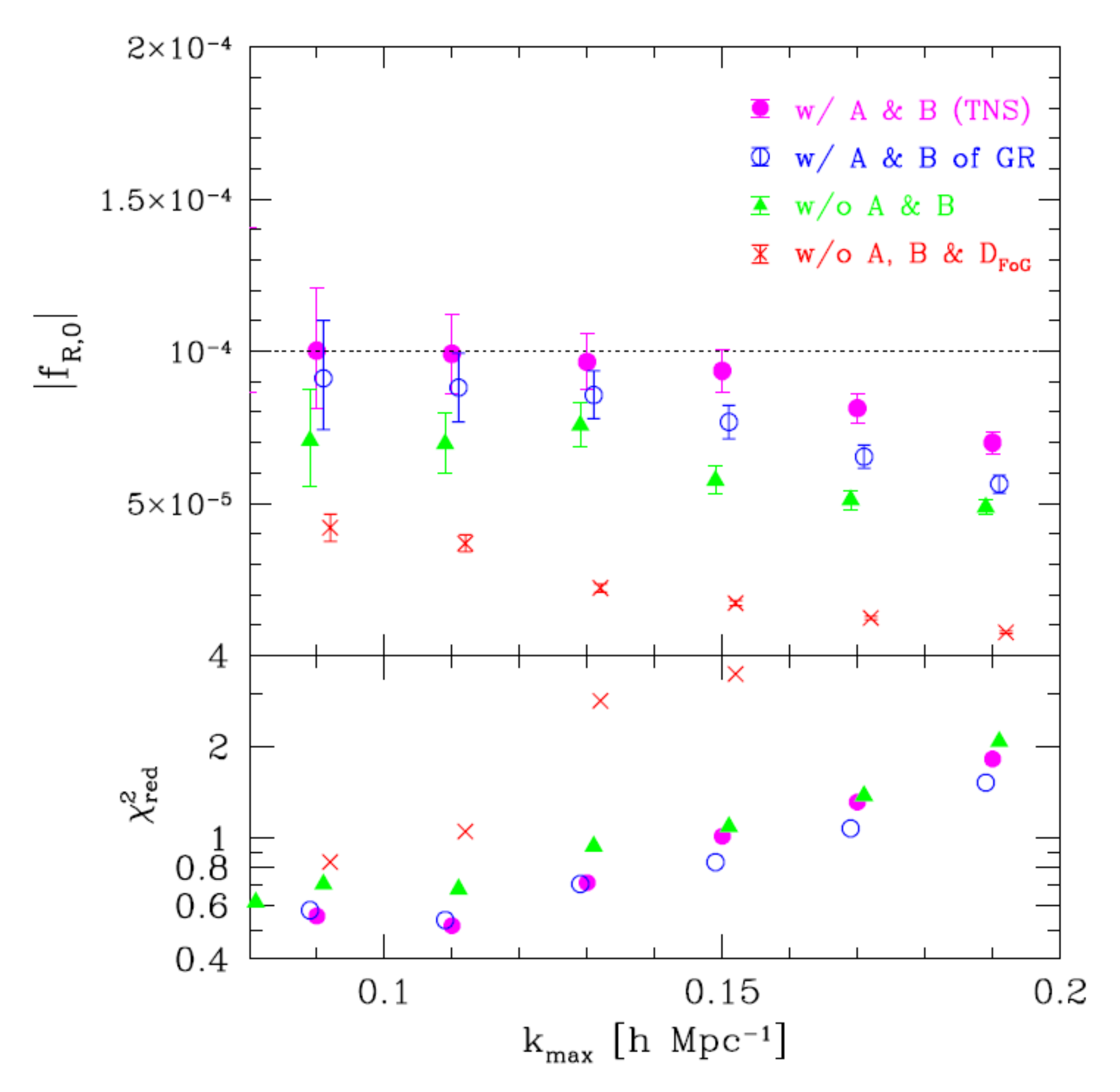}
  }
  \caption{Best-fit values of $f_{R0}$ as a function of the maximum wavenumber $k_{\rm max}$. We assume the cosmic variance limited survey of the volume $V=10 h^{-3}$Gpc$^3$, we fit the PT template to the N-body simulation with $|f_{R0}|=10^{-4}$ at $z=1$. Filled circles are the results based on (\ref{eq:TNS_model}) in $f(R)$ gravity, while filled triangles are the cases ignoring $A$ and $B$ terms. Open circles represents the results using $A$ and $B$ terms calculated in GR while crosses are the results ignoring the $A$ and $B$ terms and the damping term $D_{\rm FoG}$. From \cite{Taruya:2013quf}.}
\end{figure}
%%%%%%%%%%%%%%%%%%%%%%%%%%%%%%%%%%%%%%%%%%%%%%%%%%%%%%%%%%%%%

Fig.~11 demonstrates the importance of the non-linear corrections for an $f(R)$ gravity model described by (\ref{HS}) \cite{Taruya:2013quf}. We use the measured dark matter power spectrum in the redshift space at $z=1$ from N-body simulations for $f(R)$ gravity \cite{Jennings:2012pt} as a mock observation. We now extract the model parameter $|f_{R0}|$ by fitting the redshift space power spectrum. We decompose the redshift power spectrum into the multipole power spectra
%%%%%%%%%%%%%%%%%%%%%%%%%%%%%%%%%%%%%%%%%%%%%%%%%%%%%%%%%%%
\begin{equation}
P^{\rm(S)}_\ell(k)=\frac{2\ell+1}{2}\,\int_{-1}^{1}d\mu\,P^{\rm(S)}(k,\,\mu)
\,\mathcal{P}_\ell(\mu).
\end{equation}
%%%%%%%%%%%%%%%%%%%%%%%%%%%%%%%%%%%%%%%%%%%%%%%%%%%%%%%%%%%
Fitting the measured monopole $P_1^{(S)}$ and the dipole $P_2^{(S)}$ power spectra in simulations by the template (\ref{eq:TNS_model}) assuming that $|f_{R0}|$ and $\sigma_v$ are free parameters, we study how well we can recover the input parameter $|f_{R0}|$ in the simulation (in this case $|f_{R0}|=10^{-4}$). If we do not include the A and B terms in (\ref{eq:TNS_model}), the recovered $f_{R0}$ is biased. If we include A and B but use the $\Lambda$CDM results, again the recovered $f_{R0}$ is biased unless we cut off the power spectrum at low maximum wavenumber $k_{\rm max}<0.1h$ Mpc$^{-1}$. Only the correct template reproduces the input parameter in an unbiased way up to $k_{\rm max}= 0.15h$ Mpc$^{-1}$. This implies that we need to specify the non-linear interactions in order to constrain the modified gravity parameters accurately or it is required to use a very conservative cut-off in the wavenumber to remove the non-linear corrections. This is the major obstruction for model independent tests of gravity on linear scales. 

The same limitation applies to WL \cite{Beynon:2009yd}. If we use a conservative cut-off in the analysis to remove the non-linear scales, the discriminatory power of weak lensing is significantly reduced. However, in order to include non-linear scales, we need to accurately model the non-linear power spectrum. For weak lensing, the perturbation theory is not adequate and N-body simulations are required. 

\subsection{Non-linear scales}
In order to find fully non-linear clustering of dark matter, it is required to perform N-body simulations even in the standard $\Lambda$CDM model. In GR, the Poission equation is linear so it is possible to superpose the forces. This was used to speed up the calculations by separating the long-range and short-range forces. In modified gravity theories, the scalar field satisfies the non-linear equation. Thus it is no longer possible to superpose the forces. The non-linear scalar field equation needs to be solved directly.

There have been significant progress in developing N-body simulations \cite{Oyaizu:2008sr, Oyaizu:2008tb, Schmidt:2008tn, Schmidt:2009sg, Schmidt:2009sv, Chan:2009ew, Li:2009sy, Li:2010mqa, Li:2010re, Li:2010re, Li:2011vk, Brax:2012nk, Brax:2013mua, Li:2013nua, Barreira:2013eea,  Li:2013tda, Davis:2011pj, Llinares:2013jza, Llinares:2013jua, Hammami:2015iwa, 
Khoury:2009tk, Wyman:2013jaa, Puchwein:2013lza, Baldi:2013iza, Heidelberg:2013hfa}. Particularly, \texttt{ECOSMOG} \cite{Li:2011vk}, \texttt{ISIS} \cite{Llinares:2013jza} and \texttt{MG-GADET} \cite{Puchwein:2013lza} solve a non-linear scalar field equation on a mesh in N-body simulations using the adaptive mesh technique (see Ref.~\cite{Winther:2015wla} for a comparison between these codes). 
The \texttt{ECOSMOG} code for example has the following properties
\begin{enumerate}
\item It solves the scalar field on a mesh using the Newton-Gau{\ss}-Seidal nonlinear relaxation, which has good convergence properties. The density field on the mesh is obtained by assigning particles following the triangular-shaped-cloud scheme.
\item The mesh can be adaptively refined in high-density regions to achieve higher resolution and accuracy there, without affecting the overall performance. In order to speed up the convergence, a multigrid technique is used. 
\item It is efficiently parallelised, which makes the simulations fast.
\end{enumerate}

%%%%%%%%%%%%%%%%%%%%%%%%%%%%%%%%%%%%%%%%%%%%%%%%%%%%%%%%%%%
\begin{figure}[ht]
  \centering{
  \includegraphics[width=16cm]{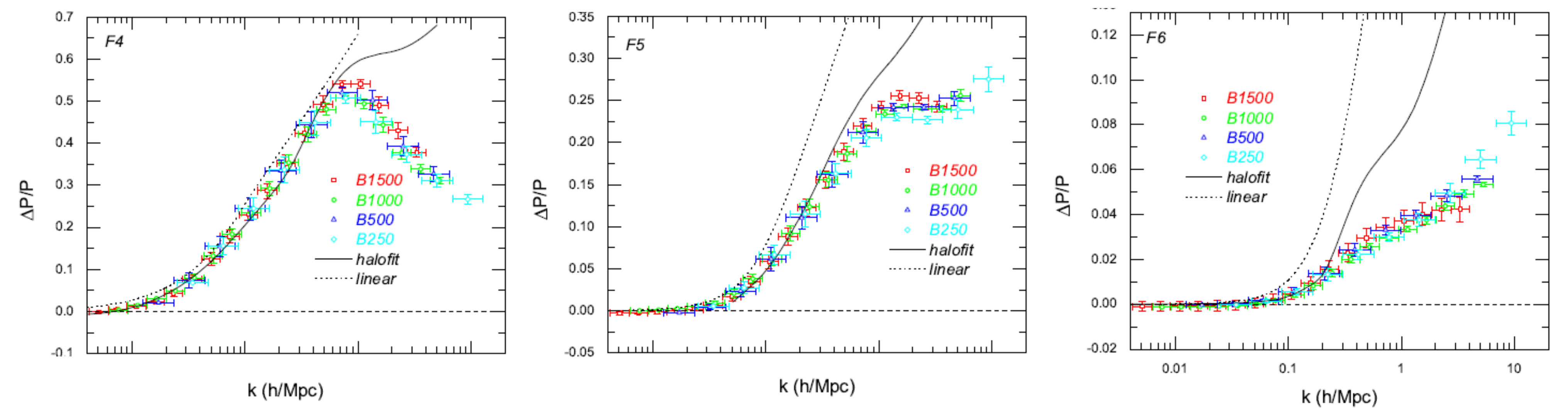}
  }
  \caption{The relative difference of the power spectra of $f(R)$ and $\Lambda$CDM models at $z=0$. See the text for details. From \cite{Li:2012by}}
\end{figure}
%%%%%%%%%%%%%%%%%%%%%%%%%%%%%%%%%%%%%%%%%%%%%%%%%%%%%%%%%%%%%
Fig.12 shows the power spectrum deviations from $\Lambda$CDM in the $f(R)$ gravity model (\ref{HS}) with three different $f_{R0}$ (F4: $|f_{R0}| =10^{-4}$, F5: $|f_{R0}| =10^{-5}$, F6: $|f_{R0}| =10^{-6}$ ) using various box sizes and resolutions obtained by \texttt{ECOSMOG} \cite{Li:2012by}. The chameleon mechanism works better for small $|f_{R0}|$ thus the power spectrum deviation from $\Lambda$CDM is more suppressed in F6. The dotted line is the linear theory prediction and the solid line is the prediction obtained by using the mapping formula ({\it halofit} \cite{Smith:2002dz}). The halofit model provides an accurate fitting formula for the non-linear matter power spectrum obtained from a suite of N-body simulations in the $\Lambda$CDM model. It provides a prediction for the non-linear power spectrum for a given linear power spectrum. The halofit works better in F4 but as soon as the screening becomes important it fails to predict the the power spectrum deviation at high wavenumber (on small scales). This is not surprising as the halofit is calibrated in the $\Lambda$CDM model and it does not capture the effect of the screening mechanism suppressing the deviation from $\Lambda$CDM on small scales. See Ref.~\cite{Zhao:2013dza} for a modification of {\it halofit} to account for the screening mechanism. This suppression of the power spectrum deviation from $\Lambda$CDM by the screening mechanism limits observational constraints. An interesting method has been proposed to enhance modified gravity effects by suppressing the contribution of the screened high-density regions in the matter power spectrum \cite{Lombriser:2015axa}. 

On small scales, the effects from baryons become important. Hydrodynamical simulations have been developed for $f(R)$ gravity and symmetron models Ref.~\cite{Puchwein:2013lza, Heidelberg:2013hfa, Hammami:2015iwa} but the degeneracy between modified gravity effects and baryonic physics remains to be understood. 

N-body simulations enable us to understand how the screening mechanisms studied in section 3 work in the large scale structure of the universe. In the next section, we will look at the effect of the screening mechanisms on the structure formation in the Universe based on N-body simulations. 

\section{Screening mechanisms in large scale structure}
In this section, we consider two screening mechanisms, the chameleon and Vainshtein mechanism that we discussed in section 3 and study how GR is recovered in the large scale structure of the Universe and how we can distinguish between the different screening mechanisms. This section is based largely on Ref.~\cite{Falck:2015rsa}.

The screening mechanisms are distinguished by how screened bodies fall in external fields \cite{Hui:2009kc}.  As a consequence of universal coupling, all un-self-screened test bodies fall in the same way and obey a microscopic equivalence principle. In the chameleon and symmetron models, screened bodies do not respond to external fields while in the Vainshtein mechanism they do, as long as those fields have wavelengths long compared to the Vainshtein radius \cite{Hui:2012jb}. These differences arise because of the non-superimposability of field solutions. Also, the peculiar structure of the non-linear interactions in the Vainshtein mechanism implies that the screening depends on the dimensionality of the system. For example, it does not work at all in one-dimensional systems. As a representative model for screening mechanism, we use the $f(R)$ gravity model (\ref{HS}) that includes the chameleon model and the normal branch DGP model (\ref{nDGP}) that exhibits the Vainshtein mechanism.  

\subsection{Screening of dark matter halos}
It is useful to construct analytic approximations for the scalar field solution inside dark matter halos. See Refs.~\cite{Schmidt:2008tn, Schmidt:2009yj, Schmidt:2010jr, Pourhasan:2011sm, Clampitt:2011mx, Li:2011uw, Lombriser:2013wta, Lombriser:2013eza, Barreira:2014zza, Falck:2014jwa, Falck:2015rsa} for analytic studies of dark matter halos in models with the above screening mechanisms. Here we mainly follow Refs.~\cite{Schmidt:2010jr, Falck:2014jwa, Falck:2015rsa}. 

We assume the dark matter density profile is described by the NFW profile~\cite{Navarro:1996gj} with a mass $M_{\Delta}$. The mass is defined as the mass contained within the radius $r=r_{\Delta}$; the density at $r_{\Delta}$ is $\rho_{\rm crit} \Delta$, where $\rho_{\rm crit}$ is the critical density of the Universe. The NFW profile is given by
\begin{equation}
\rho(r) = \rho_s f \left(\frac{r}{r_s} \right), \quad
f(y) = \frac{1}{y (1+y)^2},
\end{equation}
where $\rho_s =\rho(r_s)$ is fixed so that the mass within $r_{\Delta}$ is $M_{\Delta}$. The scale radius $r_s$ is more conveniently parameterised by the concentration $c_{\Delta} = r_{\Delta}/r_s$. By integrating this density profile, we obtain the enclosed mass within the radius $r$, $M(<r)$, as 
\begin{equation}
M(<r) = M_{\Delta} \frac{F (c_{\Delta} r/r_{\Delta})}{F(c_{\Delta})}, \quad
F(y)= -\frac{y}{1+y} + \ln(1+y).
\label{NFWm}
\end{equation}

The scalar field equation in the nDGP model, Eq.~(\ref{DGP}), can be solved analytically \cite{Schmidt:2010jr}   
\begin{equation}
  \frac{d\varphi}{dr} = \frac{G M(<r)}{r^2} \frac{4}{3 \beta} g\left(\frac{r}{r_*} \right), \quad
g(x) = x^3 \left( \sqrt{1+x^{-3}} -1 \right),
\label{halodgp}
\end{equation}
where $r_*$ is the Vainshtein radius 
\begin{equation}
r_*= \left(\frac{16 G M(<r) r_c^2}{9 \beta^2}\right)^{1/3}.
\label{r*}
\end{equation}
Inside the Vainshtein radius, the scalar force is suppressed compared with the Newtonian potential due to the non-linear derivative interactions. Outside the Vainshtein radius, the linear solution is realised where $g(r/r_*) \to 1/2$. For a larger $r_c$, the Vainshtein radius is larger thus the region in which the fifth force is suppressed becomes larger and we recover GR. 

The scalar field in $f(R)$ gravity can be approximated as \cite{Schmidt:2010jr}
\begin{equation}
  \frac{d\varphi}{dr} =\frac{1}{3} \frac{G\Big( M(<r) - M(<r_{\rm scr}) \Big)}{r^2}
  \label{halofr}
\end{equation}
for $r > r_{\rm scr}$ and $d\varphi/dr=0$ for $r < r_{\rm scr}$ if the thin shell condition (\ref{fr-thishel}) is satisfied. The screening radius for the NFW profile is obtained as \cite{Terukina:2012ji}
\begin{equation}
r_{\rm scr} = \Big( \frac{2}{3} \frac{|\Psi_{N}|(r_{\Delta})} {|f_{R0}| F(c_{\Delta})} - \frac{1}{c_{\Delta}} \Big) 
r_{\Delta}, 
\end{equation}
where the Newtonian potential is given by $|\Psi_{N}|= G M_{\Delta}/r_{\Delta}$. The screening radius is determined by the ratio between the Newtonian potential of the halo and $|f_{R0}|$, thus it is mass dependent. 

We define the ratio between the fifth force and the Newton force as
\begin{equation}
\Delta_M = \frac{F_5}{F_G}, 
\quad F_5 =\frac{1}{2} \frac{d \varphi}{ dr}, \quad F_G= \frac{d \Psi_N}{dr}.
\label{eqn:deltam}
\end{equation}
For linear solutions without screening, $\Delta_M=1/3 \beta$ in nDGP and $\Delta_M=1/3$ in $f(R)$. 

\subsection{Simulations}
We consider $N$-body simulations for nDGP and $f(R)$ models performed by the ECOSMOG code described in section 4.5 \cite{Falck:2015rsa}. To highlight the difference in the screening mechanisms, we simulate pairs of nDGP and $f(R)$ models with the $\Lambda$CDM background and identical $\sigma_8$ at $z=0$, which roughly removes the difference in screening due to the difference in linear growth. Specifically, we simulate three $f(R)$ models: F4 ($|f_{R0}|=10^{-4}$), F5 ($|f_{R0}|=10^{-5}$), and F6 ($|f_{R0}|=10^{-6}$), and three corresponding nDGP models whose parameters are listed in Table I. 

\begin{table}

\begin{center}
\begin{tabular}{c|c|c}

\hline\hline

$f(R)$                                   &       nDGP                          &                                       \\
\hline
F4: $|f_{R0}|=10^{-4}$                &   nDGP1: $H_0 r_c=0.57$              &    $\sigma_8=0.946$               \\
F5: $|f_{R0}|=10^{-5}$                &   nDGP2:  $H_0 r_c=1.20$             &   $\sigma_8=0.891$               \\
F6: $|f_{R0}|=10^{-6}$                &   nDGP3:  $H_0 r_c=5.65$             &   $\sigma_8=0.854$               \\

\hline\hline

\end{tabular}
\end{center}
\label{tab:param}
\caption{The parameters for $f(R)$ and nDGP simulations. The baseline cosmology was chosen to be the model favoured by recent Planck observations \cite{Ade:2013zuv}: $\Omega_b h^2=0.022161,~\Omega_c h^2=0.11889,~\Omega_K =0,~h=0.6777,~n_s=0.9611$, and $\sigma_8=0.841.$ 
The simulations use $256^3$ particles in a $L=64\mpcoh$ box from the initial redshift $z=49$ down to $z=0$. Each set of models ($f(R)$, nDGP, and $\Lambda$CDM) are simulated using the same initial condition, which is generated using the Zeldovich approximation, and we simulate three realisations for each model to reduce the sample variance. See \cite{Falck:2015rsa} for details. 
}
\end{table}%

%%%%%%%%%%%%%%%%%%%%%%%%%%%%%%%%%%%%%%%%%%%%%%%%%%%%%%%%%%%%
\begin{figure}[ht]
  \centering{
  \includegraphics[width=\hsize]{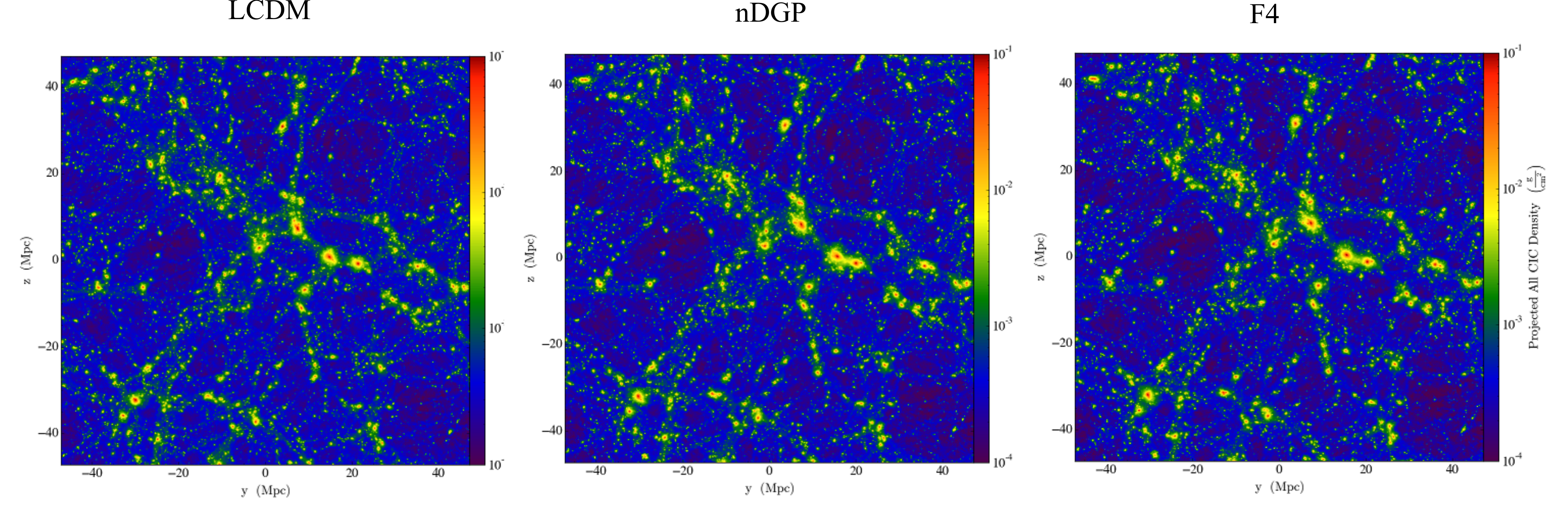}
  }
  \caption{The snapshots ($64\mpcoh\times64\mpcoh$) showing the projected density field at $z=0$ for LCDM {\it(left)}, nDGP1 {\it (middle)} and the F4 {\it (right)} models. From \cite{Falck:2015rsa} (published on 29 July 2015 \copyright SISSA Medialab Srl. Reproduced by permission of IOP Publishing.  All rights reserved).
   }
  \label{fig:snapshot}
\end{figure}
%%%%%%%%%%%%%%%%%%%%%%%%%%%%%%%%%%%%%%%%%%%%%%%%%%%%%%%%%%%%%

\begin{figure}[ht]
  \centering{
  \includegraphics[width=14cm]{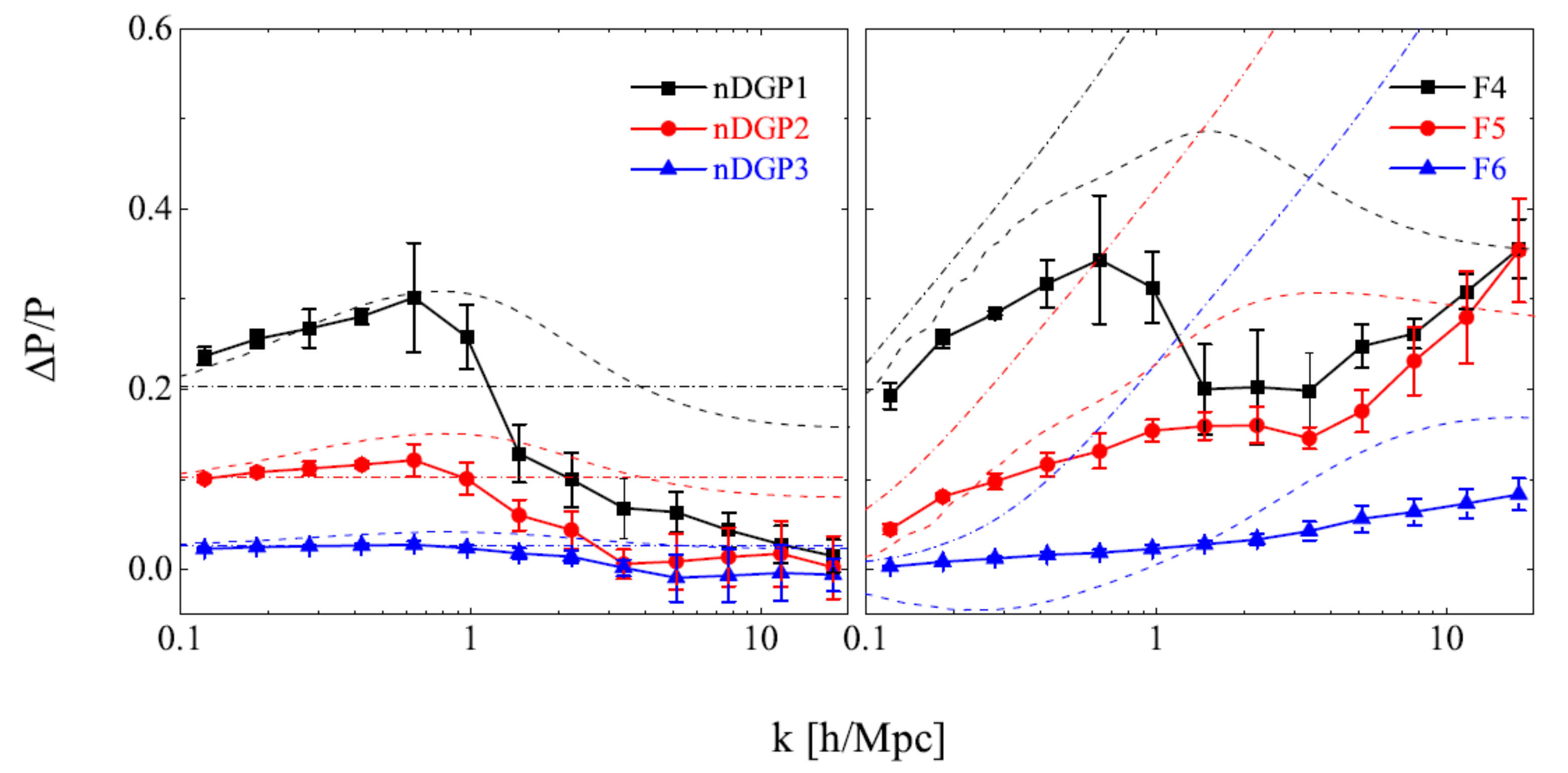}
  }
  \caption{The fractional difference in power spectrum for nDGP {\it (left)} and $f(R)$ {\it (right)} models with respect to the $\Lambda$CDM model. The data points with error bars show the simulation result, and the dashed (dash-dotted) curves show the Halofit (linear) predictions. From \cite{Falck:2015rsa} (published on 29 July 2015 \copyright SISSA Medialab Srl. Reproduced by permission of IOP Publishing.  All rights reserved).
   }
  \label{fig:power}
\end{figure}

The simulations are visualised in Fig~\ref{fig:snapshot}, where the projected density field for $\Lambda$CDM and two modified gravity models are shown at $z=0$ \cite{Falck:2015rsa}. As shown, the structures are more clustered in nDGP and $f(R)$ models due to the enhanced gravity, although the enhancement of the clustering in these models are different. To quantify the difference in clustering, we show the fractional difference of the power spectrum in $f(R)$ and nDGP models with respect to $\Lambda$CDM at $z=0$ in Fig.~\ref{fig:power}  \cite{Falck:2015rsa}. The dotted lines show the linear prediction while the dash-dotted lines show the {\it halofit} prediction~\cite{Smith:2002dz}. 

%%%%%%%%%%%%%%%%%%%%%%%%%%%%%%%%%%%%%%%%%%%%%%%%%%%%%%%%%%%%
%%%%%%%%%%%%%%%%%%%%%%%%%%%%%%%%%%%%%%%%%%%%%%%%%%%%%%%%%%%%%

The agreement of the linear prediction with simulations is better for the nDGP model. The excellent agreement with linear theory on large scales is one of the key features of the Vainshtein mechanism. This is because when the Vainshtein screening mechanism works, even if the fifth force is suppressed inside halos, these `screened' halos can still feel external scalar fields as long as those fields have wavelengths longer than the Vainshtein radius \cite{Hui:2012jb}. We will confirm this picture later by studying the velocities of dark matter particles inside halos. On small scales, the Vainshtein mechanism is very effective and the power spectrum deviation approaches zero quickly at high $k$. 

On the other hand, in the $f(R)$ model with the chameleon screening mechanism, once the fifth force is suppressed inside dark matter halos, these screened halos no longer feel the external fifth force invalidating the use of the linear theory even on large scales. The chameleon mechanism is not as efficient as the Vainshtein mechanism. Thus the power spectrum deviation does not approach zero even though it is significantly suppressed compared with the linear prediction. 

\subsection{Screening of dark matter particles}
We first study how screening operates at the level of particles \cite{Falck:2014jwa, Falck:2015rsa}.  
The cosmic web of large scale structure consists of an interconnected hierarchy of halos, filaments, walls, and voids. Here we use ORIGAMI to determine the cosmic web morphology of each dark matter particle in a simulation, which compares final positions to initial Lagrangian positions to determine whether a particle has undergone shell-crossing along a given set of axes (see Ref.~\cite{Falck:2012ai} for details). Shell-crossing denotes the formation of caustics within which the velocity field is multi-valued, called the multi-stream regime. The number of orthogonal axes along which shell-crossing has occurred corresponds to the particle's cosmic web morphology and denoted by the morphology index $M$: halo particles have crossed along three axes ($M=3$), filaments along two ($M=2$), walls along one ($M=1$), and void particles are in the single-stream regime ($M=0$). 

We quantify the screening by calculating the deviation of the fifth force to gravitational force ratio from the linear relation ($\Delta_M=1/3 \beta$ in nDGP and $\Delta_M=1/3$ in $f(R)$)
\begin{equation}
\Delta F = \frac{F_5}{\Delta_M F_G} - 1,
\end{equation}
which ranges between $\Delta F = -1$ when screening is working to $\Delta F = 0$ when it is not. For some particles $\Delta F$ can be greater than 0 due to numerical noise, especially for low values of $F_5$ and $F_G$ (see, e.g.,~\cite{Moran:2014zxa}).

\begin{figure}[ht]
  \centering{
  \includegraphics[width=15cm]{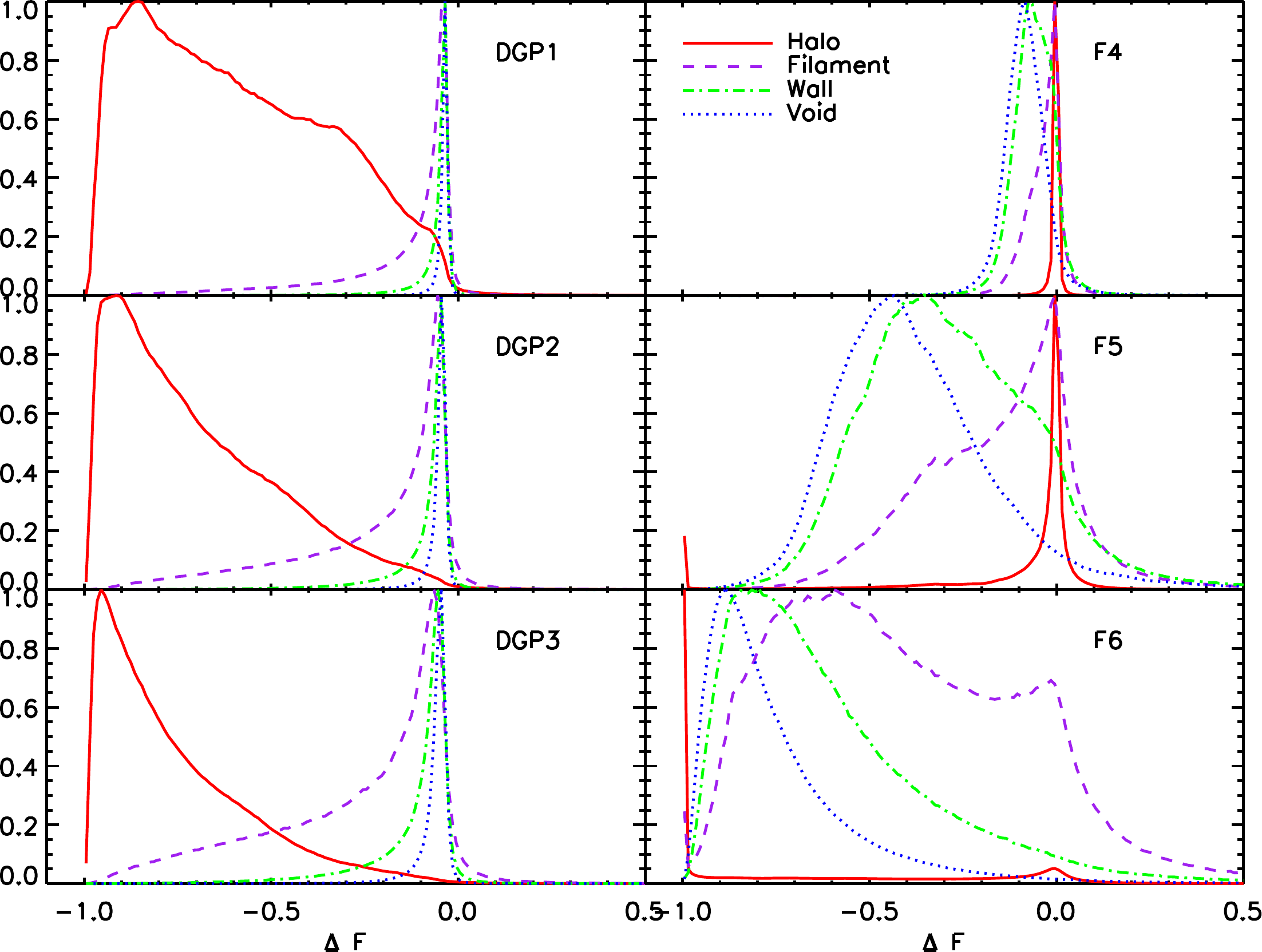}
  }
  \caption{Cosmic web morphology dependence of the Vainshtein {\it (left)} and chameleon {\it (right)} screening mechanisms, given by histograms of $\Delta F$, the deviation from the linear relation of the fifth force to gravitational force ratio. In DGP models with the Vainshtein mechanism, there is a clear difference between halo particles (solid, red line) and filament, wall, and void particles, while no such distinction exists for $f(R)$ models with the chameleon screening mechanism. From \cite{Falck:2015rsa} (published on 29 July 2015 \copyright SISSA Medialab Srl. Reproduced by permission of IOP Publishing.  All rights reserved).
  }
  \label{fig:deltaF}
\end{figure}

In Figure~\ref{fig:deltaF} we show histograms of $\Delta F$ split according to ORIGAMI morphology for each modified gravity model \cite{Falck:2015rsa}. Note that the histograms are normalised to peak at unity so that the shape of each can be seen, so they do not reflect the relative abundances of the particles according to their cosmic web morphology. It is clear that, as for all 3 nDGP models the halo particles are screened while the filament, void, and wall particles are unscreened. This reflects the dimensionality dependence of the Vainshtein mechanism~\cite{Brax:2011sv, Bloomfield:2014zfa, Falck:2014jwa}. The non-linear term in Eq.~(\ref{DGP}), $(\partial^2 \varphi)^2 - (\partial_i \partial_j \varphi)^2$, vanishes for a one-dimensional system. The distribution of $\Delta F$ for halo particles is still broad, reflecting the fact that screening becomes weaker beyond the virial radius, so particles in the halos' outer edges can have larger values of $\Delta F$.

%%%%%%%%%%%%%%%%%%%%%%%%%%%%%%%%%%%%%%%%%%%%%%%%%%%%%%%%%%%%
%%%%%%%%%%%%%%%%%%%%%%%%%%%%%%%%%%%%%%%%%%%%%%%%%%%%%%%%%%%%%

On the other hand, there is no such morphology dependence of the screening mechanism for the chameleon models on the right side of Figure~\ref{fig:deltaF}. For the halos, screening is not very effective in F4 and most particles follow the linear relation; in F5 the halo particles have a small peak at $\Delta F = -1$ as the most massive halos become screened; and in F6 most of the halo particles are in screened halos, while some remain unscreened. Note that since the histograms are for halo particles and not halos themselves, massive halos are weighted more heavily, resulting in many screened halo particles in F6, while the smaller bump of unscreened halo particles is due to halos with low mass. We will look at the screening of halos in the next section. The wall and void histograms notably do not peak at $\Delta F = 0$ in F4, and the fifth force is further suppressed in F5 and especially F6. This is because the background Compton wavelength is quite short, $\sim~1$Mpc in F6, and the scalar field does not propagate beyond this length, providing a blanket screening for particles that are sparsely distributed. The filament distribution develops a double peak in F6: a narrow peak of unscreened filament particles, which have large forces and are in relatively dense environments, and a broader peak of low $\Delta F$ filament particles that are blanket screened. 

\subsection{Dark Matter Halos}
\label{sec:halos}

We now turn to a comparison of the screening of halos in the Vainshtein and chameleon mechanisms \cite{Falck:2015rsa}. 

\subsubsection{Screening - mass dependence}
\label{sec:mass}

\begin{figure}[ht]
  \centering{
  \includegraphics[width=13cm]{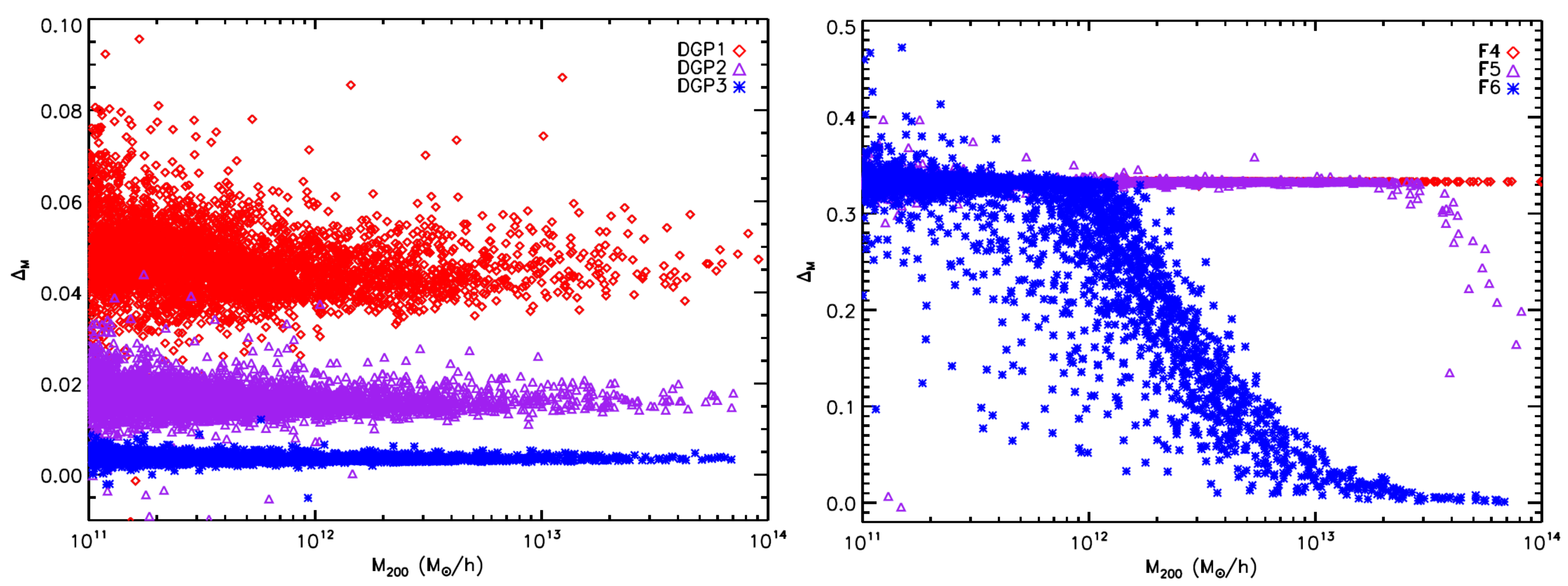}
  }
  \caption{Ratio of the fifth force to gravitational force as a function of halo mass for nDGP {\it (left)} and $f(R)$ {\it (right)} models. From \cite{Falck:2015rsa} (published on 29 July 2015 \copyright SISSA Medialab Srl. Reproduced by permission of IOP Publishing.  All rights reserved).
  }
  \label{fig:dm_mass}
\end{figure}

As with the dark matter particles, to determine whether screening is effective we calculate the ratio of the fifth force to gravitational force, $\Delta_M$. The value of $\Delta_M$ for each halo is given by the average $\Delta_M$ of all the particles in the halo within the halo's virial radius, $R_{200}$. This is plotted as a function of halo mass, $M_{200}$, for all nDGP and $f(R)$ simulations in Figure~\ref{fig:dm_mass} \cite{Falck:2015rsa}.

%%%%%%%%%%%%%%%%%%%%%%%%%%%%%%%%%%%%%%%%%%%%%%%%%%%%%%%%%%%%
%%%%%%%%%%%%%%%%%%%%%%%%%%%%%%%%%%%%%%%%%%%%%%%%%%%%%%%%%%%%%

For the DGP models, the $\Delta_M$ of the halos depends on the model parameter and is independent of mass \cite{Schmidt:2010jr, Falck:2014jwa}; as the model parameter changes to make the deviation from $\Lambda$CDM stronger, $\Delta_M$ increases. However, note that all halos are screened in the Vainshtein mechanism: the linear values of $\Delta_M$ for nDGP1, nDGP2, and nDGP3 are 0.20, 0.11, and 0.03, respectively, well above the corresponding values in Figure~\ref{fig:dm_mass}. As we will see in the next section, the Vainshtein suppression gradually reduces (and $\Delta_M$ increases) outside the virial radius, but even including these particles in the calculation of $\Delta_M$ only increases $\Delta_M$ by $\sim 50 - 70$\% and the halos remain screened overall. 

In contrast to the Vainshtein mechanism, Figure~\ref{fig:dm_mass} shows there is a clear dependence on both mass and model parameter in the chameleon mechanism \cite{Schmidt:2010jr, Falck:2015rsa}. When the deviation from $\Lambda$CDM is high, in F4, screening becomes ineffective for all halos; in F5, screening is effective only for high mass halos; and in F6, there is a population of unscreened small halos and a transition to screened large halos. Including particles outside the virial radius in the calculation of $\Delta_M$ has a very small effect, increasing $\Delta_M$ for some halos but not changing the overall trends; we will show in the next section that the radius of transition from screened to unscreened parts of the halo depends on the halo mass in the chameleon mechanism.

\subsubsection{Screening profiles}
\label{sec:profiles}

%%%%%%%%%%%%%%%%%%%%%%%%%%%%%%%%%%%%%%%%%%%%%%%%%%%%%%%%%%%%
\begin{figure}[ht]
  \centering{
  \includegraphics[width=13cm]{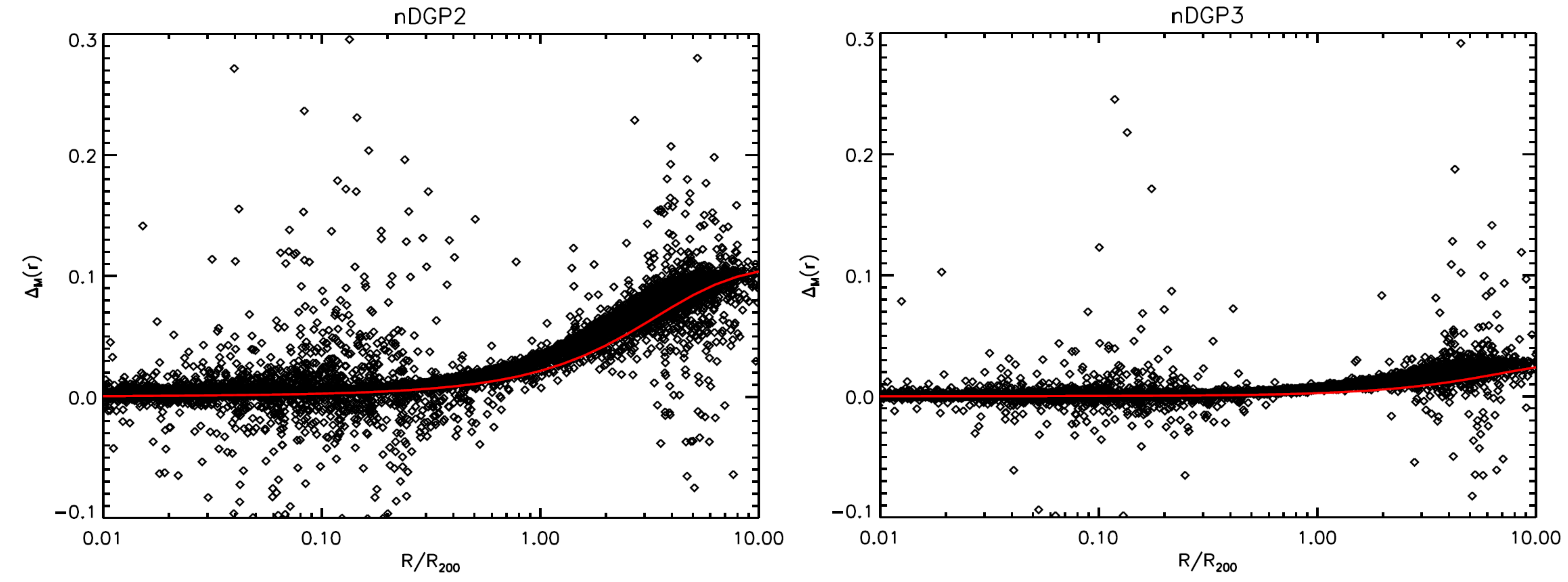}
  }
  \caption{Profiles of $\Delta_M$ in logarithmic bins of normalised radius, for nDGP2 {\it (left panel)} and nDGP3 {\it (right panel)}, with the analytic prediction for an NFW profile plotted in red. Screening profiles are independent of halo mass, and Vainshtein screening suppresses the fifth force within the virial radius. From \cite{Falck:2015rsa} (published on 29 July 2015 \copyright SISSA Medialab Srl. Reproduced by permission of IOP Publishing.  All rights reserved).
  }
  \label{fig:dgp_profile}
\end{figure}
%%%%%%%%%%%%%%%%%%%%%%%%%%%%%%%%%%%%%%%%%%%%%%%%%%%%%%%%%%%%%

%%%%%%%%%%%%%%%%%%%%%%%%%%%%%%%%%%%%%%%%%%%%%%%%%%%%%%%%%%%%
\begin{figure}[ht]
  \centering{
  \includegraphics[width=13cm]{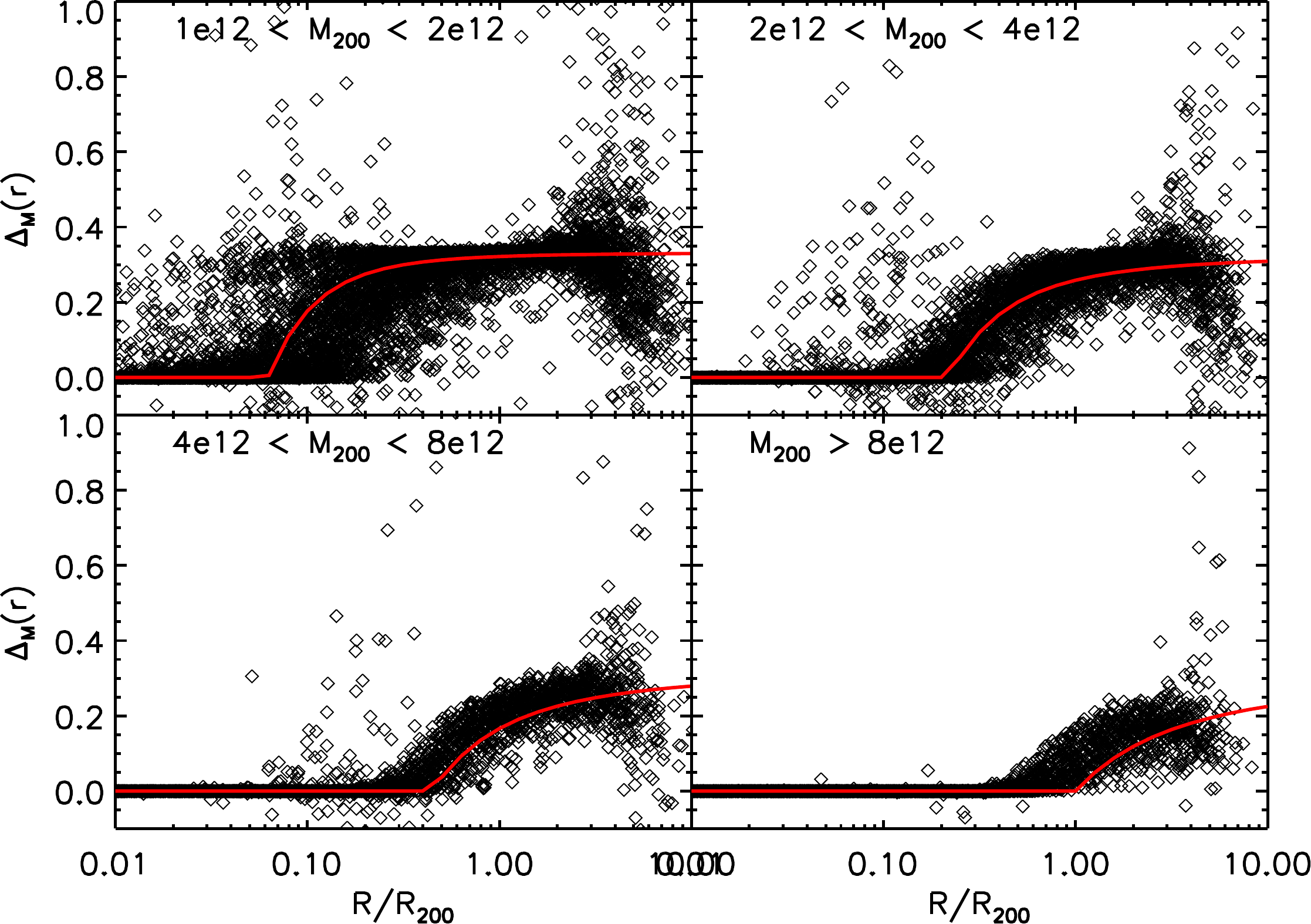}
  }
  \caption{Profiles of $\Delta_M$ in logarithmic bins of normalised radius for the F6 model, split into four bins of halo mass, with the analytic prediction calculated for each mass bin plotted in red. Screening profiles in the chameleon mechanism depend on halo mass. From \cite{Falck:2015rsa} (published on 29 July 2015 \copyright SISSA Medialab Srl. Reproduced by permission of IOP Publishing.  All rights reserved).
  }
  \label{fig:fr_profile}
\end{figure}
%%%%%%%%%%%%%%%%%%%%%%%%%%%%%%%%%%%%%%%%%%%%%%%%%%%%%%%%%%%%%

To determine the radial dependence of the screening within the halos, here we calculate the ratio between the fifth force and Newtonian force as a function of normalised radius $R/R_{200}$ by averaging $\Delta_M$ of the halo particles in logarithmic radial bins. $\Delta_M$ profiles are plotted in Figure~\ref{fig:dgp_profile} for nDGP2 and nDGP3  \cite{Falck:2015rsa}. It is clear that regardless of halo mass, the Vainshtein screening profiles of dark matter halos are roughly the same and correspond quite well to the spherically symmetric analytic solution for an NFW profile (\ref{halodgp}) \cite{Falck:2014jwa}. For both models, the fifth force is mostly suppressed within the virial radius and $\Delta_M$ increases outside the virial radius, and the magnitude of this increase is greater for models with a stronger enhancement to gravity, nDPG2 (and nDGP1, not shown).

In $f(R)$, Figure~\ref{fig:dm_mass} shows that chameleon screening does depend on mass, so in Figure~\ref{fig:fr_profile} we split up the $\Delta_M$ profiles into four different mass bins for the F6 model  \cite{Falck:2015rsa}. There is a much sharper and less gradual transition from the chameleon screened inner regions of the halo to the unscreened outer regions compared to the Vainshtein mechanism, and the radius of this transition depends on the halo mass. There is again quite a good agreement with the analytic predictions Eq.~(\ref{halofr}) \cite{Schmidt:2010jr, Falck:2015rsa}. 

Both Figure~\ref{fig:dgp_profile} and Figure~\ref{fig:fr_profile} highlight the importance of probing galactic halos in unscreened regions beyond their virial radii in order to detect deviations from $\Lambda$CDM, for all halo masses if the Vainshtein mechanism is operating and for high mass halos in the chameleon mechanism. For example, Refs~.\cite{Lam:2012by, Lam:2013kma} proposed a method to measure the velocity field by stacking redshifts of surround galaxies around galaxy clusters from a spectroscopic sample. It was shown that order unity deviations from $\Lambda$CDM show up both in $f(R)$ gravity and the nDGP model. On the other hand, the effect on the interior mass profile, which is measurable through stacked weak lensing is much less affected by a modification of gravity (see also \cite{Lombriser:2011zw}). Ref.~\cite{Zu:2013joa} proposed to use the 2D galaxy velocity distribution in the cluster infall region by applying the galaxy infall kinetics model developed by Ref.~\cite{Zu:2012sc}. 

\subsection{Velocities}
\label{sec:velocity}
The non-linear derivative interaction that is responsible for the Vainshtein mechanism enjoys the Galilean symmetry (\ref{gal-sym}), $\partial_{\mu} \phi \to \partial_{\mu} \phi + c_{\mu}$, which means that if the external fields have wavelengths that are long compared to the Vainshtein radius, Eq.~(\ref{r*}), we can regard the gradient of these  external fields as constant gradients in the vicinity of an object. We can always add these constant gradients to the internal field generated by the object, and thus the internal field can superimpose with external fields. Even if the internal field is suppressed by the Vainshtein mechanism, the object still feels the fifth force generated by the external fields \cite{Hui:2012jb}. On the other hand, in screening mechanisms that rely on non-linearity in the potential or the coupling function to matter, such as the chameleon and symmetron mechanisms, the internal field generated by an object does not superimpose with an external field. Therefore, the field inside the object loses knowledge of any exterior gradient and the fifth force generated by the external field, and thus once the object is screened, it does not feel any fifth force \cite{Hui:2009kc, Hu:2009ua, Hui:2012jb}. 
%%%%%%%%%%%%%%%%%%%%%%%%%%%%%%%%%%%%%%%%%%%%%%%%%%%%%%%%%%%%
\begin{figure}[ht]
  \centering{
  \includegraphics[width=13cm]{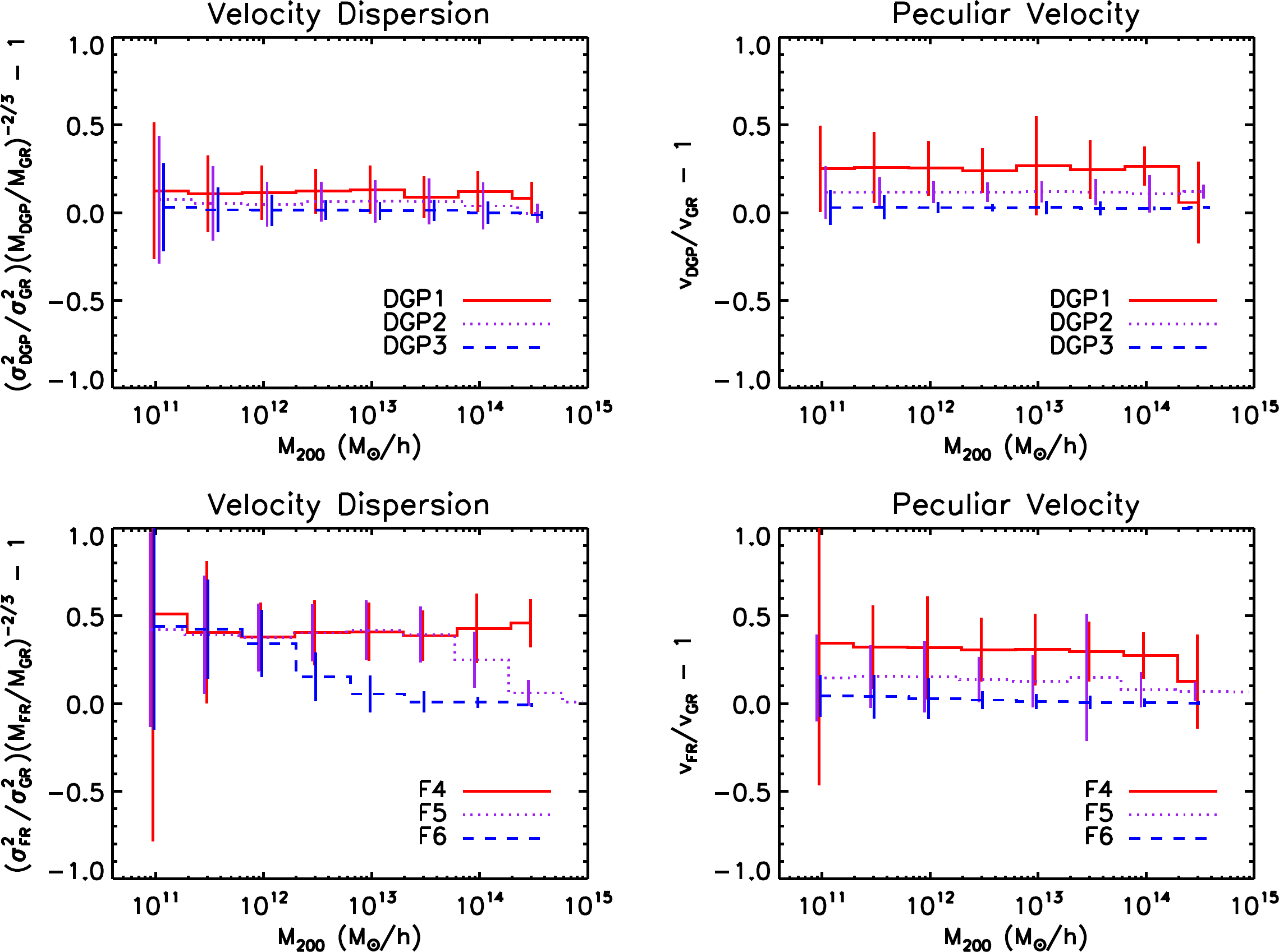}
  }
  \caption{Velocity dispersion {\it (left)} and peculiar velocity {\it (right)} ratios of nDGP {\it (top)} and $f(R)$ {\it (bottom)} models with respect to matched halos in the $\Lambda$CDM simulations, as a function of halo mass. Lines indicate the average value in each mass bin, and error bars indicate the scatter. From \cite{Falck:2015rsa} (published on 29 July 2015 \copyright SISSA Medialab Srl. Reproduced by permission of IOP Publishing.  All rights reserved).
  }
  \label{fig:velocity}
\end{figure}
%%%%%%%%%%%%%%%%%%%%%%%%%%%%%%%%%%%%%%%%%%%%%%%%%%%%%%%%%%%%%

In Figure~\ref{fig:velocity}, we plot the velocity dispersion ratios and peculiar velocity ratios for all DGP and $f(R)$ models \cite{Falck:2015rsa}. The velocity dispersion ratios are scaled by the virial expectation, $\sigma^2\propto M^{2/3}$, to remove the standard mass dependence \cite{Schmidt:2010jr}. The lines are the average values in bins of mass, and the error bars show the $1 \sigma$ standard deviation. Both the velocity dispersion and peculiar velocity ratios show no dependence on mass for the Vainshtein mechanism~\cite{Schmidt:2010jr,Falck:2014jwa}, but the peculiar velocity ratios deviate from 0, especially in the nDGP1 model, due to the effect of external fields. 

The $f(R)$ models, on the other hand, do show a mass dependence in the velocity dispersion ratios: in F4, where screening is not effective, the velocity dispersion is enhanced; in F5, the velocity dispersion is only suppressed at high masses; and in F6, the transition between enhanced and suppressed velocity dispersion occurs at an even lower mass. These trends mimic the mass dependence of the screening ratio, $\Delta_M$, in Figure~\ref{fig:dm_mass}. However, the peculiar velocity ratios do not show this mass dependence; they are suppressed in F6, somewhat enhanced in F5, and further enhanced in F4. Since most of the particles in a halo are located near the centre, the peculiar velocity effectively probes the halo centre and is suppressed because the halo core is screened, as seen in the $\Delta_M$ profiles for F6 in Figure~\ref{fig:fr_profile}. Unlike in the Vainshtein mechanism, once screened by the chameleon mechanism a halo does not feel the effect of external fields and so its peculiar velocity is suppressed. 

This casts doubt on the effectiveness of using redshift space distortions to detect $f(R)$ gravity at the level of F6. The Vainshtein mechanism is better suited to tests in the linear regime, since even though halos are screened, they can still feel the effect of external fields induced by large scale structure.

\subsection{Screening - environment dependence}
\label{sec:environment}

In the previous sections, we have shown that chameleon screening depends on both the mass of the object and the model parameter, while Vainshtein screening is independent of halo mass \cite{Schmidt:2010jr}. In Section~5.3 we showed that Vainshtein screening of dark matter particles depends on their cosmic web morphology \cite{Falck:2014jwa}; halo particles are screened while filament, void, and wall particles are not, reflecting the dimensionality dependence of the Vainshtein mechanism~\cite{Brax:2011sv, Bloomfield:2012ff}. However, it was found that the Vainshtein screening of halos themselves does not depend on their large scale cosmic web environment ~\cite{Falck:2014jwa}, and the chameleon mechanism has no cosmic web dependence for either particles or halos, so here we use a different definition of halo environment. 

We use a density-based definition of environment developed in Ref.~\cite{Haas:2011mt},
\begin{equation}
D_{N,f} \equiv \frac{d_{N, M_N/M \geq f}}{r_N}.
\label{D}
\end{equation}
This is the distance $d_N$ to the $N$th nearest neighbour having mass, $M_N$, at least $f$ times as large as the halo mass, $M$, scaled by the virial radius of the neighbouring halo, $r_N$. $D_{1,1}$ (hereafter just $D$) is almost uncorrelated with the halo mass~\cite{Haas:2011mt}, and several studies have found that both chameleon and symmetron screening mechanisms correlate with this environment parameter~\cite{Zhao:2011cu, Cabre:2012tq, Winther:2011qb,Gronke:2014gaa,Moran:2014zxa}. In particular, while massive halos can be self-screened and in general smaller halos can be unscreened, small halos that live in dense environments (where $D$ is small) can be environmentally screened. Subhalos, especially those within the virial radius of their host halo with $D < 1$, are usually environmentally screened. We use the AHF halo finder~\cite{Knollmann:2009pb} to study the environmental dependence of chameleon and Vainshtein screening \cite{Falck:2015rsa}.  

\begin{figure}[ht]
  \centering{
  \includegraphics[width=13cm]{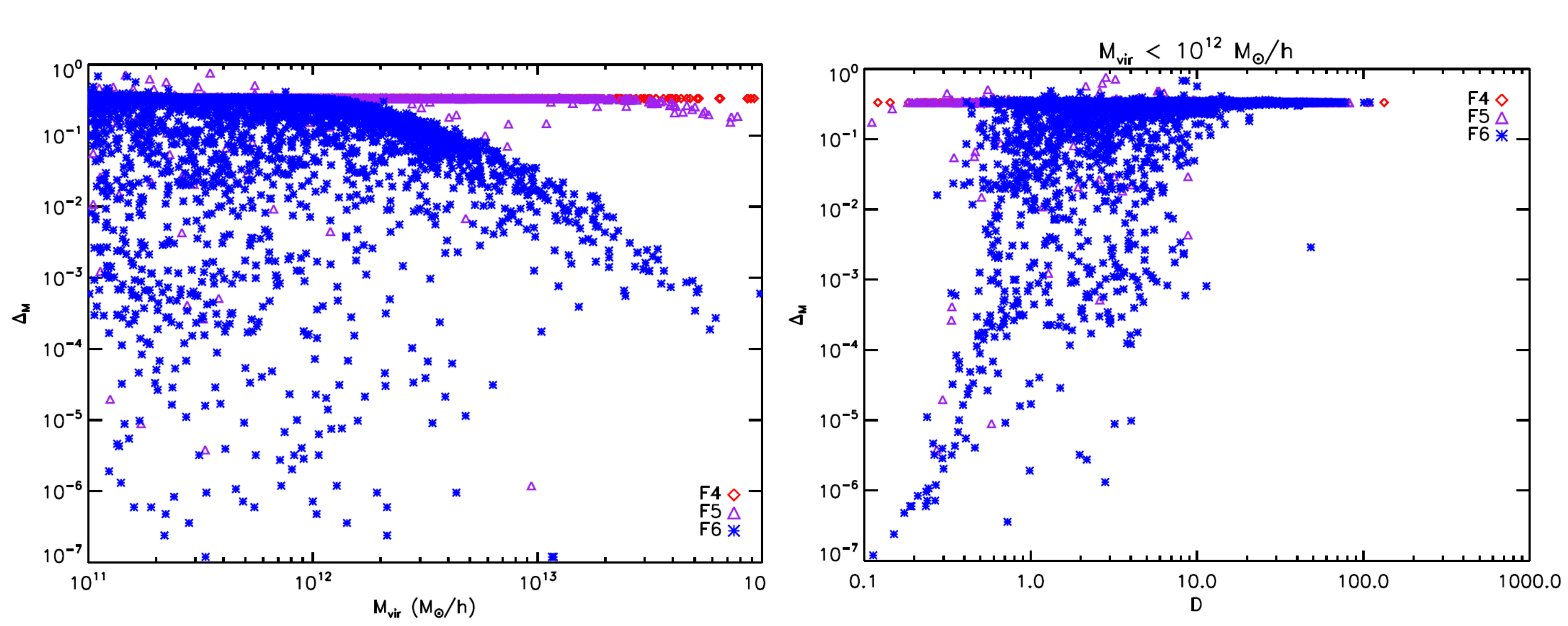}
  }
  \caption{The ratio of the fifth force to gravitational force for AHF halos in $f(R)$ models as a function of mass {\it (left panel)} and environment $D$ {\it (right panel)}. Only halos with mass below $10^{12}\,h^{-1}\,{\rm M}_{\astrosun}$ are shown in the right panel to remove halos that are massive enough to be self-screened. Chameleon screening displays both a mass and environmental dependence. From \cite{Falck:2015rsa} (published on 29 July 2015 \copyright SISSA Medialab Srl. Reproduced by permission of IOP Publishing.  All rights reserved).
  }
  \label{fig:fr_env}
\end{figure}

In Figure~\ref{fig:fr_env} we show the results for chameleon screening \cite{Falck:2015rsa}. In the left panel we plot $\Delta_M$ vs. halo mass, $M_{\rm{vir}}$, with a logarithmic scaling of $\Delta_M$ instead of the linear scaling of ORIGAMI values in Figure~\ref{fig:dm_mass}. The same trend in mass dependence of screening is seen, except AHF halos can have much lower values of $\Delta_M$, which is due to the presence of subhalos. In the right panel of Figure~\ref{fig:fr_env} we show the environmental dependence of chameleon screening, after first removing halos with mass above $10^{12}\,h^{-1}\,{\rm M}_{\astrosun}$ that are self-screened. Halos in dense environments, and especially those with $D<1$, tend to be screened while those in under dense environments, especially those with $D > 10$, are unscreened \cite{Zhao:2011cu}.

%%%%%%%%%%%%%%%%%%%%%%%%%%%%%%%%%%%%%%%%%%%%%%%%%%%%%%%%%%%%
%%%%%%%%%%%%%%%%%%%%%%%%%%%%%%%%%%%%%%%%%%%%%%%%%%%%%%%%%%%%%

\begin{figure}[t]
\includegraphics[width=13cm]{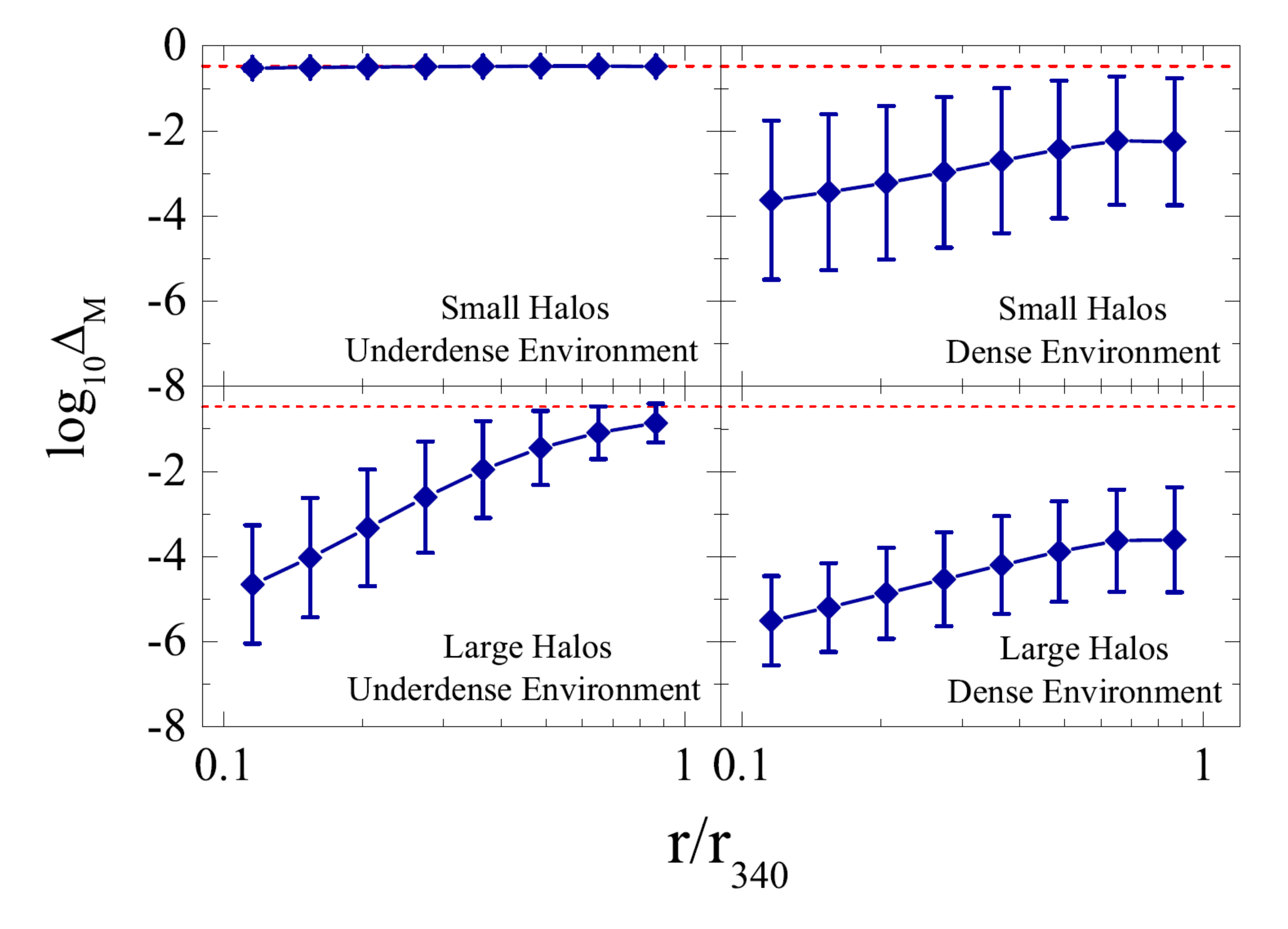}
\caption{The profile of log$_{10}\Delta M$ as a function of the
rescaled halo radius $r/r_{340}$ for the $|f_{R0}|=10^{-6}$ model.
We show the profile with $1-\sigma$ error bars for the halos
divided into four categories as illustrated in the legend. The red
dashed line shows $\Delta_M(r_{340})=1/3$. From \cite{Zhao:2011cu}.} 
\label{fig:prof}
\end{figure}

We show the profile of the mass difference as a function of
rescaled radius for the $|f_{R0}|=10^{-6}$ model in Fig
\ref{fig:prof} \cite{Zhao:2011cu}. To see the environmental effect on the profile, we show the result for the samples selected according to both their mass and $D$.  As we can see, the
small halos in the under dense region are almost not screened at
any radius, while the halos with similar mass in the dense region
are efficiently screened, and the screening effect is stronger in
the core of the halos. For the large halos, the innermost part is
well screened regardless of external environment due to the high
matter density there, but the part close to the edge shows a clear
environmental dependence, and the difference can be as large as 3
orders of magnitude in $\Delta_M$ in different environments. This
is because in this region the external environment plays an
important role.

%%%%%%%%%%%%%%%%%%%%%%%%%%%%%%%%%%%%%%%%%%%%%%%%%%%%%%%%%%%%
%%%%%%%%%%%%%%%%%%%%%%%%%%%%%%%%%%%%%%%%%%%%%%%%%%%%%%%%%%%%

%%%%%%%%%%%%%%%%%%%%%%%%%%%%%%%%%%%%%%%%%%%%%%%%%%%%%%%%%%%%
\begin{figure}[ht]
  \centering{
  \includegraphics[width=13cm]{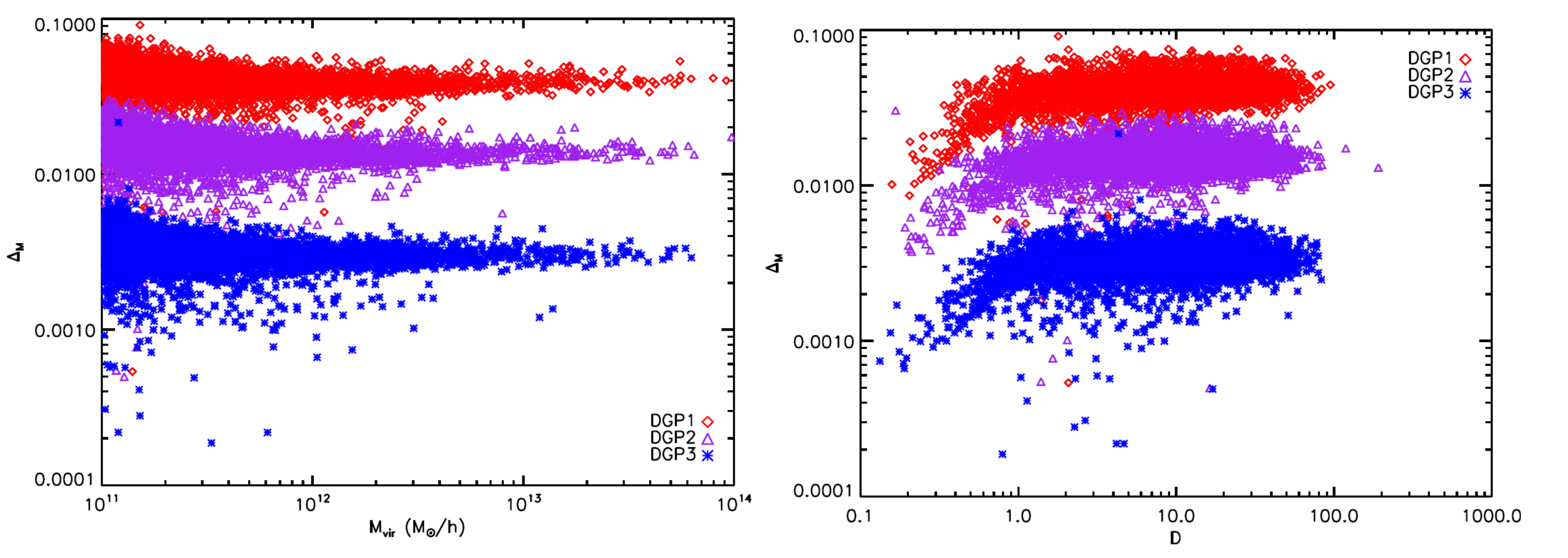}
  }
  \caption{The ratio of the fifth force to gravitational force for AHF halos in nDGP models as a function of mass {\it (left panel)} and environment $D$ {\it (right panel)}. There is no mass or environmental dependence of screening in the Vainshtein mechanism, and sub-halos within their host virial radius (with $D < 1$) have fifth forces that are even further suppressed. From \cite{Falck:2015rsa} (published on 29 July 2015 \copyright SISSA Medialab Srl. Reproduced by permission of IOP Publishing.  All rights reserved).
  }
  \label{fig:dgp}
\end{figure}
%%%%%%%%%%%%%%%%%%%%%%%%%%%%%%%%%%%%%%%%%%%%%%%%%%%%%%%%%%%%%

The results for the Vainshtein screening mechanism are shown in Figure~\ref{fig:dgp} \cite{Falck:2015rsa}. The left panel shows the screening as a function of mass, with again no mass dependence, though there is a lot of scatter in $\Delta_M$ for low mass halos. The right panel shows $\Delta_M$ as a function of environment: in general there is no environmental dependence, but halos within their host halo with $D < 1$ are screened even further by their host halo. Note again that all halos are screened in the Vainshtein mechanism: the linear values of $\Delta_M$ for nDGP1, nDGP2, and nDGP3 are 0.20, 0.11, and 0.03, respectively, well above the corresponding values in Figure~\ref{fig:dgp}. The Vainshtein mechanism is therefore very efficient at screening halos: there is no dependence on mass~\cite{Schmidt:2010jr,Falck:2014jwa}, cosmic web morphology~\cite{Falck:2014jwa}, or density of their local environment.

\section{Astrophysical tests of gravity}
The character of the screening mechanism leaves distinct signatures in non-linear structures as we discussed in the previous section. In this section we review novel astrophysical tests of gravity that have been developed recently. These tests are model dependent as opposed to the model independent tests of gravity on linear scales. We need to specify the screening mechanism to discuss the tests.

\subsection{Chameleon mechanisms}
We first consider the chameleon mechanism. The chameleon screening of dark matter halos depends on their mass and environment. Thus we need to carefully find the places where gravity is modified. From Fig.~21, the places where gravity is modified are at the outskirt of large dark matter halos and small dark matter halos in the under dense environment. 

\subsubsection{Clusters}
The abundance of clusters already gives strong constraints on $|f_{R0}|$ \cite{Ferraro:2010gh, Lombriser:2010mp, Cataneo:2014kaa}. For large $|f_{R0}|>10^{-5}$ the screening is ineffective and stronger gravity creates much more massive dark matter halos as is seen in Fig.~23. On the other hand, for $|f_{R0}|<10^{-5}$ the screening is effective for massive halos and the relative enhancement of the mass function disappears. 
%%%%%%%%%%%%%%%%%%%%%%%%%%%%%%%%%%%%%%%%%%%%%%%%%%%%%%%%%%%%
\begin{figure}[h]
  \centering{
  \includegraphics[width=10cm]{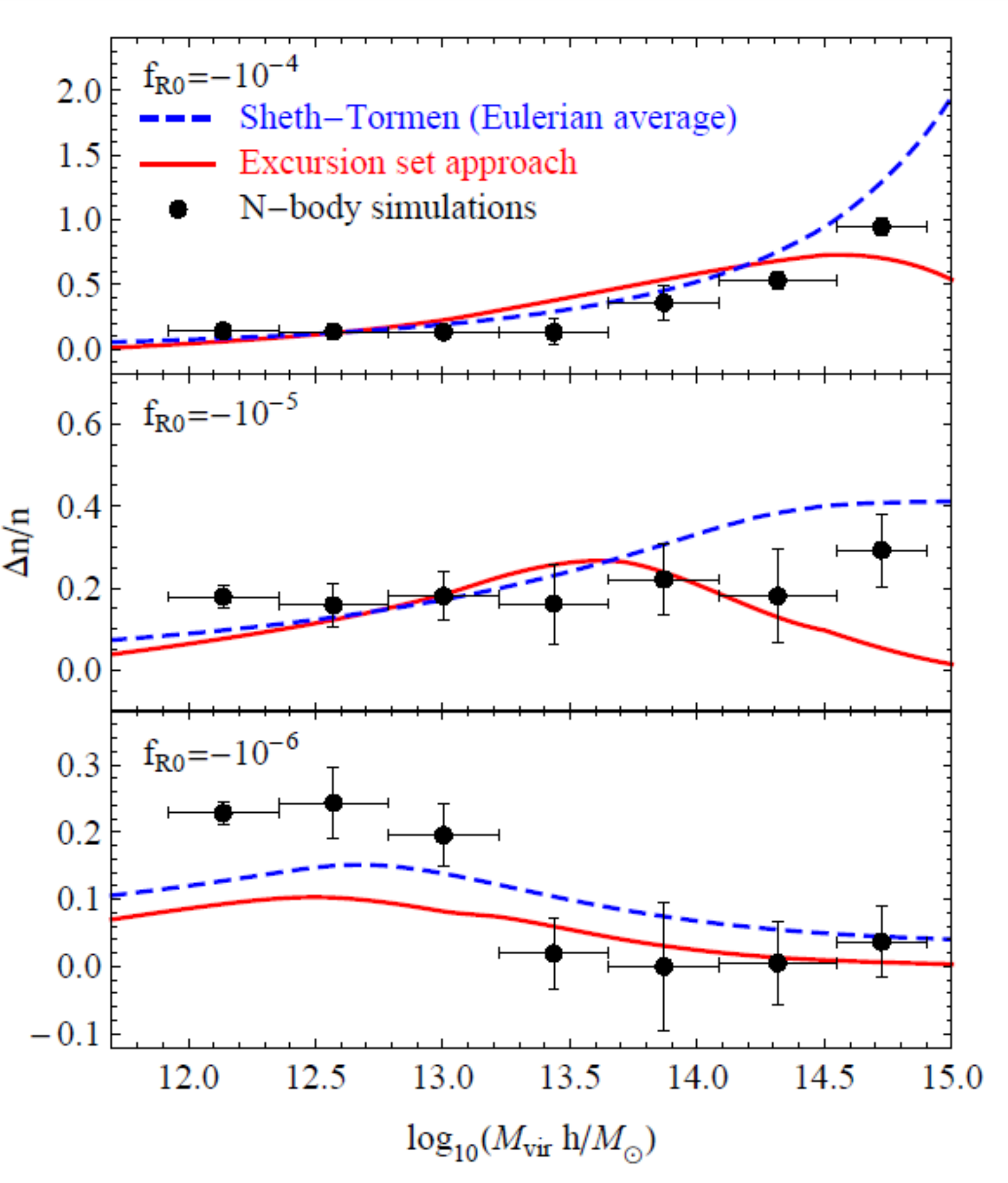}
  }
  \caption{The relative enhancement of the halo mass function in $f(R)$ gravity models compared with $\Lambda$CDM. The solid line and dashed line are analytic predictions (see \cite{Lombriser:2013wta} for the detail). From \cite{Lombriser:2013wta}.  
  }
\end{figure}
%%%%%%%%%%%%%%%%%%%%%%%%%%%%%%%%%%%%%%%%%%%%%%%%%%%%%%%%%%%%%
The latest constraints on $|f_{R0}|$ from cluster abundance is obtained as $|f_{R0}|<2.6 \times 10^{-5}$ for $n=1$ in Eq.~(\ref{HS}) \cite{Cataneo:2014kaa}. 

If the screening length is smaller than the radius of the dark matter halo, we will see the transition from the screened and unscreened region through the difference between the lensing and dynamical masses \cite{Terukina:2012ji} . We assume that a cluster of galaxies satisfies hydrostatic equilibrium 
\begin{equation}
\frac{d P_{tehrmal}}{d r} = - \frac{G M(<r)}{r} - \frac{\beta}{\mpl} \frac{d \phi}{dr},
\label{equi}
\end{equation}
where $P_{thermal}$ is the thermal pressure of the gas. The enclosed mass $M(<r)$ agrees with the lensing mass as the scalar field does not affect the lensing. The last term in Eq.~(\ref{equi}) is the contribution from the fifth force. This changes the dynamical mass, which can be inferred from the pressure gradient of the gas. Fig.~24 shows the enclosed mass reconstructed from X-ray observations using the hydrostatic equilibrium equation with the chameleon force for the Coma cluster \cite{Terukina:2013eqa}. It also shows the measurement of the lensing mass for the Coma cluster \cite{Okabe:2010ue}. Since the chameleon gravity adds an additional contribution to the dynamical mass, the lensing mass estimated from the X-ray observation decreases. In this way, we can put constraints on the chameleon gravity. Using the Coma cluster \cite{Terukina:2013eqa} and the stacked 58 clusters between $0.1 < z <1.2$ using X-ray data from XMM Cluster Survey and weak lensing data from the CFHTLenS \cite{Wilcox:2015kna}, the constraint on $f(R)$ gravity was obtained $|f_{R0}| < 6 \times 10^{-5}$.
%%%%%%%%%%%%%%%%%%%%%%%%%%%%%%%%%%%%%%%%%%%%%%%%%%%%%%%%%%%%
\begin{figure}[h]
  \centering{
  \includegraphics[width=11cm]{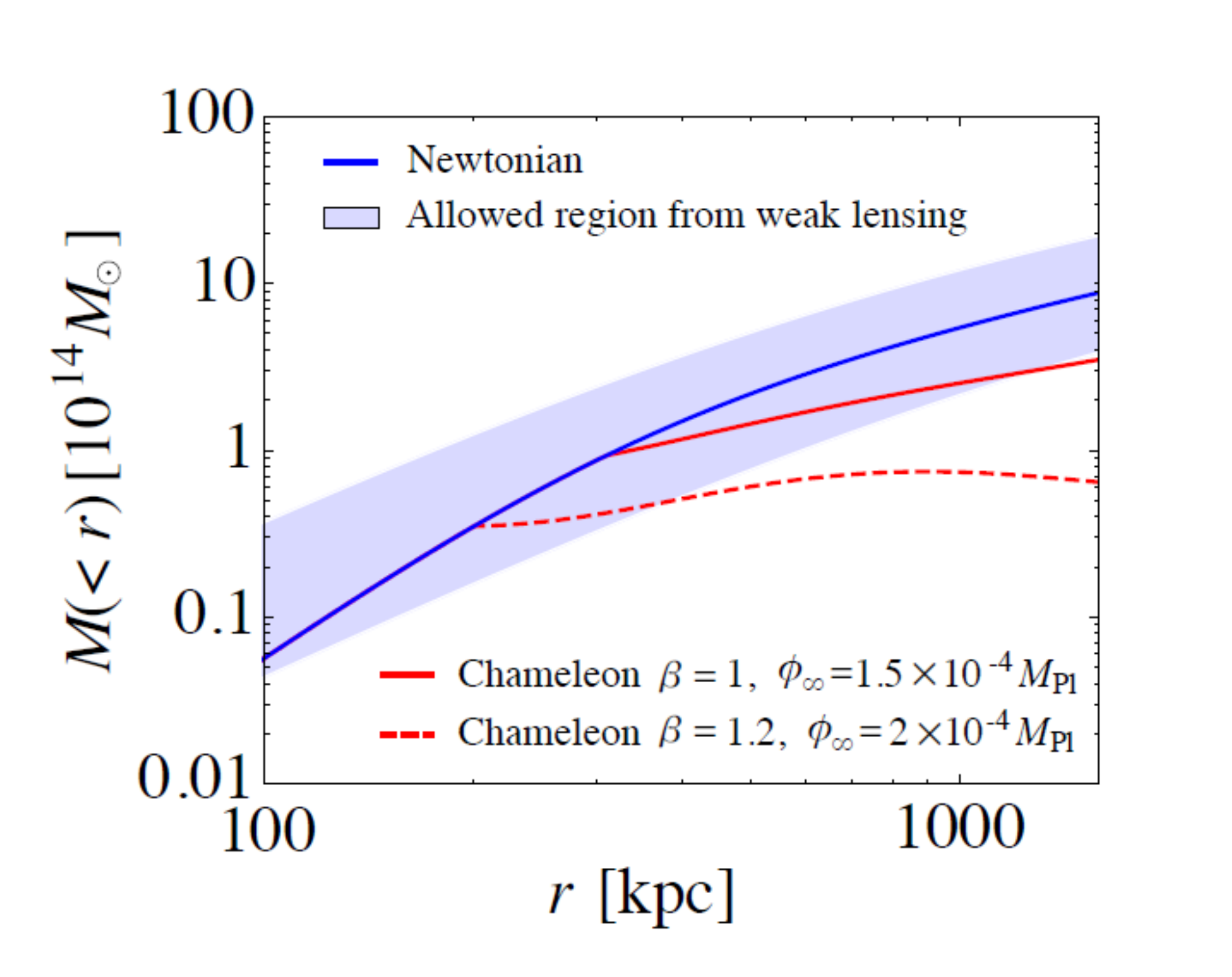}
  }
  \caption{Radial mass profile of the Coma cluster. The shaded region is the observationally allowed $1\sigma$ region of the WL observation. The blue solid curve is the thermal mass component estimated from the X-ray observations. The red solid and dashed curves are the combination of the termal and chameleon mass components. From \cite{Terukina:2013eqa} (published on 14 April 2014 \copyright SISSA Medialab Srl.  Reproduced by permission of IOP Publishing.  All rights reserved).
  }
\end{figure}
%%%%%%%%%%%%%%%%%%%%%%%%%%%%%%%%%%%%%%%%%%%%%%%%%%%%%%%%%%%%%

\subsubsection{Dwarf galaxies}
As is clear from Fig.~20 and Fig.~23, for $|f_{R0} |<10^{-6}$, cluster are effectively screened and we need to find objects with a shallower potential. 
Another place where gravity is strongly modified is provided by dwarf galaxies in voids. Dwarf galaxies have a potential $|\Psi_N| \sim 10^{-7}$ thus they are not self-screened even with $|f_{R0}| =10^{-6}$. We need to carefully identity dwarf galaxies that are not environmentally screened. One way is to use the $D$ parameter introduced in the previous section (Eq.~(\ref{D})). Another environmental measure is defined in terms of the sum of gravitation potentials of neighbouring galaxies \cite{Cabre:2012tq}
\begin{equation}
\Phi_{\rm ext} = \sum_{d_i < \lambda_c + r_i} \frac{G M_i}{d_i},
 \end{equation}
where $d_i$ is the 3D distance to the neighbouring galaxy with mass $M_i$ and the virial radius $r_i$, $\lambda_C$ is the background Compton wavelength of the scalar field. Using the thin shell condition (\ref{fr-thishel}), we select the environmentally unscreened galaxies 
\begin{equation}
\Phi_{\rm ext} < \frac{3}{2} |f_{R0}|.
\label{cut1}
\end{equation} 
This parameter is more suited to identify the environment of galaxies when uncertainties in the estimation of mass and distance and the incompleteness of the samples are included compared with $D$ \cite{Cabre:2012tq}. Using this criteria a {\it screening map} has been created using SDSS galaxies within 200 h$^{-1}$Mpc \cite{Cabre:2012tq} (see Fig.~25 and 26). We also introduce the self-unscreening condition 
\begin{equation}
\Phi_{\rm int} < \frac{3}{2}|f_{R0}|,
\label{cut2}
\end{equation}
where $\Phi_{\rm int} = G M/R$, $M$ is the mass and $R$ is the virial radius of a halo. 
%%%%%%%%%%%%%%%%%%%%%%%%%%%%%%%%%%%%%%%%%%%%%%%%%%%%%%%%%%%%
\begin{figure}[h]
  \centering{
  \includegraphics[width=16cm]{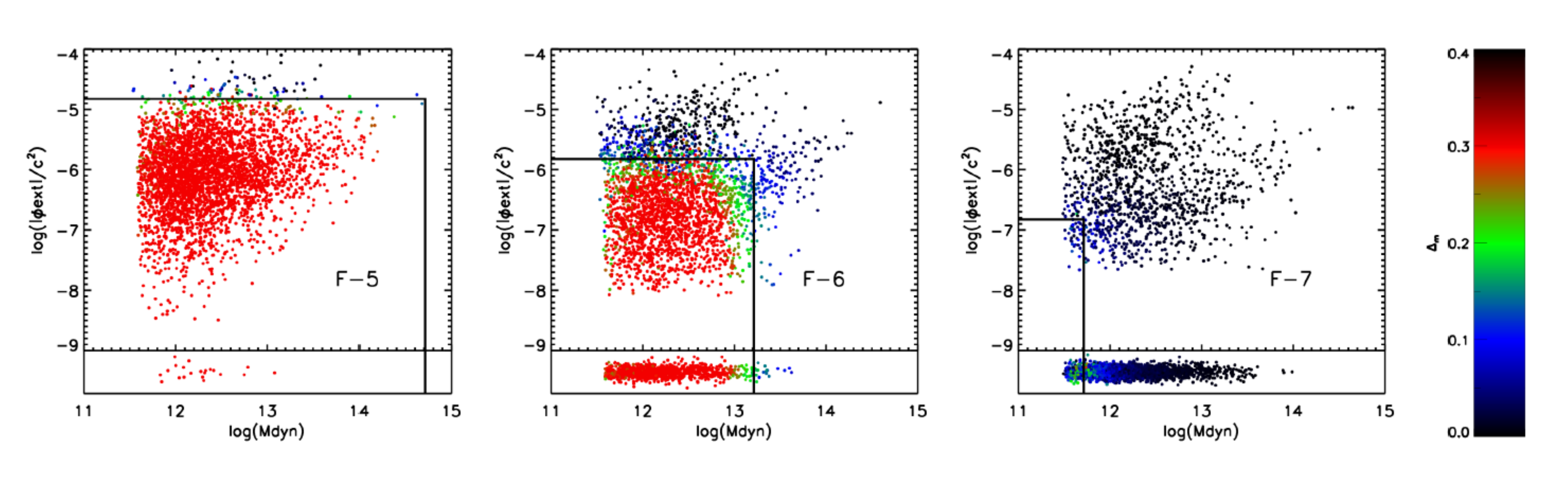}
   }
   \caption{Scatter plot for $\Phi_{\rm ext}$ vs. the dynamical mass $M_{\rm dyn}$ (equivalent to $|\Phi_{\rm int}|$), coloured accordingly to the difference between the dynamical and lensing mass $\Delta_M$. A simple cut in $\Phi_{\rm ext}$  and $M_{\rm dyn}$ given by Eqs.~(\ref{cut1}) and (\ref{cut2}) seem to work to separate screened from unscreened galaxies. From \cite{Cabre:2012tq} (18 July 2012 \copyright SISSA Medialab Srl.  Reproduced by permission of IOP Publishing.  All rights reserved).} 
\end{figure}
%%%%%%%%%%%%%%%%%%%%%%%%%%%%%%%%%%%%%%%%%%%%%%%%%%%%%%%%%%%%%
 
%%%%%%%%%%%%%%%%%%%%%%%%%%%%%%%%%%%%%%%%%%%%%%%%%%%%%%%%%%%%
\begin{figure}[h]
  \centering{
  \includegraphics[width=16cm]{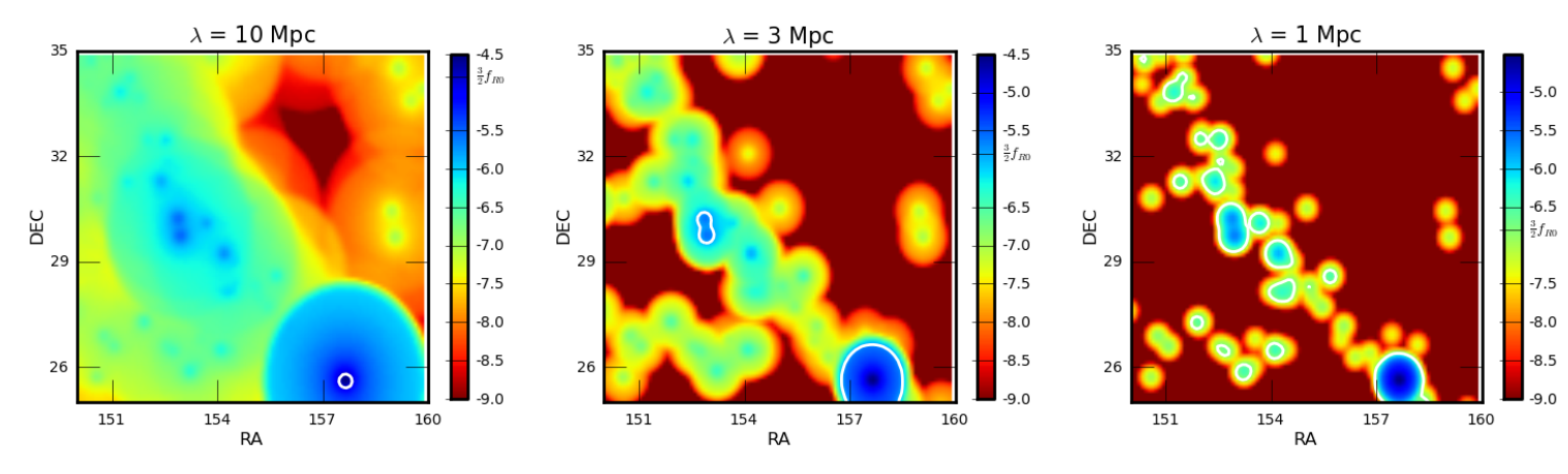}
  }
  \caption{The environmental screening map generated in the SDSS region at $210$Mpc for $10 \times 10$ degree of survey area, approximately $38 \time 38$ Mpc, with 2 arcmin resolution. The external potential  $\Phi_{\rm ext}$ is shown with the screening condition evaluated using Eq.~(\ref{cut1}) for models with $|f_{R0}|=10^{-5}, 10^{-6}$ and $10^{-7}$ (left, middle and right panels). The cut in screening classification $3/2 |f_{R0}|$ is shown in the map as a white contour line (regions inside the contour are screened). From \cite{Cabre:2012tq}  (18 July 2012 \copyright SISSA Medialab Srl.  Reproduced by permission of IOP Publishing.  All rights reserved).
  }
  \label{fig:dgp_env}
\end{figure}
%%%%%%%%%%%%%%%%%%%%%%%%%%%%%%%%%%%%%%%%%%%%%%%%%%%%%%%%%%%%%

There are a number of effects that we expect to see in unscreened galaxies 
summarised in \cite{Vikram:2013uba}. 

\begin{itemize}
\item
Offset between stellar and gaseous component.\\
In the presence of the external scalar field, unscreened components such as dark matter or neutral hydrogen (HI) will fall faster along the external field than will stars that are self-screened. This difference may cause an offset between the centroid of the stellar disk and the gaseous disk. 

\item
Warping of the stellar disk. \\
As the dark matter halo moves along the external potential gradient, the screened stellar component lags behind the dark matter halo. If the potential gradient is aligned with the axis of rotation, this may introduce a U-shaped warp in the stellar disk. The warp is expected to align with the potential gradient. 

\item 
Asymmetry in stellar rotation curve. \\
The offsets between stellar and halos centres can perturb the stellar disk and cause an asymmetry in the stellar rotation curve as the dominant force on the stellar disk is from the potential of dark matter halos.

\item
Mismatch of the rotation velocities of a gaseous and stellar disk\\
The rotation velocity of an unscreened gaseous disk 
is enhanced relative to the screened stellar disk by a factor of $\sqrt{1+ \Delta G/G}$. This test requires us to derive the stellar rotation curves from stellar absorption lines instead of using H$\alpha$ or the 21 cm line as these lines prove the unscreened gaseous components of the galaxies. 

\end{itemize}

All these tests have been carried out recently \cite{Vikram:2013uba,Vikram:2014uza}. The absence of these effects give the strong constraint $|f_{R0}| < 10^{-6}$. These constraints can be improved by on-going surveys such as Sloan Digital Sky Survey (SDSS)IV Mapping Nearby Galaxies at APO (MaNGA) which will find many unscreened galaxies even if $|f_{R0}| =10^{-6}$ \cite{MaNGA-g}. 

There are also interesting implications for dwarf galaxies in the Milky way if the screening radius $r_{\rm scr}$ of the Milky way is $60$ kpc \cite{Lombriser:2014nfa}.  

\subsubsection{Stars}
Another astrophysical object which has a shallow potential is a post main-sequence star whose potential is $|\Psi_{N}|\sim 10^{-7}$. The stellar structure is again determined by the hydrostatic equilibrium. The effect of the fifth force can be approximated by Eq.~(\ref{halofr}). Due to the enhanced gravity, stars are expected be more luminous \cite{Chang:2010xh, Davis:2011qf}.
This modified equation has been implemented in the stellar structure code MESA \cite{Sakstein:2015oqa}. Assuming that a star is in an unscreened galaxy, the evolution of a 1 solar mass star in the Hertzprung-Russell diagram is shown in Fig.~27 in the case of $f(R)$ gravity. For $|f_{R0}|=0.67 \times 10^{-7}$, the main-sequence stars are screened and the modified gravity affects only red giants. For  $|f_{R0}|=0.67 \times 10^{-6}$ and $3.35 \times 10^{-6}$, main sequence stars are also affected. 

%%%%%%%%%%%%%%%%%%%%%%%%%%%%%%%%%%%%%%%%%%%%%%%%%%%%%%%%%%%%
\begin{figure}[h]
  \centering{
  \includegraphics[width=13cm]{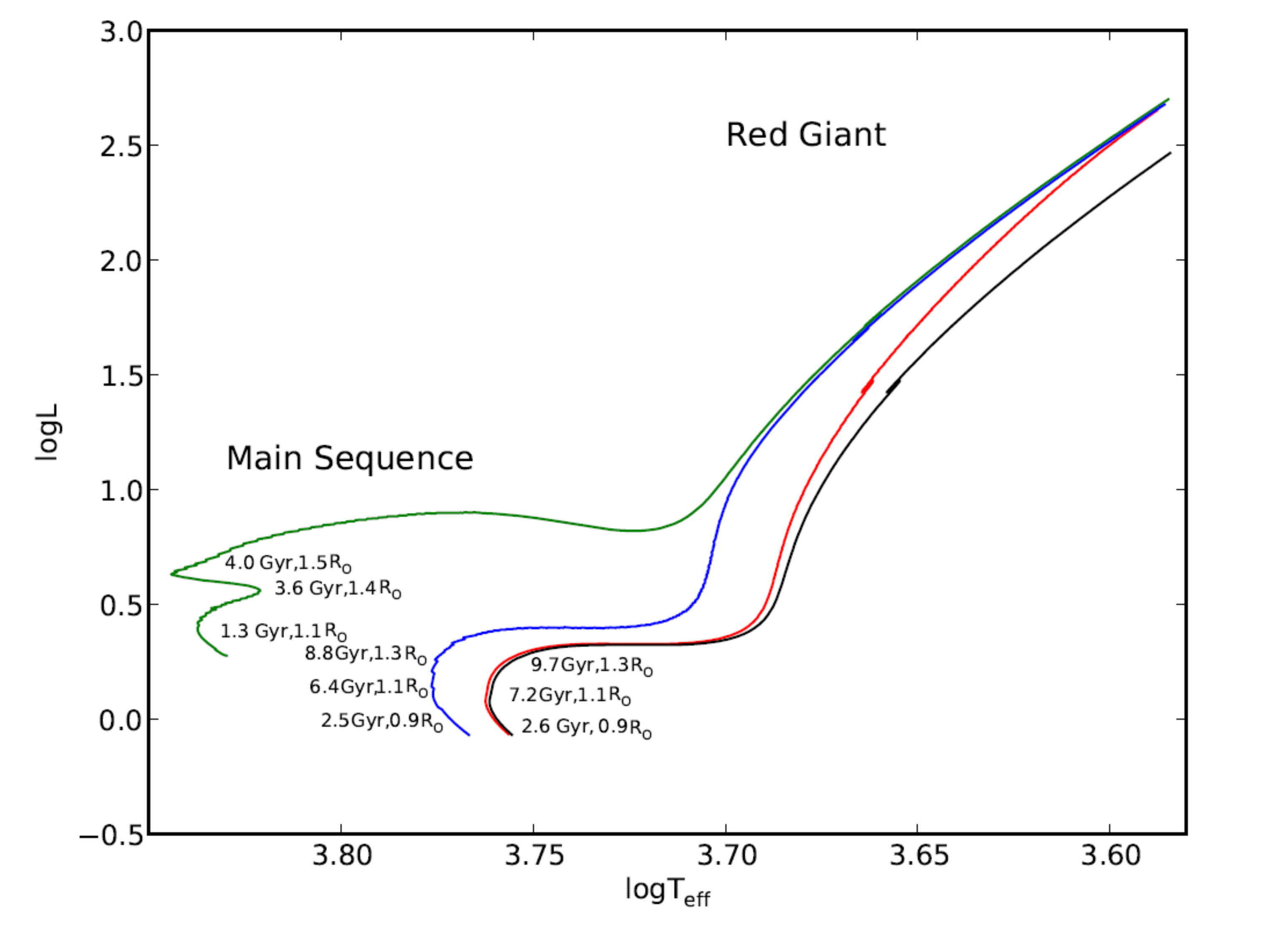}
  }
  \caption{The Hertzprung-Russell diagram for stars of one solar mass with initial metallicity $Z=0.02$ in $f(R)$ gravity. The black line shows the tracks for stars in GR while the red, blue and green tracks correspond to stars in modified gravity models with the chameleon mechanism with $|f_{R0}| =0.67 \times 10^{-7}, 0.67 \times 10^{-6}$ and $3.35 \times 10^{-6}$. From \cite{Davis:2011qf}.}
  \label{fig:dgp_env}
\end{figure}
%%%%%%%%%%%%%%%%%%%%%%%%%%%%%%%%%%%%%%%%%%%%%%%%%%%%%%%%%%%%%
This modification of the stellar structure was used to put a strong constraint on the chameleon mechanism \cite{Jain:2012tn}. Cepheid variable stars with $5-10 M_{\astrosun}$ pulsate with a known period-luminosity relation once they go off the main sequence and situated in a narrow temperature gap in the HR diagram. This property has been used to measure the distance to the Cepheids. The period $\tau$ is determined by the strength of gravity $\tau \propto G^{-1/2}$. This relation is calibrated using local group stars that are screened. Hence if one uses this relation to estimate the distance, then the derived distance will be incorrect. If gravity is stronger than GR, the period of Cepheids becomes shorter. It is found that the change in the distance is given by 
\begin{equation}
\frac{\Delta d}{d} = - 0.3 \frac{\Delta G}{G}.
\end{equation}    
If we measure the distance using Cepheid assuming GR, the measured distance will be shorter than the actual distance. In order to test this, we need another distance indicator that is insensitive to gravity and measure the correct distance. The tip of the red giant branch (TRGB) distance is such a distance indicator as the absolute magnitude of red giants that go off the red giant branch after a helium flash is insensitive to gravity. The comparison between the two distances has been performed and compared against the prediction of the chameleon mechanisms. This led to the stringent constraint $|f_{R0}| < 5 \times 10^{-7}$.   

\subsubsection{Summary of constraints}
Fig.~28 summarises the constraints on chameleon gravity in the case of $f(R)$ gravity. A detailed summary of constraints is available in Ref~\cite{Lombriser:2014dua}. 

%%%%%%%%%%%%%%%%%%%%%%%%%%%%%%%%%%%%%%%%%%%%%%%%%%%%%%%%%%%%
\begin{figure}[h]
  \centering{
  \includegraphics[width=12cm]{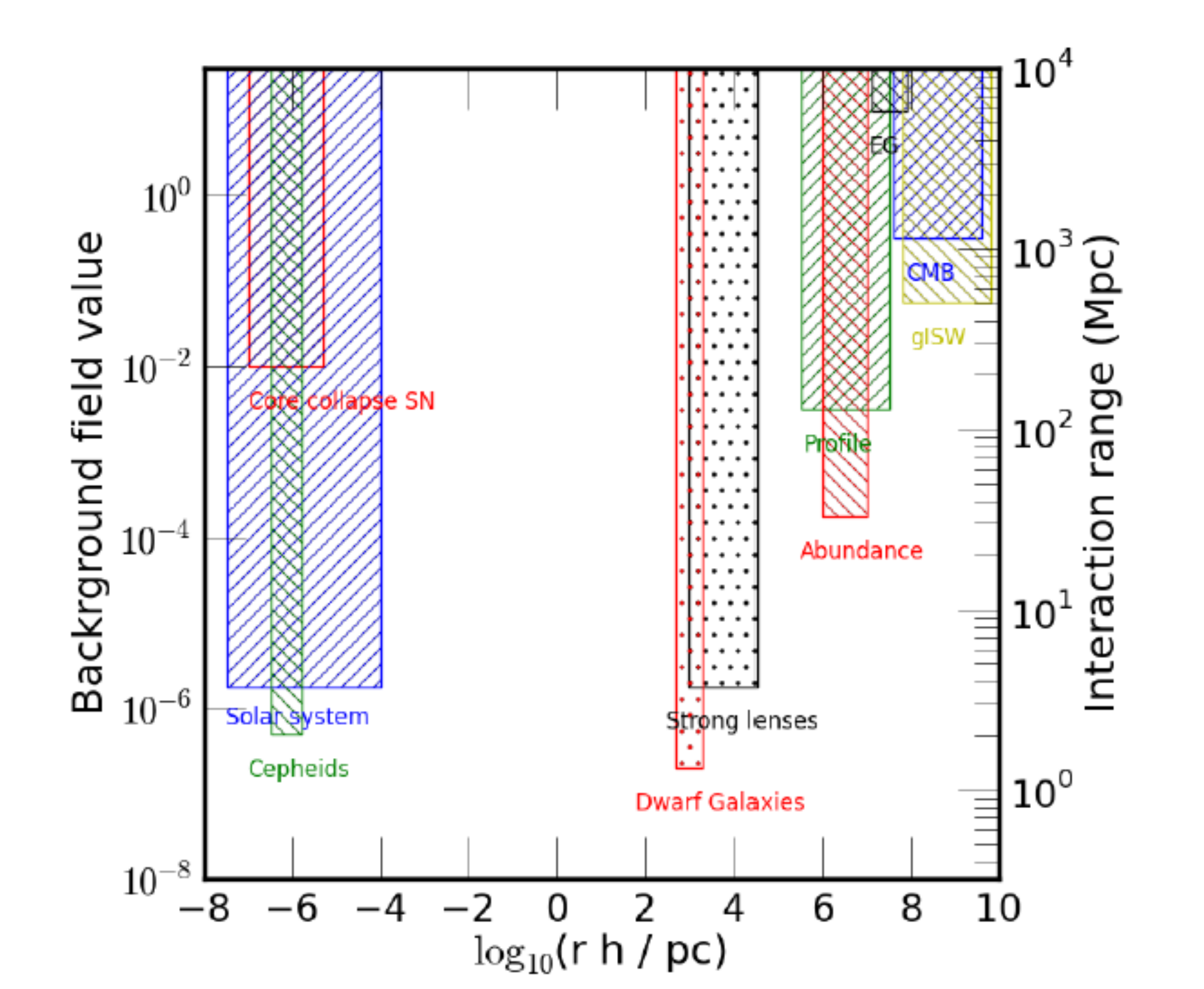}
  }
  \caption{Astrophysical and cosmological limits on chameleon theories. The spatial scale on the x-axis gives the range of the length scales probed by particular experiments. The parameter on the y-axis is the background field $|f_{R0}|$ value or the range of the interaction for an $f(R)$ gravity model.
  The rectangular region gives the exclusion zone. From \cite{Jain:2013wgs}.  }
  \label{fig:dgp_env}
\end{figure}
%%%%%%%%%%%%%%%%%%%%%%%%%%%%%%%%%%%%%%%%%%%%%%%%%%%%%%%%%%%%%

Constraints from cosmological probes such as CMB and large scale structure are relatively weak $|f_{R0}| < 10^{-2} - 10^{-4}$. Clusters give an order of magnitude stronger constraint $|f_{R0}| < 10^{-5}$ which requires the understanding of non-linear structure formation. In order to satisfy the Solar System constraints, the Milky Way needs to be screened. This condition depends on whether the Local Group gives the environmental screening and the constraint is given by $|f_{R0}| <10^{-4} - 10^{-6}$. Constraints from dwarf galaxies and Cepheid are much stronger $|f_{R0}| < 10^{-7}$. All the constraints except for the Solar System constraint have been obtained in the last 10 years illustrating impressive development of new tests of gravity on cosmological and astrophysical scales.  
 
\subsection{Vainshtein mechanism}
Unlike the chameleon mechanism, it is very difficult to test models with the Vainshtein mechanism on small scales. The screening of dark matter halos does not depend on mass nor environment and all dark matter halos are screened inside the virial radius. In order to find modifications of gravity, we need to go beyond the virial radius of halos. 

An interesting test using black holes at the centre of galaxies has been proposed \cite{Hui:2012jb}. As we have seen in section 5.5, even if a dark matter halo is ``screened" and the fifth force is suppressed within the dark matter halo, it still feels the external gradient of the scalar field as it still carries a scalar charge. On the other hand, black holes do not carry a scalar charge due to the no-hair theorem. Thus the central black hole will lag behind the dark matter halos and there may be a displacement of the central black holes. The magnitude of this effect depends on the external field and the strength of the fifth force and it is estimated as \cite{Hui:2012jb}
\begin{equation}
r= 0.1 \; \mbox{kpc}  \times 2 \beta^2 
\left(   
\frac{|\nabla \Psi_{\rm ext} |}{20 (km/s)^2 /\mbox{kpc}} \right)
\left( \frac{0.01 \mbox{Mpc}^{-3}}{\rho_0} \right),
\end{equation}
where $\rho_0$ is the central density of the galaxies. 

In beyond Horndeski theories discussed in section 2.5, the Vainshtein mechanism is broken inside matter. The Poisson equation within the Vainshtein radius is given by
\begin{equation}
\frac{d \Psi(r)}{dr} =
\frac{G M(<r)}{r^2} + \frac{\Upsilon}{4} G \frac{d^2 M(<r)}{dr^2}, 
 \end{equation}
where $\Upsilon$ characterises the deviation from the Horndeski theory. The last term modifies GR even within the Vainshtein radius. This for example changes the structure of stars, rotation velocities of galaxies and lensing \cite{Koyama:2015oma, Saito:2015fza}. 

Observational tests have not been carried out yet and there remains to be seen if we can get tighter constraints on the model parameter such as $r_c$ compared with cosmological observations and the Solar System constraint from these astrophysical tests.  

\section{Conclusion}
The discovery of the accelerated expansion of the Universe has come
relatively late in our study of the cosmos, but in showing that gravity can act repulsively, it has opened up many new questions about the nature of gravity and what the Universe might contain. Is the acceleration being
driven by dark energy?  Or is general relativity itself in error,
requiring a modification at large scales to account for the late
acceleration? 

Modified gravity models have provided many interesting new ideas to solve the cosmological constant problem by modifying the way in which the vacuum energy gravitates. The expansion of the Universe can accelerate without the cosmological constant due to the modification of gravity on cosmological scales. Gravitons may have a mass and provide an effective cosmological constant realising the technical naturalness of the smallness of the cosmological constant. Einstein equations are the only second-order local equations of motion for metric derivable from the action in 4D. If we modify GR, a new degree of freedom appears in the graviton, modifying gravity even in the Solar System. One of the most important recent developments are the screening mechanisms that enable us to modify gravity significantly on cosmological scales while satisfying the stringent Solar System constraints. Despite extensive study of modified gravity models, we still do not have a model that can be an alternative to the $\Lambda$CDM model. Our quest to find modified gravity models still continue. 

Structure formation in our Universe can be different even if
the geometry of the homogeneous and isotropic universe is the same in these two classes of models, offering a possibility to distinguish between them observationally. It is in principle possible to construct non-parametric consistency tests of GR on cosmological scales by combining various probes of large-scale structure, because the Einstein equations enforce a particular relation between observables. On large scales, there are two functions of time and space that are required to describe these relations. One describes the relation between the density perturbation and the Newton potential, which determines the motion of non-relativistic objects such as galaxies; the other parametrises the relation between the density perturbation and the lensing potential, which determines the geodesics of photons. In the next five years, new surveys such as DES and eBOSS will be able to constrain several parameters at the 5-10$\%$ level. This shows that upcoming surveys will make cosmological tests of gravity a reality in the next five years. Surveys such as Euclid will further make cosmological tests of gravity a precision test providing constraints on these parameters at the 1$\%$ level. 

It is possible to construct model independent tests on linear scales, but the bulk of information is available on non-linear scales. Also in order to extract the information of the linear growth rate of structure, it is required to model the non-linear corrections accurately.  The non-linear structure formation is complicated by the screening mechanisms that restore GR on small scales. Perturbation theory and N-body simulations have enabled us to understand how structure formation is affected by the screening mechanisms. The screening mechanisms leave distinct signatures in the non-linear structure formation. This has led to the developments of novel astrophysical tests of gravity. In the case of the chameleon mechanism, the astrophysical tests using dwarf galaxies and stars provide stringent constraints on the model that are even stronger than the Solar System constraints.  

There are remaining issues in developing theoretical frameworks for probing gravitational physics with upcoming cosmological surveys and exploiting the constraints expected to be obtained. The questions that we need to address include: 

\begin{itemize}
\item What theoretical models should we be looking at? 
\item	Is there a better, more physically meaningful parameterisation of modifications to GR on cosmological scales? 
\item	How do we include non-linear physics either numerically or analytically? 
\item	How do we combine various cosmological observations to test gravity? 
\item	How limited are we to using linear scales? 
\item	Can we develop novel tests of gravity on non-linear scales?   
\end{itemize}

As we have seen in this review, we have started to answer these questions and have already obtained a number of new constraints on gravity on astrophysical and cosmological scales in the last ten years. The next ten years will be an exciting time for cosmological tests of gravity thanks to the on-going and future surveys. We may be able to detect the breakdown of GR that is responsible for the later acceleration of the Universe or GR will be confirmed as the theory of gravity on cosmological scales. 

\section*{Acknowledgement}
I would like to thank the authors of Figure. 1, 2, 6, 7 and 27 for granting permission to reproduce these figures in this article. I would like to thank all my collaborators who helped me form my thinking on this wide ranging topic. I thank  Marco Crisostomi, Bridget Falck, Lucas Lombriser, Gianmassimo Tasinato and Gong-bo Zhao for useful comments and Matthew Hull and Benjamin Bose for a careful reading of the manuscript. KK was supported by the European Research Council starting grant ``Modified Gravity as an Alternative To Dark Energy" (2008-2013). KK is supported the European Research Council consolidator grant ``Cosmological Tests of Gravity" (646702) and the UK Science and Technology Facilities Council (STFC) grants ST/K00090/1 and ST/L005573/1.

\section*{References}

\bibliography{reviewV2}{}
\bibliographystyle{unsrt}

\end{document}